**TRACING AND SEGMENTATION OF MOLECULAR PATTERNS IN 3-DIMENSIONAL**

**CRYO-ET/EM DENSITY MAPS THROUGH ALGORITHMIC IMAGE PROCESSING**

**AND DEEP LEARNING-BASED TECHNIQUES**

by

Salim Sazzed
B.S. October 2009, University of Dhaka, Bangladesh
M.S. November 2011, University of Dhaka, Bangladesh

A Dissertation Submitted to the Faculty of
Old Dominion University in Partial Fulfillment of the
Requirements for the Degree of

DOCTOR OF PHILOSOPHY

COMPUTER SCIENCE

OLD DOMINION UNIVERSITY
December 2023

Approved by:

Jing He (Director)

Willy Wriggers (Member)

Jiangwen Sun (Member)

**ABSTRACT**

TRACING AND SEGMENTATION OF MOLECULAR PATTERNS IN 3-DIMENSIONAL
CRYO-ET/EM DENSITY MAPS THROUGH ALGORITHMIC IMAGE PROCESSING AND
DEEP LEARNING-BASED TECHNIQUES

Salim Sazzed
Old Dominion University, 2024
Director: Dr. Jing He

Understanding the structures of biological macromolecules is highly important as they are closely associated with cellular functionalities. Comprehending the precise organization of actin filaments is crucial because they form the dynamic cytoskeleton, which offers structural support to cells and connects the cell's interior with its surroundings. However, determining the precise organization of actin filaments is challenging due to the poor quality of cryo-electron tomography (cryo-ET) images, which suffer from low signal-to-noise (SNR) ratios and the presence of missing wedge, as well as diverse shape characteristics of actin filaments. To address these formidable challenges, the primary component of this dissertation focuses on developing sophisticated computational techniques for tracing actin filaments. In particular, three novel methodologies have been developed: i) BundleTrac, for tracing bundle-like actin filaments found in Stereocilium, ii) Spaghetti Tracer, for tracing filaments that move individually with loosely cohesive movements, and iii) Struwwel Tracer, for tracing randomly orientated actin filaments in the actin network. The second component of the dissertation introduces a convolutional neural network (CNN) based segmentation model to determine the location of protein secondary structures, such as helices and $\beta$-sheets, in medium-resolution (5-10Å) 3-dimensional cryo-electron microscopy (cryo-EM) images. This methodology later evolved into a tool named DeepSSETracer. The final component of the dissertation presents a novel algorithm, cylindrical fit measure, to estimate image/structure match at helix regions in medium-resolution cryo-EM images. Overall, my dissertation has made significant contributions to addressing critical research challenges in structural biology by introducing various computational methods and tools.

To my wife and parents, your love and support have been my driving force throughout this academic journey. This dissertation is a tribute to the strength and inspiration you've provided. Thank you for being my unwavering pillars

# ACKNOWLEDGMENTS

I would like to express my heartfelt gratitude to several individuals who have played pivotal roles in my Ph.D. journey at Old Dominion University. First and foremost, I extend my sincere appreciation to my Ph.D. supervisor, Dr. Jing He, and my project supervisor, Dr. Willy Wriggers. Their unwavering guidance, encouragement, and consistent support throughout my doctoral years have been instrumental in shaping my research trajectory. Their advice spanned various facets of my academic endeavors, encompassing critical analysis of existing literature, development of my research approach, and the art of scholarly paper composition and presentation. Engaging in discussions with them has always been enlightening, and I consider myself extremely fortunate and honored to have had the privilege of earning my degree under their guidance. In addition to my research aspects, my supervisor, Dr. Jing He supported me a lot during the challenging time of my Ph.D when two of my babies were born. Dr. He was always very supportive during that difficult time and allowed me the necessary time off to take care of the babies. Dr. Willy Wriggers helped me a lot during my job search by introducing me to many people through in-person communication, emails,etc. I am indebtful to both of them for their support. I am also deeply thankful to my other Ph.D. committee member, Dr. Jiangwen Sun. The expertise of Dr. Jinagwan Sun in deep learning helped me a lot throughout the protein secondary structure project through stimulating discussions and suggestions. I extend special thanks to my collaborators, Dr. Manfred Auer, Dr. Julio Kovacs, and Jun Ha Song. Especially, I would like to remember Junha Song, who suddenly passed away in 2023. He was a great soul and helped me a lot in the actin filament tracing project (particularly in BundleTrac project). He will be always remembered. Last but certainly not least, I wish to express my heartfelt gratitude to my wife, Toma Islam, my parents, and other family members for their unwavering support and encouragement throughout the years. Their steadfast belief in me has been a constant source of motivation. To all these individuals and groups, I extend my deepest thanks for their pivotal roles in my academic journey.

# TABLE OF CONTENTS

# LIST OF TABLES

# LIST OF FIGURES

# CHAPTER 1

# INTRODUCTION

A precise quantitative description of the characteristics of the biological specimens is often key to understanding their functionalities. This is particularly true for dynamic supramolecular assemblies, such as filaments or protein secondary structures. My dissertation focuses on determining and comprehending the presence and organizations of these two important biological structures: i) actin filament and ii) protein secondary structure in low-resolution cryo-electron tomography (cryo-ET) and medium-resolution cryo-electron microscopy (cryo-EM) density images, respectively.

## 1.1 TRACING ACTIN FILAMENT IN LOW-RESOLUTION CRYO-ELECTRON TOMOGRAPHY (CRYO-ET) IMAGE

Actin filaments have roles in various functions such as sensing environment, cell motility, cell integrity, cytokinesis, tissue formation, and maintenance [1]. For example, our senses of hearing and balance rely on the proper functioning of hair cells [2, 3]. Mechanoelectrical transduction occurs at the hair bundle located at the apical pole of the hair cell. The hair bundle comprises of individual organelles known as stereocilia. Stereocilia are actin bundle-filled membrane protrusions of the apical hair cell surface that develop simultaneously with the maturation of hair cells during the development or regeneration of the inner ear [4]. Researchers are still exploring how each stereocilium adopts its defined length and width and how they organize into hair bundles with species and organ-specific characteristic shapes. All these research problems are incredibly challenging to figure out as over one hundred proteins affect hair bundle formation, function, and maintenance [5, 6].

Tracing actin filaments in the tapper region of the Stereocilium is even more challenging as the spacing of actin filaments maintained by actin–actin crosslinkers disappear in the taper region, where most filaments terminate on or near the plasma membrane in a systematic way, forming the taper [7].

The actin filaments in the actin network that forms the Filopodia are highly complex to trace since they are highly dynamic in nature (i.e., random orientations and length). The finger-like extensions of the cell surface of actin networks are involved in sensing the environment [8]. Filopodia have crucial roles in cell migration, neurite outgrowth, and wound healing and serve as precursors for dendritic spines in neurons [9].

The recent advances in high-throughput cryo-electron tomography (cryo-ET) have enabled the acquisition of biological complexes in their native state. However, the 3D cryo-ET images exhibit low resolution (3-5 nm) and considerable anisotropic noise and artifacts because of the fixed electron dose and the limited tilt series that masks out a wedge in Fourier space. Given the noise, missing wedge artifacts [10], and the enormous amount of information present in cryo-ET reconstructions, tracing cytoskeletal filaments is a non-trivial task.

The manual procedure of annotation of the cryo-ET map is a highly labor-intensive process; it may take over 60 hours or more human labor (depending on the map resolution and complexity) to annotate filaments in a dataset [11]. In addition, the annotation scheme of the structural features in the reconstructed tomograms is highly complicated due to crowded cellular systems and low SNR. Besides, the ambiguity of the filamentous structures in the tomograms makes the interpretation highly subjective, likely to be biased by the comprehension and expertise of the annotator(s) as actin filaments are often minimally visible (or even non-distinguishable) in the 3D tomograms. Hence, automatic or semi-automatic approaches for the detection and tracing of actin filaments are highly desirable for model building and understating the formation of complex macromolecules such as actin filaments.

This dissertation focuses on introducing novel computational techniques for tracing filaments in Stereocilia and Filopodia. Due to the heterogeneity in shape characteristics of actin filaments, it is highly unlikely that a single method functions satisfactorily for all cases; thus, specialized techniques are needed to exploit the biological characteristics of various types of filaments. Accordingly, to deal with actin filaments of diverse shapes, multiple computational techniques and tools have been developed. Besides, in addition to the experimental map, simulated maps were utilized to assess the performance of the tracing methods as it provides a better ground truth.

The main contributions of this dissertation related to actin filament tracing are the development of the following computational methodologies-

1) **BundleTrac**: A computational tool, BundleTrac, is developed to trace actin filaments that mimics hexagonal bundle patterns. The proposed algorithm employs a hexagonal filter that imitates the relative orientations of 7 filaments within a neighborhood. It was demonstrated that the proposed algorithm can effectively trace actin filaments in an actin bundle and yield high agreement with the manual annotation.

2) **Spaghetti Tracer**: Spaghetti Tracer, another computational tool, is proposed for tracing actin filaments in the taper region of stereocilia. Unlike the BundleTrac, which requires a hexagonal-shaped bundle representing multiple filaments to trace the center one, the Spaghetti Tracer traces individual filaments separately. SpaghettiTracer is specially designed for tracing filaments in the taper region, where filament movements may not be collective. Utilizing a path-based density accumulation algorithm, Spaghetti Tracer generates initial filament segments and then fuses those to yield the final filament traces.

3) **Struwwel Tracer**: Finally, a robust and generalizable multi-directional filament tracing tool, Struwwel Tracer, is introduced, capable of tracing filaments in a highly dynamic actin network. Unlike the BundleTrac or Spaghetti Tracer, which assume a dominant directional movement of filaments, Struwwel Tracer can trace filaments moving in any direction. It follows a similar principle to Spaghetti Tracer; it employs a locally-optimized dynamic programming (DP) algorithm for outlining filament segments in various directions.

## 1.2 DETECTION AND QUALITY ASSESSMENT OF PROTEIN SECONDARY STRUCTURE IN CRYO-EM IMAGE

In addition, this dissertation focuses on introducing specific methodologies for the detection and quality assessments of protein secondary structure.

The first component of this project concentrates on the detection of protein secondary structures from medium-resolution cryo-electron microscopy (cryo-EM) images. Although the cryo-EM technique has been successfully used to derive atomic structures for many proteins, it is still challenging to derive atomic structures when the resolution of cryo-EM density maps is below 5Å. My research presents a deep learning-based approach to segment secondary structure elements as helices and $\beta$-sheets from medium-resolution density maps. An effective optimization function is developed and employed to deal with the imbalance class issue present in the dataset. The proposed 3D convolutional neural network was trained and tested using a pool of 1350 3D images extracted from experimentally derived cryo-EM density maps. A test using 33 cases shows overall residue-level F1 scores of 0.76 and 0.60 to detect helices and $\beta$-sheets, respectively. The chains were selected from three different quality bins: Bin 1, Bin 2, and Bin 3, with a distribution of 8, 13, and 12 cases, respectively. The results show that the F1 score of helix detection for cases in the highest-quality bin differs by about 0.1 on average from cases in the next quality bin.

The second task focuses on the quantification of the quality of the helix, the most prominent type of protein secondary structure, in the medium-resolution cryo-EM image. Cryo-electron microscopy (cryo-EM) density maps at medium resolution (5–10 Å) reveal secondary structural features such as $\alpha$-helices and $\beta$-sheets. However, they lack the side chain details that would enable a direct structure determination. Among the more than 800 entries in the Electron Microscopy Data Bank (EMDB) of medium-resolution density maps that are associated with atomic models, a wide variety of similarities can be observed between maps and models. To validate such atomic models and to classify structural features, a local similarity criterion, the F1 score, is proposed and evaluated in this study. The F1 score is theoretically normalized to a range from zero to one, providing a local measure of cylindrical agreement between the density and atomic model of a helix. A systematic scan of 30,994 helices (among 3,247 protein chains modeled into medium-resolution density maps) reveals an actual range of observed F1 scores from 0.171 to 0.848, suggesting that the cylindrical fit of the current data is well stratified by the proposed measure.

# CHAPTER 2

# BACKGROUND AND RELATED WORKS

This chapter discusses essential concepts and existing research on actin filaments and protein secondary structure. Section 2.1 covers a comprehensive analysis of actin filaments, including their diverse structures, functionalities, significance, image acquisition process, and related studies. In section 2.2, various aspects of protein secondary structure, such as their structures, importance, and the cryo-EM image acquisition technique, are discussed.

## 2.1 ACTIN FILAMENTS IN CYTO-SKELETON

The cytoskeleton is a complex and dynamic network of interlinking protein filaments [12] found in the cytoplasm of all cells, excluding bacteria and archaea. It extends from the cell nucleus to the cell membrane and is composed of similar proteins in various organisms. The cytoskeleton plays a prominent role in numerous cellular physiological processes such as cell growth and migration [13, 14], proliferation and differentiation [15], and apoptosis [16]. Understanding the cytoskeleton dynamics is of prime importance for revealing mechanisms involved in cell adaptation. For example, as a skeleton of the cell, the cytoskeleton is responsible for giving structure and support to the cell.

In animal cells, the cytoskeleton is made up of three main types of proteins[12, 17]:

- Microfilament or Actin Filaments: The actin cytoskeleton is a highly dynamic structure that polymerizes and depolymerizes in response to various intracellular and extracellular stimuli. It is composed of a set of actin filaments, also known as microfilaments or F-acin, organized in a complex three-dimensional network spanning within the cell and is anchored to the extra-cellular matrix via transmembrane proteins and focal-adhesion-related proteins.

- Microtubules : Microtubules are the largest protein filaments found in the cytoskeleton. Microtubules are made of a protein called tubulin which has two subunits: alpha-tubulin and beta-tubulin. These protein subunits string together to form long protofilaments. To form a hollow straw-like structure of the microtubule, thirteen protofilaments come closer. Microtubules can grow and shrink by the addition or removal of tubulin proteins. The two ends of the microtubule are known as the plus (+) end and minus (-) end. Mechanical stimulation can cause significant variations in the geometry of the cells, triggering actin filaments polymerization/depolymerization to balance the applied extra-cellular forces.

- Intermediate Filaments: Intermediate filaments (IFs) are cytoskeletal structural components found in the cells of vertebrates and many invertebrates [17]. Homologs of the IF protein have been discovered in an invertebrate, the cephalochordate Branchiostoma. Intermediate filaments comprise a family of related proteins sharing common structural and sequence features. The term *intermediate* was assigned because the average diameter (10 nm) is between those of narrower actin microfilaments (7 nm) and wider myosin filaments (25 nm) found in muscle cells [18]. Animal intermediate filaments can be subcategorized into six types based on similarities in amino acid sequence and protein structure [19]. Most types are cytoplasmic, but one type, Type V is a nuclear lamin. Unlike microtubules, IF distribution in cells shows no good correlation with the distribution of either mitochondria or the endoplasmic reticulum [20].

### 2.1.1 Actin Filament: Structures and Functions

Actin filaments, also known as microfilaments or F-Actin, are slender protein filaments present in the cytoskeleton. These filaments consist of monomeric actin subunits that come together to form elongated fibers with a diameter of around 7 nm. Arranged like long spiral chains, they can create linear bundles, two-dimensional networks, and three-dimensional gels. Actin filaments possess two distinct ends, namely the plus(+) end and the minus(-) end. Studies suggest that actin filaments can exist in various states. Generally, an actin filament exhibits a total rise of 27.3 Å between subunits on adjacent strands and a rotation of 166.15° around the axis [21]. Additionally, these filaments are flexible, exhibit a helical repeat every 37 nm, and have a diameter ranging from 5 to 9 nm. Notably, there are 13 actin subunits between each cross-over point [21].

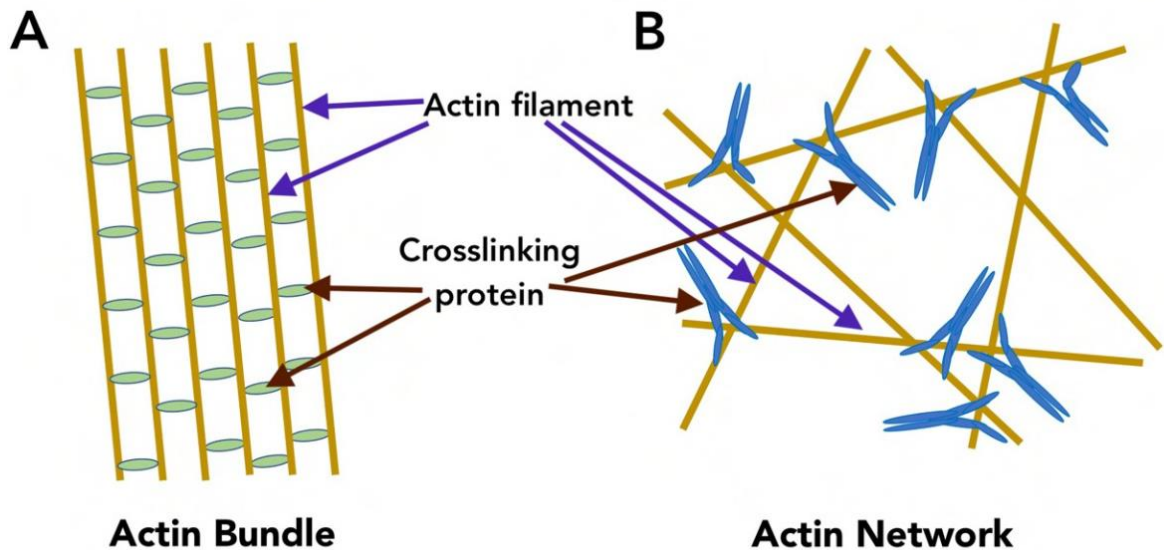

**Fig. 1.** Structures of actin bundle and actin network.

Individual actin filaments are typically assembled into two general types of structures, namely actin bundles and actin networks (Fig 1), and perform different roles in the cell. The varied structures and patterns of actin filaments are controlled by various actin-binding proteins that crosslink actin filaments. For example, actin filaments are crosslinked into closely packed parallel arrays to create the actin bundle, while in an actin network, the actin filaments are loosely crosslinked in orthogonal arrays that form three-dimensional meshworks.

Actin filaments play essential roles in various cellular processes. Some of the key roles of actin filaments are:

- Cell shape and structure: Actin filaments provide mechanical support to cells and help maintain their shape and structure.

- Cell division: Actin filaments are involved in cytokinesis, the process by which the cell divides into two daughter cells.

- Cell motility: Actin filaments are crucial for cell motility, which enables cells to move and change shape. This is important in a variety of biological processes, such as wound healing and immune cell response.

- Muscle contraction: Actin filaments, along with myosin, are responsible for muscle contrac-

tion in both smooth and striated muscles.

- Intracellular transport: Actin filaments are involved in the intracellular transport of organelles and vesicles within cells.

- Signal transduction: Actin filaments are involved in signal transduction pathways that regulate a variety of cellular processes, including gene expression and cell growth.

### 2.1.2 Cryo-Electron Tomography

Electron cryotomography (CryoET) is an imaging technique used to construct three-dimensional views of samples, typically of the biological macromolecules and cells, within a range of 2–4 nm [22, 23]. CryoET is a specialized application of transmission electron cryomicroscopy (CryoTEM) in which samples are imaged as they are tilted, resulting in a series of 2D images combined to produce a 3D reconstruction, similar to a CT scan of the human body.

### Image Acquisition

The CryoET technique involves immobilizing samples in non-crystalline ("vitreous") ice, which are then subjected to cryogenic conditions (below -150 $^\circ$ C). This approach enables imaging without the risks of dehydration or chemical fixation, both of which could otherwise cause disruptions or distortions to biological structures.

In transmission electron microscopy (TEM), the resolution is constrained by the thickness of the sample due to the strong interaction of electrons with matter. Additionally, tilting the sample increases its thickness, and excessive thickness can obstruct the electron beam, resulting in a dark or entirely black image. Hence, CryoET requires samples to be under 500 nm thickness for achieving "macromolecular" resolution of about 4 nm. As a result, the majority of ECT investigations have focused on purified macromolecular complexes, viruses, and small cells typically found in various species of Bacteria and Archaea. [22].

Larger cells, and even tissue samples, can be prepared for CryoET by employing thinning techniques, which involve either cryo-sectioning or focused ion beam (FIB) milling. In the cryo-sectioning method, frozen cell or tissue blocks are meticulously sliced into thin samples using a cryo-microtome [24]. In FIB milling, samples that have been rapidly frozen are subjected to a precise stream of focused ions, usually gallium, which methodically removes material from both the top and bottom surfaces of the sample. This process results in a slender lamella that is well-suited for ECT imaging [25].

The substantial interaction between electrons and matter also leads to an anisotropic resolution effect. When the sample is tilted for imaging, the electron beam engages with a proportionally larger cross-sectional area at steeper tilt angles. In practical terms, tilt angles surpassing around 60–70° offer limited information and are consequently excluded. This phenomenon creates a "missing wedge" in the final tomogram, causing a reduction in resolution along the direction of the electron beam [26].

**Tomogram Enhancement**

By leveraging repetitious features present in the tomograms, a higher resolution can be obtained by averaging several copies of structures of interest. The iterative alignment and averaging can increase the SNR and resolution. In each iteration, the relative rotations and shifts between sub-tomograms and the reference structure are determined by a cross-correlation similarity metric. Applying these orientations yields a new reference that may be exploited to refine the sub-tomogram orientations in the subsequent phase. Refinement is performed continuously until it convergence, or the number of iterations reaches a specific threshold. To avoid overfitting, the reference structure is filtered between iterations to its estimated resolution by comparing resolution shells of two independently reconstructed volumes obtained by splitting the dataset into two halves. In addition, maximum likelihood methods further reduce the risk of overfitting by allowing each sub-tomogram to contribute simultaneously to different orientations in a weighted manner. In order to mitigate compositional and conformational variations and enhance the effectiveness of averaging outcomes, it becomes essential to categorize the sub-tomograms into more uniform subgroups.

Several software packages exist for sub-tomogram analysis such as Dynamo [27], EMAN2 [28], emClarity [29], Protomo [30], PyTom [31], RELION [32]. These packages include functionalities for different sub-tomogram averaging steps; some also provide data management, allow users to design highly customizable workflows, or they guide sub-tomogram analysis along a predefined path.

### 2.1.3 Existing Study Related to Actin Filament Tracing

Various automated and semi-automated computational tools have been developed to recognize filamentous and line-shaped objects in medical and biological imaging [33, 34]. However, studies particularly focusing on cytoskeletal filament detection are rather limited. Moreover, these limited studies considered images acquired through multiple techniques such as fluorescent microscopy, confocal microscopy, cryo-electron microscopy, or electron tomography [35]. Each of these meth-ods produces images with a different spatiotemporal resolution, signal-to-noise ratio, and contrast properties. Studies targeting particular aspects of cytoskeleton structure and function must first choose a microscopy method capable of providing that information. For instance, a study aiming for a very high resolution can resort to an electron microscopy method, which can achieve a spatial resolution of around 5 nm. Electron microscopy images, however, typically have a very low signal-to-noise ratio and low contrast, which makes visualization, processing, and analysis of these images difficult, especially in 3D [36, 35].

**Tracing Filament in Cryo-ET image**

A limited number of studies especially concentrated on identifying actin filaments from the cryo-ET image. Fully-automated and semi-automated filament-tracing methods for single tomograms have been integrated into several different software packages, such as Sculptor and Situs [37, 38], and AMIRA [39]. Sculptor and Situs are non-commercial and freely available, while AMIRA requires a paid license. Rusu et al. [40] developed an automatic method (available in Sculptor and Situs) to trace filaments from the cryo-electron tomograms using cylindrical templates. Weber et al. [41] developed an automated tracing method using modified cylinders that account for the missing wedge artifacts in electron tomography density data. Redemann et al. [42] traced microtubules to build a model of a spindle. Further, segmented filaments can be quantified to understand the organization of a large number of filaments [43].

Kovacs et al. [44] developed a template-based deconvolution method to correct the missing-wedge artifacts prior to the tracing; however, this approach required expensive numerical techniques (up to a week of run time) and was likely to interpret the high-density voxels in entire tomogram (including membranes, other biomolecules, and noise) as filaments, leading to false-positive predictions.

Loss et al. [45] developed a framework for the automatic segmentation and analysis of filamentous objects in 3D electron tomography that allows the quantification of filamentous networks in terms of their compositional and morphological features. Their proposed approach consists of three steps: (i) local enhancement of filaments by Hessian filtering; (ii) detection and completion (e.g., gap filling) of filamentous structures through tensor voting; and (iii) delineation of the filamentous networks. The authors first validated the approach using a set of specifically designed synthetic data. Afterward, they applied the segmentation framework to tomograms of plant cell walls undergoing different chemical treatments for polysaccharide extraction.

Recently a number of deep learning-based approaches were proposed for the segmentation of diverse biological assemblies. For example, Chen et al. [46] presented a deep learning-based segmentation model for pixel-level annotation of shapes of interest, which was integrated with the EMAN [28]. However, these segmentation tools are generic in nature and were not particularly developed considering filamentous structures. Besides, they require manual annotation provided by users to train the deep learning model further, all of which could be a laborious process. Besides, as supervised classification methods, algorithms may require a large amount of manually annotated training data for deep model training, or in case of the pre-trained model, the models need to be trained again to apply to a new dataset. Besides, these segmentation approaches only provide pixel-level segmented results instead of individual filament traces, thus, can not be considered as a complete filament tracing framework.

.

**Filament Tracing in Other Types of Images**

Previous studies also explored the segmentation of cytoskeletons in other types of images, including fluorescence images [47] and in confocal microscope images [48, 49]. Alioscha et al. [47] performed segmentation on the fluorescence microscopy image to segment actin filament networks. Their method begins with an image decomposition operation, which yields a cartoon image and a noise/texture image component. The cartoon component is then used as input for the computation of a multiscale line-response image (via a method proposed originally by [61], where each pixel holds a score belonging to a line). The authors then applied thresholding to the response image using a local adaptive thresholding method for separating the actin filaments. Another work that utilized fluorescent microscopy images has been performed by [50]. The proposed method uses multiple Stretching Open Active Contours (SOACs) that are automatically initialized at image intensity ridges and then evolve along the centerlines of filaments in the network. SOACs can merge, stop at junctions, and reconfigure with others to allow smooth crossing at junctions of filaments. This SOACs-based approach is usually applicable to images of curvilinear networks with low SNR.

## 2.2 SEGMENTATION AND QUALITY ASSESSMENT OF PROTEIN SECONDARY STRUCTURES

### 2.2.1 Protein Structure

Proteins are large and complex molecules that are essential for the structure, function, and regulation of cells, tissues, and organs in living organisms. Proteins are made of chains of smaller molecules called amino acids linked successively in a specific sequence to form long, folded chains that give proteins their unique three-dimensional structures.

Proteins perform diverse roles in living organisms. They function as enzymes, catalyze chemical reactions in cells, and provide structural support to maintain cellular integrity. Proteins also play a role in transport and storage, facilitating the movement of molecules across cell membranes and within the body. Additionally, proteins are involved in cell signaling, immune response, and gene expression, among other functions. Overall, proteins are fundamental to the proper functioning of cells and are critical for various physiological processes in living organisms.

Protein secondary structure refers to the regular, repeating patterns of folding that occur within a protein chain. There are two main types of protein secondary structure: alpha helices and beta sheets (Fig 2). Alpha($\alpha$) helices are coiled structures that resemble a spring. The backbone of the protein chain forms the inner part of the helix, while the amino acid side chains extend outward. The helix is stabilized by hydrogen bonds between the carbonyl group of one amino acid and the amide group of another, located four residues away in the sequence. $\beta$-sheets are extended structures that resemble a pleated sheet or a series of accordion folds. The backbone of the protein chain forms the edges of the sheet, while the amino acid side chains extend alternately above and below the sheet. The sheet is stabilized by hydrogen bonds between adjacent strands.

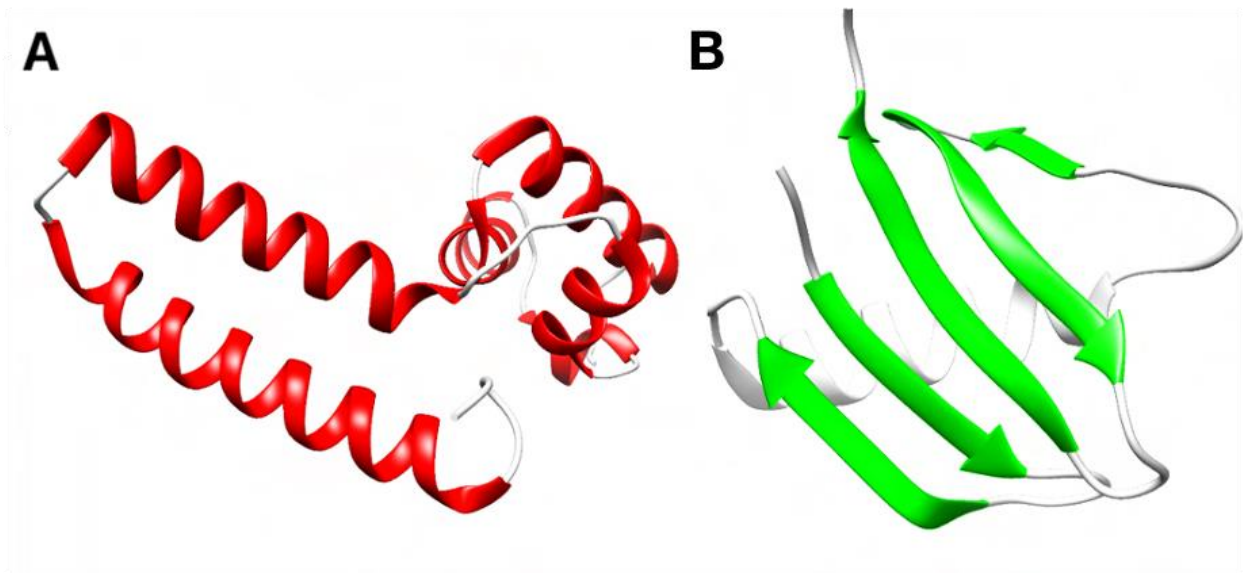

**Fig. 2.** Structures of $\alpha$-helix (A) and $\beta$-sheet (B)

Proteins may incorporate a combination of alpha helices and $\beta$-sheets, along with additional structural features like loops and turns. The specific organization of these secondary structures within a protein plays a critical role in shaping its overall three-dimensional structure, which, in consequence, governs its functional characteristics

### 2.2.2 Cryo-EM Image Acquisition

Cryo-electron microscopy (cryo-EM) is a technique used for the high-resolution imaging of biological molecules such as proteins, viruses, and complexes. Cryo-EM has become an increasingly popular technique for studying the structures of biological molecules, particularly those that are difficult to study using other methods, such as X-ray crystallography. In cryo-EM, samples are flash-frozen in a thin layer of vitrified ice and then imaged with an electron microscope.

The cryo-EM imaging process typically involves the following steps:

**Sample preparation:** The sample is purified and prepared in a way that allows it to be imaged in a thin layer of vitrified ice. This involves removing any impurities or contaminants, as well as optimizing the sample concentration and buffer conditions.

**Grid preparation:** The sample is applied to a thin carbon film supported by a copper or gold grid. The grid is then blotted with filter paper to remove excess liquid, leaving behind a thin layer of sample-containing solution.

**Freezing:** The grid is rapidly plunged into liquid ethane or propane, which rapidly freezes the sample in a thin layer of vitrified ice. This process is critical to preserve the sample in a near-native state and prevent any structural changes that might occur during sample preparation.

**Imaging:** The grid is loaded into a cryo-electron microscope, which uses a beam of electrons to generate images of the sample. The microscope is equipped with a low-temperature stage to maintain the sample in a frozen state throughout the imaging process. The electron beam interacts with the sample, generating a projection image that is captured by a detector.

**Data processing:** The raw data collected from the microscope is processed using specialized software to generate a three-dimensional reconstruction of the sample. This involves aligning and averaging multiple two-dimensional images to generate a high-resolution 3D model of the sample.

### 2.2.3 Existing Studies Related Protein Secondary Structure Detection and Validation

## Detection of Protein Secondary Structure

Various approaches have been proposed to detect the location of secondary structures from cryo-EM maps of medium resolution since the introduction of the first method Helixhunter [51]. However, precise detection of secondary structures is still challenging. Since helices longer than one turn often have a cylindrical feature in medium-resolution maps, early approaches target helix detection using different ideas. In Helixhunter, cross-correlation with a cylindrical template was used to detect the initial location of helices [51]. In Helixtracer, a gradient graph was used for the initial segmentation of voxels in helix regions [52]. In Voltrac, a genetic algorithm was used in bidirectional search with a cylindrical template [40]. The density features of a $\beta$-sheet are often less distinctive than those of a helix, due to factors such as lower density and a wide range of shapes for $\beta$-sheets. Nevertheless, a thin layer of density could be observed in parts of a $\beta$-sheet and skeletonization is an effective method for detecting it [53, 54]. SSEhunter and SSETracer are two software tools to detect both helices and $\beta$-sheets incorporating skeleton information in the detection of these structures [54, 55, 56]. A limitation of the aforementioned image processing methods is that carefully selected parameters are needed, which is challenging for a typical user. A few machine learning-based methods using either SVM or K nearest-neighbor classifiers were proposed in SSELearner ( [55] and RENNSH [57]. SSELearner uses pre-determined features of local structure tensors and local thickness in training [55].

Recent approaches based on deep convolutional neural networks (CNNs) have shown promising potential for the secondary structure detection problem [58, 59]. Li et al. [58] introduced a CNN-based model consisting of three inception blocks, two reduction blocks, and three deconvolution levels for secondary structure detection. Haslam et al.[56] described a patch-based U-Net architecture for the segmentation of protein secondary structures. Nevertheless, these two network frameworks were primarily tested using simulated 3D images, rather than cryo-EM images derived experimentally. Emap2sec, another deep learning-based architecture proposed by Maddhuri et al. [59] utilizes patches of size 11 Å cube. Since secondary structure regions often occupy a small portion of the entire 3D image of a protein chain, the use of patches during training reduces the impact of class imbalance. However, automatic cropping of a rectangular box from a 3D image produces artifacts to some helices and $\beta$-sheets, since they are likely to be cropped.

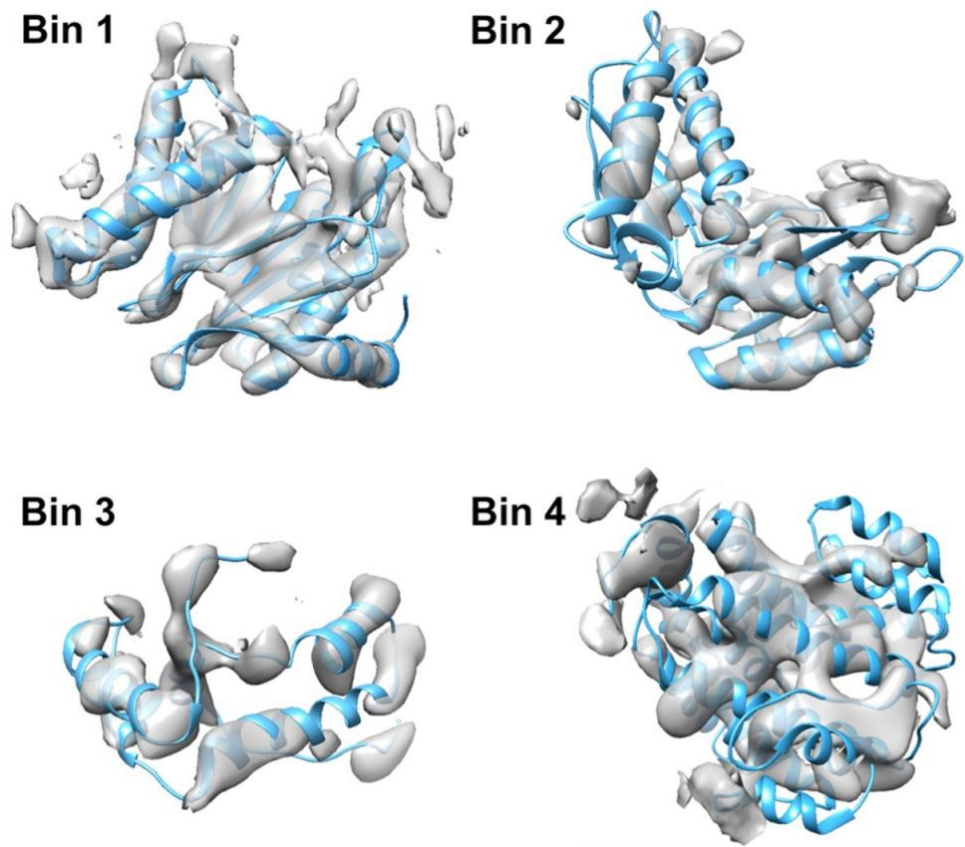

**Fig. 3.** Cryo-EM density maps of variable map-to-model agreements. Bin 1 represents Chain H of EMDB ID: 1733, 6.8 resolution; Bin 2 represents Chain C of EMDB ID: 4075, 5.35 resolution; Bin 3 represents Chain Q of EMDB ID: 8129, 7.80 resolution; Bin 4 is Chain W of EMDB ID: 6286, 8.30 resolution.

A challenge in the secondary structure detection problem is that capturing geometric patterns of helices and $\beta$-sheets requires 3-dimensional image information; 2D slices cropped along different axes unlikely to provide enough information to determine secondary structures. Another challenge is the labeling of ground truth. There is no human-labeled ground truth for cryo-EM images. Existing labels were derived automatically using the guidance of atomic structures. This method assumes a perfect helix/$\beta$-sheet volume [55].The current labeling method does not consider defects that are often seen in experimentally-derived images. The quality of cryo-EM density maps varies significantly in medium resolutions [60, 61]. In some density maps, secondary structure density characters are clearly visible (Fig 3 A and B), but not well distinguishable in others (Fig 3 C and D). For example, labeling using atomic structure (blue ribbon in Fig 3D) would not align with the density character observed for some helices. The lack of a precise labeling method makes it challenging to utilize a large number of cryo-EM data.

**Validation of Protein Secondary Structure**

The reproducibility and discriminative ability of a local similarity score are important during the model-building process as a global score, such as the cross-correlation (CC), varies with the size of the map volume under consideration. Consequently, an earlier application of CC scores in the PHENIX program required a local region mask to evaluate individual residues. An alternative tool, EM-Ringer, evaluated side chains only and produced a single score to represent the quality of the fit; nevertheless, the approach did not generalize to the secondary structure level. The most related approach, in terms of applicability to a secondary structure, was the Z-score. A Z-score was calculated from 27 CC scores using ± zero or one voxel shifts in the x, y, and z directions, respectively. The approach did not take specific shape information into account, and by design, it depended on the granularity of the map. In contrast, the new F1 score proposed below uses a simple geometric metric and is invariant when zooming in to the level of residues.

# CHAPTER 3

# BUNDLETRAC: TRACING ACTIN BUNDLE IN THE SHAFT REGION OF STERIOCILLUM

The actin bundle in stereocillia primarily consists of primarily parallel actin filaments in close proximity, frequently connected by cross-linking proteins. To identify such straight and parallel filaments, a novel computational tool, BundleTrac, is introduced that can trace bundle-like features from 3D tomograms.

## 3.1 DATASET

The dataset represents a simplified volumetric model of the actin core consisting of the tip, shaft, and taper regions of stereocilia and collected from utricular sensory epithelia of murine Pls1-/- mice [11]. Here, filaments exhibit straightness and are parallel to one another along the primary stereocilia axis. The actin-actin spacing at different locations of the actin core in the shaft region is around $12.6 \pm 1.2$ nm (mean $\pm$ SD; n=2803 and does not vary much along the stereocilium shaft. Distributions fit well with a single Gaussian model, which refers to the uniformity of crosslinking through the stereocilium.

The process of obtaining stereocilia adhered to an electron microscope grid involved blotting the mouse's inner ear utricle sensory epithelium onto a lacey-carbon support film on the grid [9]. Subsequently, the lacey-coated gold grids, with stereocilia attached, were subjected to immersion-plunge freezing in liquid ethane at the temperature of liquid nitrogen. The whole-mount intact stereocilia were then imaged using a 300-kV FEI Krios microscope equipped with a K2 camera (Gatan Inc., Pleasanton, CA, USA), which was operated under low-dose conditions in its traditional integrating mode. Cryo-tilt series were collected from -60° to +60°, with 2° increments, at a voxel size of 0.947 nm. These images were then aligned and reconstructed using the IMOD software [62]. Finally, the orientation of the actin bundle was adjusted to align with one of the X, Y, and Z coordinate directions.

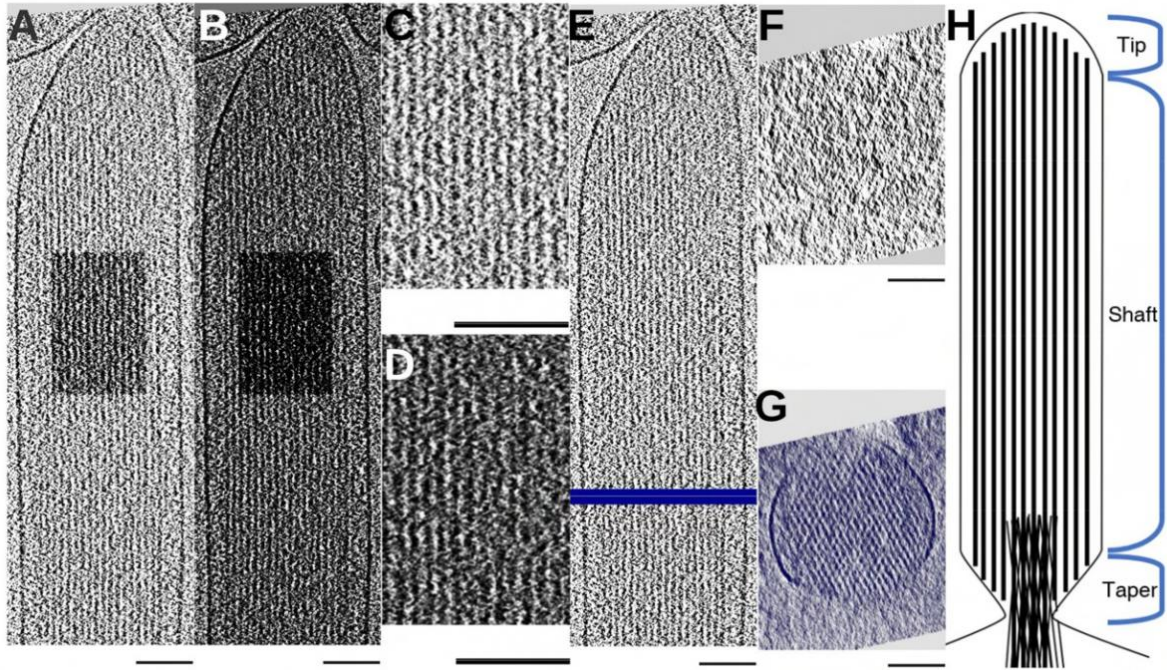

**Fig. 4.** Cryo-ET density map for the shaft of a Stereocilium and its membrane-enclosed actin bundle. The longitudinal view of a Stereocilium is shown for a slab about 1 nm thick in (A) and about 8 nm thick in (B). (C,D) are sub-regions (box) of (A,B), respectively. (E) shows the location of cross-section slices of (F) (white line) and (G) (blue line) respectively. Cross-sections about 1 nm thick and 30 nm thick are shown in (F,G), respectively. (H) shows the schematic of a stereocilium. All scale bars are 100 nm. (Reproduced from [63])

Fig 4 shows the tomographic reconstruction of Stereocilia, with actin bundle orientation aligned approximately with the Y (vertical) axis. Figure 4 (A and B) show a 1-nm single slice and an 8-nm thick slab of the bundle, respectively. The actin bundle consists of largely parallel actin filaments in close proximity, frequently connected by cross-linking proteins. Fig 4 (panel F and G), represents the cross-sectional views of the Stereocilium of a 1-nm single slice and a slab with a thickness of 30-nm respectively. The effect of the missing wedge is visible in Fig 4G, with only about one third of the Stereocilia membrane being clearly visible and the top and bottom regions of the Stereocilia being blurred. Throughout the entire cross-sectional profile, the missing wedge artifact manifests in the elongation of the circular density in the cross-section of an actin filament. The resulting ellipsoids can cause the filaments to merge in the cross-section in the Z-direction (Fig 4G), which makes fully-automated computational tracing approaches rather challenging.

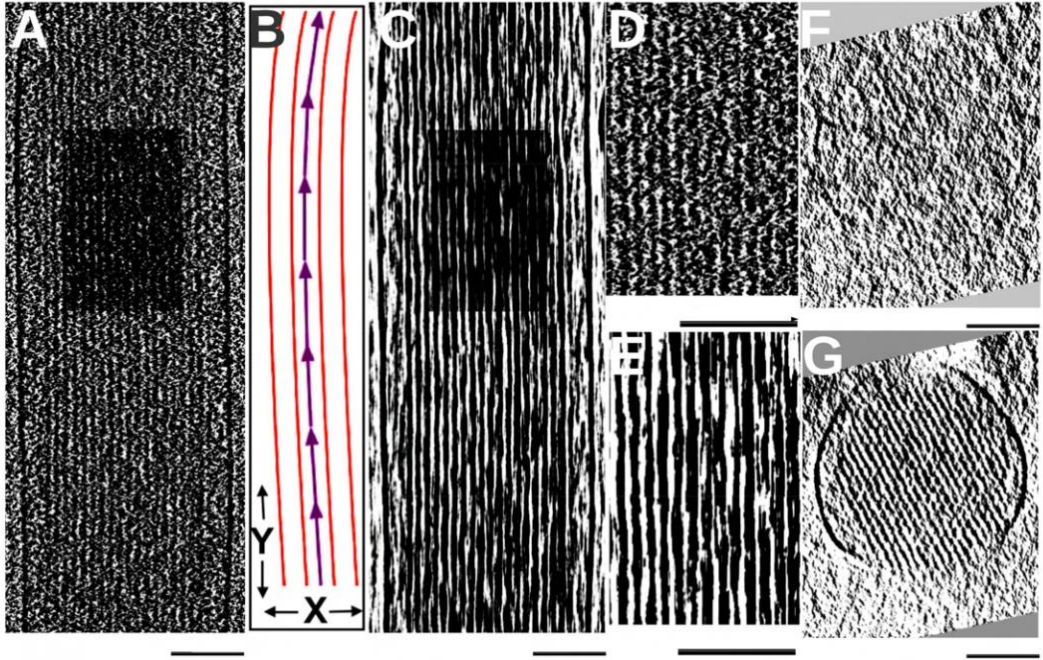

**Fig. 5.** Longitudinal averaging along the direction of the bundle. (A) A slab of the stereocilium map 8 nm thick; (B) An illustration of the bundle direction (purple) calculated using 2D cross-correlation of cross-sections; (C) An 8-nm slab of the longitudinally averaged map at the same position as in (A); (D,E) Zoomed-in views of (A,B), respectively, in the same sub-region (dark box in (A,B)); (F) A 1-nm cross-section of the original map (G) The 1-nm cross-section of the longitudinally-averaged map at the same position as in (F). Note that the cross-sectional view of the averaged map has high clarity; all scale bars represent 100 nm.(Reproduced from [63])

## 3.2 METHODOLOGY

BundleTrac contains two main steps: (1) the detection of the bundle axis and longitudinal averaging; (2) filament tracing using 2D convolution optimization.

### 3.2.1 Longitudinal Average Along an Estimated Direction of the Bundle

A simple image enhancement method based on the filamentous nature of the bundle is designed. The idea is to perform averaging along the filament direction. A critical step in this averaging is determining the overall orientation of the bundle with a two-dimensional (2D) cross-correlation of cross-sections. The cross-sections are sampled at different spots along the bundle to capture the gradual change in the direction of the bundle. Using 2D cross-correlation to identify the axis of a cylindrical object is a quicker alternative to 3D template matching, as it does not involve sampling the translational and rotational space of the object; Instead, it directly calculates the orientation from a trace of the 2D cross-correlation peak shifts. Once the orientation of the bundle axis is determined, the average density along the filament axis is calculated. Each voxel is averaged with those within ±15 nm along the direction of the bundle axis. Longitudinal averaging appears to effectively enhance signals along filaments, as the signals are more visible in the longitudinally averaged density slab (Figure 5C,E) than in the original map (Figure 5 A,D). This enhancement is clearly visible in a 1 nm-thick cross-sectional slice but undetectable in the original map (Figure 5 F). Density maps displaying a low signal-to-noise ratio make it challenging to directly detect the direction of each individual filament. However, it is possible to detect the direction of a bundle that comprises multiple filaments. Although the direction of the bundle may not be identical to the direction of each individual filament, the former is a close estimate of the latter if the filaments do not suddenly change direction. The direction of the bundle is used as an initial estimate for all filaments in the signal enhancement process.

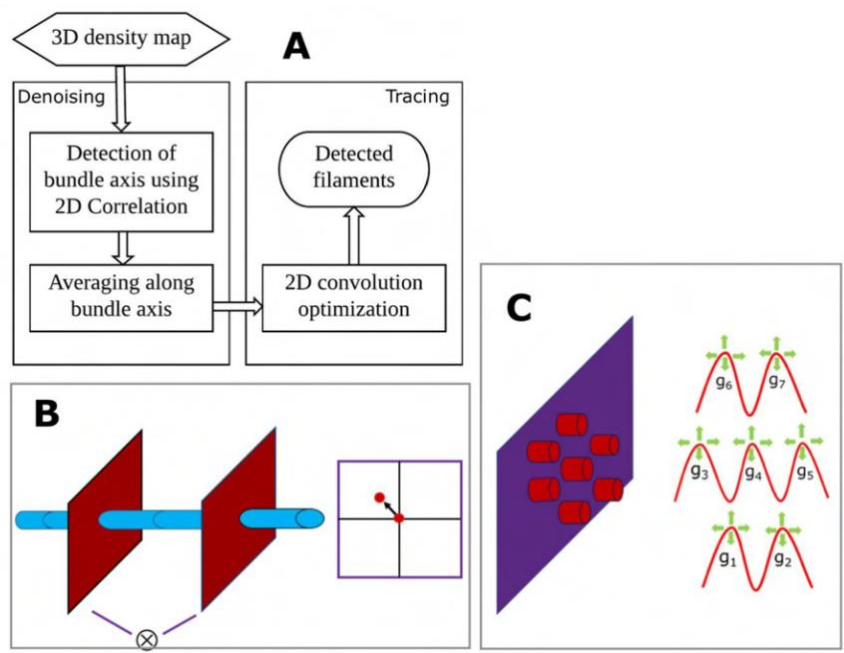

**Fig. 6.** The design of BundleTrac. (A) overall steps; (B) detection of the bundle axis using 2D cross-correlation; (C) 2D convolutional optimization using seven hexagonally arranged Gaussian peaks.(Reproduced from [63])

### 3.2.2 Filaments Tracing Employing Cross-Correlation

Bundle Trac exploits the fact that the filaments in the actin bundle are roughly parallel and change their direction gradually. While it is hard to detect individual filaments in a single cross-sectional slice 1 nm thick (Figure 5 F), a longitudinal average of 30 cross-sectional slices (of 30 nm thickness) significantly enhances the signal and clearly reveals the hexagonal arrangement of the actin filaments.

Due to the long persistence length of actin filaments, they can be traced by placing markers a certain distance apart along each filament. Without the use of a 3D template, the tracing relies on a 2D convolution of a kernel constructed using seven Gaussian peaks that resemble a local bundle of seven filaments. The seven-peak 2D convolutional optimization method is used to trace the filaments in the density map of a stereocilium.

The proposed method utilizes seed points provided by an expert on a 30-nm cross-sectional slab. It then traced the entire filament length, constrained by the maximum spacing of filaments (between 11 and 15 nm). The filaments detected using BundleTrac represent locally optimized positions of roughly seven filaments. The underlying assumption, which seems reasonable and is confirmed by manual and semi-automated detection, is that the orientations of the actin filaments only change slowly throughout the 3D volume. Within a 30-nm window along the filament, changes are gradual, so averages provide strong signals.

As the direction of the bundle is roughly parallel to the Y-axis of the image, a series of X-Z slices (approximate cross-sections of the bundle) is used to calculate the precise direction of the bundle axis. Two cross-sections (the red planes in Figure 6B) were sampled in increments of 55 slices along the Y-axis. This spacing is chosen to capture the gradual change in the direction of the bundle. A 2D cross-correlation was then performed using the two cross-sections to calculate the peak shift. If the bundle is perfectly parallel to the Y-axis, the cross-correlation peak will be at the center of the cross-section, and there will be no peak shift. The degree of the peak shift represents the vector of the bundle at this local region. This process starts from one end of the shaft and extends to the tip region. The overall direction of the bundle comprises a series of vectors (Figure 5B), each generated from 2D cross-correlations, but it is important to note that the use of 2D correlation to detect the bundle axis assumes the architectural rigidity of the bundle. This appears to be the case for the shaft region of stereocilia in the current resolution range. Once the bundle direction was derived, the original 3D density map was averaged along the bundle axis. Specifically, each voxel along the bundle axis was averaged with 15 slices up and down the filament. This denoising step appears to be effective, as the averaging reveals the hexagonal pattern of the filaments (Figure 5G).

## 3.3 RESULTS AND DISCUSSION

### 3.3.1 Evaluation Metric

In order to evaluate the performance of BundleTrac, computationally detected filaments are compared with the ground truth filaments (i.e., manual annotation). The average cross-distance is calculated for all the filaments. The cross-distance measures the projection distance between two lines that may not be straight and may have relative shifts [64]. Computationally derived markers are finely interpolated such that, for each visually derived marker, it is possible to find the closest computationally derived point on the corresponding filament.

The cross-distance ($CD$) of a filament is calculated by applying the following equation-

$$CD = \frac{\sum_{i=1}^{n}\left(D(P_i - Q_i)\right)}{n} \tag{1}$$

$P_i$ is a point on the $i$'th cross-section (the X-Z plane) of the manually annotated filament, $Q_i$ is the point on the computationally derived filament on the same cross-section, $n$ is the total number of points along the filament, $D(P_i - Q_i)$ is the distance between two 3-dimensional vectors, $P_i$ and $Q_i$. Note that the cross-distance does not reflect the longitudinal distance. The same length was used for a manually detected filament and its corresponding computationally detected filament.

### 3.3.2 Discussion

The effectiveness of the BundleTrac is analyzed in five different settings.

**Table 1.** The effect of longitudinal averaging, isotropic Gaussian filter, seven-peak 2D convolution, and one-peak 2D convolution on overall accuracy. (Reproduced from [63])

| Implementation | Avg_L [a] | Gauss [b] | 7 Peaks [c] | 1 Peak [d] | AvgError [e] |
|---|---|---|---|---|---|
| Trace_L_G_7 | ✔ | ✔ | ✔ | ✘ | 1.300 |
| Trace_L_7 | ✔ | ✘ | ✔ | ✘ | 1.345 |
| Trace_G_7 | ✘ | ✔ | ✔ | ✘ | 1.523 |
| Trace_L_1 | ✔ | ✘ | ✘ | ✔ | 2.157 |
| Trace_L_G_1 | ✔ | ✔ | ✘ | ✔ | 2.420 |
| Improvement in the seven-peak convolution [f] | | | | | 46.28% |
| Improvement in the longitudinal average [g] | | | | | 14.62% |

The one-peak 2D convolution method traces a filament based on its own density. It is the best method to trace individual filaments if the strength of the signal is sufficient and the artifact of the data is minimized. The seven-peak 2D convolution finds the best superposition of seven hexagonally-arranged Gaussian distributions on the density map at a cross-section of the bundle. It is more robust than the one-peak convolution, as it is less affected by the artifacts and noise in individual filaments. The use of seven-peak convolution reduces the average cross-distance error by 42.3%, from 2.42 voxels for the one-peak convolution to 1.3 voxels for the seven-peak convolution (Table 1). In fact, the most effective technique in our comparison is the seven-peak convolution (46.3%), followed by longitudinal averaging (14.6%). Applying a Gaussian filter to the original map only slightly enhances the accuracy—by 3.3%, from a cross-distance of 1.345 voxels without the filter to 1.3 voxels with it. This is to be expected, because Gaussian filters are isotropic and are not tuned to the artifacts and resolution anisotropy of the experimental data.

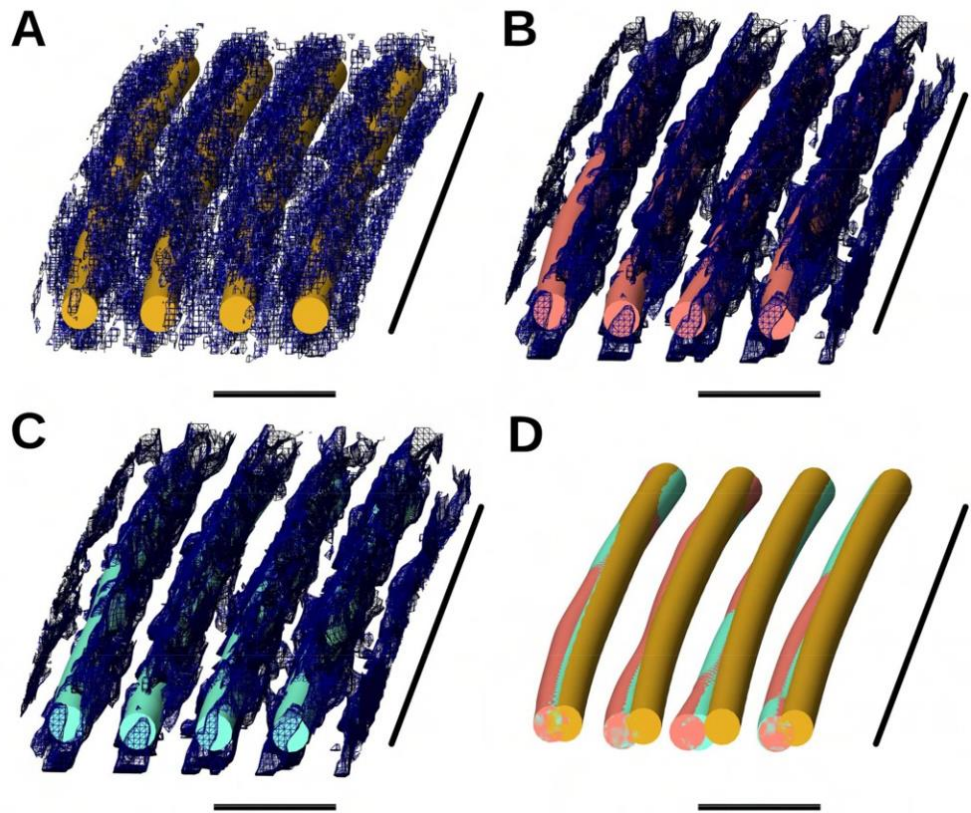

**Fig. 7.** A local view of the filaments built with various implementations. (A) Filaments built manually (gold) overlaid with the original map; (B) filaments built using the one-peak method (salmon) overlaid with the longitudinally averaged density map; (C) filaments built using the seven-peak method (aquamarine) overlaid with the same map as in (B); (D) superposition of the three sets of filaments in (A–C). For (A–D), the horizontal scale bar is 20 nm, and the vertical scale bar is 630 nm.(Reproduced from [63])

Fig 7 shows the visual comparison between three sets of filament models: those detected manually (gold), those detected using one-peak convolution (salmon) and those detected using seven-peak convolution (aquamarine). All three sets fit the density well overall, but the filaments detected using seven-peak convolution best align with the manually obtained set. The four filaments for which a portion is shown in Fig 7 have an average cross-distance between 0.36 and 0.39 voxels (3.44 Å–3.67 Å) when seven-peak convolution was applied. For the sake of viewing clarity, only an example of a local region of four single-layer actin filaments are shown. The length of the four filaments is 630 nm, which is about 1/20 the entire length of the filaments. The longitudinal aver-age appears to increase the signal-to-noise ratio, as the filaments are more obvious in the averaged map (Fig 5 B) than in the original map (Fig 5 A). In fact, longitudinal averaging reduces the cross-distance error by 14.6%, from 1.5 voxels to 1.3 voxels (Table 1).

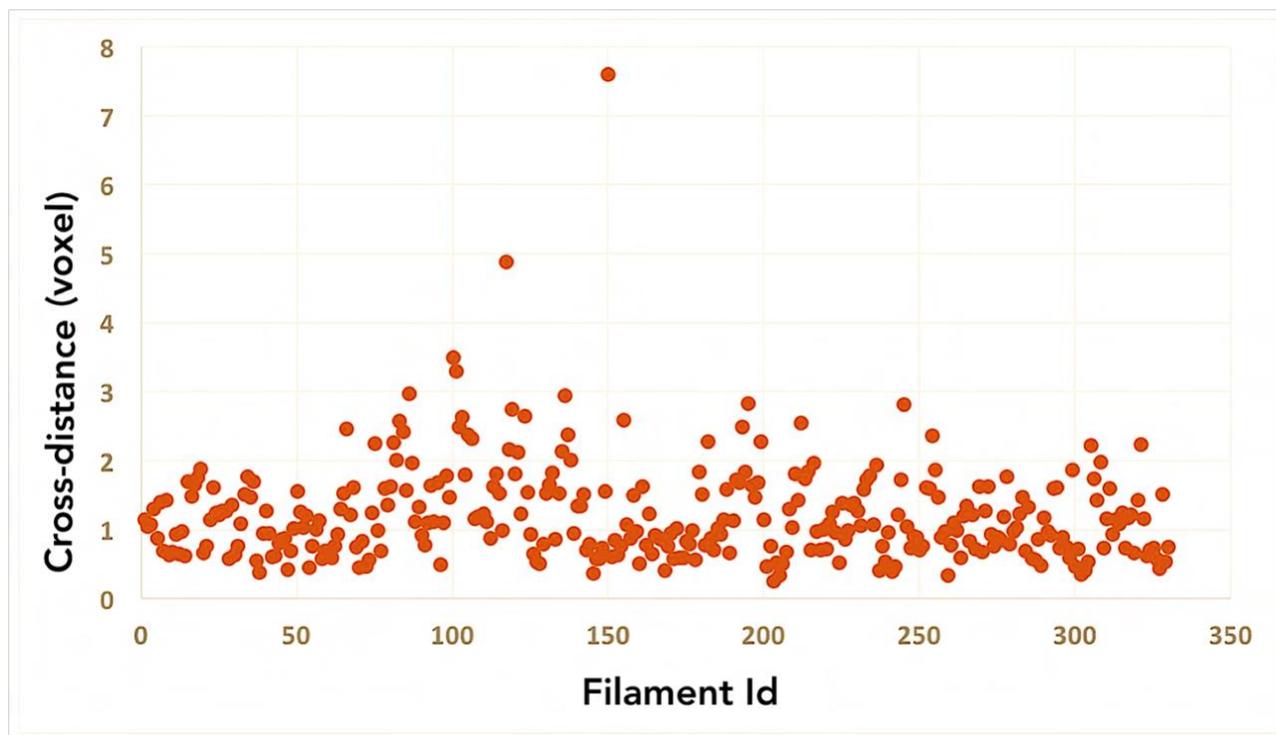

**Fig. 8.** The average cross-distance between BundleTrac predictions and the corresponding manual annotations for 330 filaments.(Reproduced from [63])

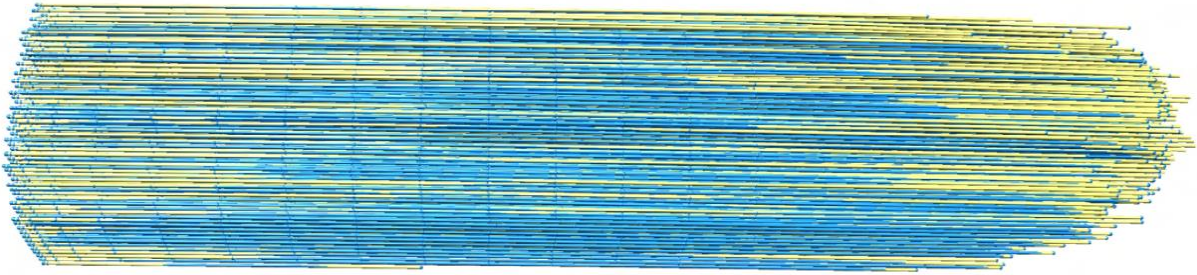

**Fig. 9.** Filaments detected by BundleTrac (blue) overlayed with corresponding manually detected filaments (yellow)

Tracing with a convolution of seven Gaussian peaks appears to be effective for 328 of 330 filaments. Of the 330 filaments, 326 have an average cross-distance less than or equal to three voxels, which is well within the roughly 11 to 13 voxels distance between two neighboring actin filaments. Two detected filaments significantly deviating from the ground truth with a 5–8-voxel cross-distance (Figure 8). In these two cases, it appears that the filament traces jumped over to the neighboring filament density and were affected by the high degree of local density discontinuities and distortions at certain points along the actin filaments. Additional control may need to be considered in such cases of filament ambiguity. Overall, the results of the comparison suggest that the two sets of filaments agree with each other very well overall.

### 3.3.3 Limitation

BundleTrac is designed to trace filaments that collectively and gradually change directions. Although most filaments experience only a minor change in direction, it is observed more dynamic characteristics of certain filaments such as joining to neighboring filament at certain spots along their length. BundleTrac likely to be less accurate in the case of such irregular filaments. In another larger tomogram of a hair cell stereocilium ($500 \times 2400 \times 500$ voxels), it is observed that multiple such filaments. In that tomogram, BundleTrac yields an overall cross-distance of 2.772 voxels for 337 filaments. This stereocilium is roughly twice as long as the first one ($348 \times 1194 \times 483$ voxels), in which no "irregular" filaments are observed. In a test of the sub-region of the second stereocilium with none of the "irregular" filaments included, BundleTrac identified 69 filaments with an average cross-distance of 1.519 voxels, which is close to the accuracy obtained from the first stereocilium. Although not currently implemented in BundleTrac, it is possible to detect "irregular" filaments in a post-processing step if they occur in small numbers. The inter-filament spacing of detected filaments may be used to detect such irregularities in the actin bundle. These spots can be flagged, allowing an expert to determine the reason for the failure, which may have less to do with the algorithm and more to do with the underlying density maps. In such cases, a local (possibly manual) re-tracing may be done by exploring alternative paths.

# CHAPTER 4

## SPAGHETTI TRACER: TRACING SEMIREGULAR ACTIN FILAMENTS IN 3D TOMOGRAMS

In the shaft region of hair cell stereocilia, the filaments form dense hexagonally packed bundles that can be detected by the BundleTrac [63] software utilizing the hexagonal-shaped templates. However, the taper region of hair cell stereocilia [11] imposes additional challenges as individual filaments can deviate from the dominant direction of the regular bundle; as a result, Bundletrac is not effective for such kind of scenario. To deal with such a deviation of filaments existing approaches rely on convolution or deconvolution methods with rotating shape templates [43]. These operations are computationally expensive, as they have a large search space due to considering various angles, lengths, and thicknesses of the filaments and may take days of computing time. Unlike them, our proposed approach performs significantly faster tracing in dense filaments by combining denoising with tracing in a single conceptual framework. Instead of slow-rotating templates, the dominant orientation of the filaments are considered while allowing individual traces to deviate from the main direction with a certain freedom.

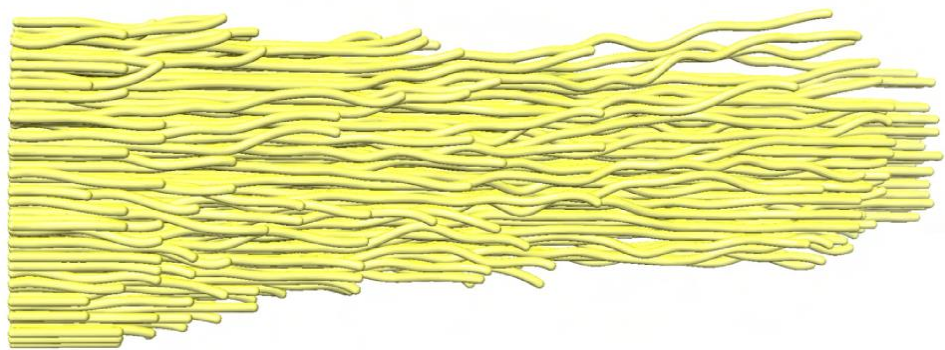

**Fig. 10.** Manually traced filaments in the stereocilia taper region. The manual spaghetti model is consistent with the experimentally observed density of the actin filaments [11]. The true position of actin filaments in the experimental map is not known with complete certainty, but manual annotation can serve as a ground truth for testing our algorithms when tomograms are simulated based on it. All molecular graphics figures in the present paper have been prepared with UCSF Chimera [65] and oriented with the $+Y$ direction to the right.(Reproduced from [66])

## 4.1 DATASET

To test the framework, a method for the realistic simulation of tomograms is developed based on an existing tracing of filament cores (Fig 10). The ground truth traces used for the simulation can be the manual annotation of an experimental tomogram through visual inspection or an automated tracing generated by the current (or another) computational approach. The simulation aims to mimic the noise and missing wedge artifacts present in experimental maps. The procedure does not include the non-filamentous biological features, such as membranes, in the simulated tomogram. Thereby, the simulated tomogram provides a known ground truth that is specifically designed to validate the efficacy of any filament tracing framework.

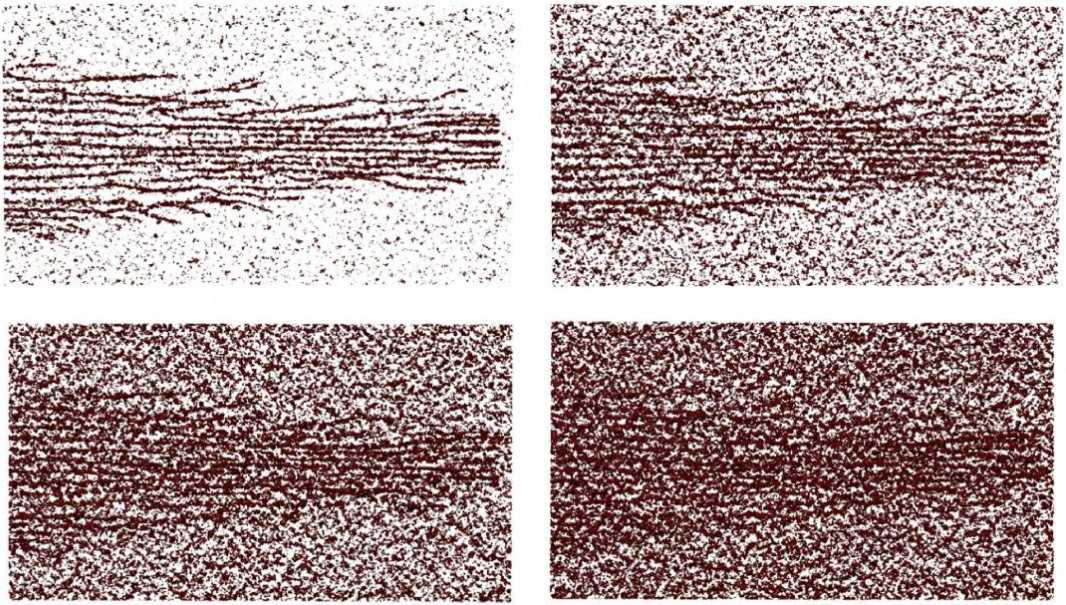

**Fig. 11.** Illustration of the simulated density maps at various noise levels relative to noise in the experimental map [11]. (**A**) noise level 0.40; (**B**) noise level 0.60; (**C**) noise level 0.80; (**D**) noise level 1.00. For the illustration, a 10-voxel-thick slab (corresponding to $Z$-indices 120–129, using experimental map voxel spacing 0.947 nm [11]) is used, with an isocontour density threshold of mean plus two times the standard deviation.(Reproduced from [66])

The simulation approach starts by interpolating the existing model filaments and rasterizing them onto the cubic grid of an experimental map corresponding to the start model. The grid indices $i$, $j$, and $k$ correspond to the $X$, $Y$, and $Z$ axes, respectively. The size and dimensions of the experimental reference map grid in our simulations is retained. The projected filament traces are then enlarged by convolving the voxel densities $D(i, j, k)$ with a Gaussian-shaped kernel whose dimensions (full width at half maximum 5 nm with a voxel spacing of 0.947 nm, $\sim$ 2.01 voxels) were matched to the width of an actin filament. Color-filtered noise is added from a noise map that matched the radial power spectral density (noise color) and the signal-to-noise ratio of an experimental reference tomogram [67].In the second step, the noise level in the experimental map is visually matched by applying an amplification factor of 1.85 to the filament voxels before volumization. This additional manual amplification can be seen as subjective, but it helps provide a better visual match to the experimental data than our present automated noise matching, as described in [67]. The manual adjustment accounted mainly for discrepancies between the model and the experimental map. For example, the manual tracing may not perfectly match the experimental tomogram positions (which is referred to "alignment error" in [67]) or the tomogram might exhibit an inhomogeneous density distribution across the volume due to gaps or helical twist in the actin [11] (whereas our simulated filament densities were perfectly homogeneous along the filament length). In such situations, the automatically calibrated filament signal strength can be weaker than a visual inspection suggests. After noise is added, a wedge is masked out in Fourier space to emulate the missing views from the limited tilt range of the microscope specimen holder.

To validate the performance of our proposed tracing framework, simulated tomograms with varying degrees of noise are considered, ranging from 0.4 to 1.0 (Figure 11). The manually amplified noise level (factor 1.85) was used as an upper bound (worst-case experimental noise), which was normalized as a noise level of 1.0 in the current paper. Lower levels were used as an additional scale factor. For example, the 0.6 noise level would be the closest to the automatically matched experimental noise. The lowest 0.4 level might help identify improvements that could be afforded by better quality data in future work.

Using simulated maps brings following advantages: (i) it provides us with needed ground truth for validation, and (ii) it is free from the above subjective amplification uncertainty (the manual matching is only used once to set the reference level 1.0). Consequently, a reasonable range of noise levels is considered, which is expected to be encountered in experimental maps. The noise level (global scale factor) is not affected by the noise color calibration or missing-wedge modeling.

## 4.2 METHODOLOGY

The proposed tracing framework operates in several stages, as shown in Fig. 12. The pre-processing step, which is also referred to as the path-based density modification of tomograms, strengthens the density values of the filaments to make them more prominent and distinguishable. Based on automatically determined seed points, candidate filament segments (CFSs) are then evolved from the seeds, which are iteratively refined and fused to form the final filaments.

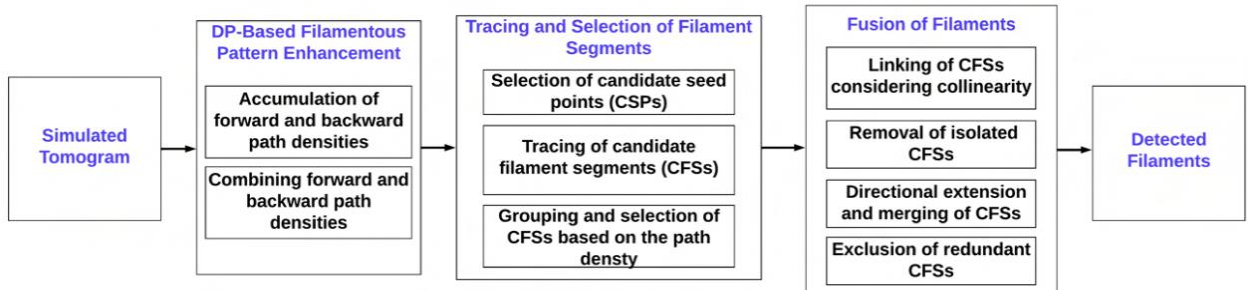

**Fig. 12.** Overall steps of Spaghetti Tracer.(Reproduced from [66])

### 4.2.1 Filamentous Pattern Enhancement

Because of the low radiation dose allowed for each image in a tilt series and the limited view directions, the reconstructed 3D cryo-ET maps typically exhibit a high noise level and missing Fourier wedge artifacts. The pre-processing step uses a path-based density filter to strengthen the underlying filamentous features so that they can be better visualized and automatically extracted.

Our approach assumes that filaments have a mean direction, which, in the current work, is in the $Y$ direction, the same as the experimental map [11]. Individual filaments may deviate from the mean direction, so up to a 45°deviation from the dominant axis are allowed. For each voxel ($i, j, k$) in the map, the preprocessing step assigns the path density values accumulated following a search window, starting from ($i, j, k$). This search window originating from ($i, j, k$) has a pyramidal shape in both forward and backward directions constrained by the 45°limiting angle (described in next sections).

**Pyramidal Search Window and Maximum Path Density Selection**  The forward path density $FPD(i, j, k; l)$ (Fig. 13) and the backward path density $BPD(i, j, k; l)$ (Fig 13) are computed for paths of fixed length $l$ that originate at each voxel ($i, j, k$). The forward and backward directions are determined by the dominant coordinate system axis.  In this work, the filaments are aligned mostly in the $Y$ direction [11], but generally, the dominant axis is assumed to be one of either the $X$, $Y$, or $Z$ coordinate axes. To allow for an up to a $45°$ deviation of filaments from the dominant directions, a pyramidal search window is considered (Fig. 13).

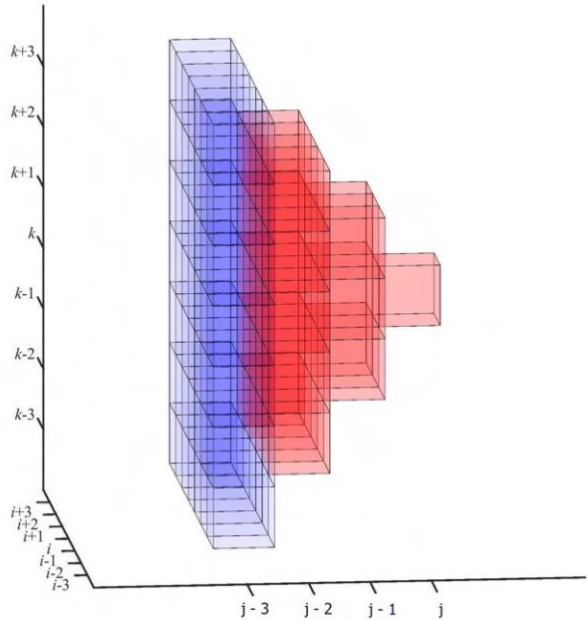

**Fig. 13.** Pyramidical search window for backward tracing starting from voxel $(i, j, k)$. (Reproduced from [66])

Voxel densities $D(i, j, k)$ (normalized to a range from 0 to 1) are accumulated in the search cones in the forward and backward directions, yielding a $PD$ at the front ($j + l$; blue in Fig. 13) and rear ($j − l$; Fig. 13) slices of the pyramids, from which maximum values are chosen:

$$FPD(i, j, k; l) = \max_{\substack{i-l \leq i' \leq i+l \\ k-l \leq k' \leq k+l}} PD(i', j+l, k'), \text{ and} \tag{2}$$

$$BPD(i, j, k; l) = \max_{\substack{i-l \leq i' \leq i+l \\ k-l \leq k' \leq k+l}} PD(i', j-l, k'), \tag{3}$$

where $PD(i, j, k)$ is initialized as

$$PD(i, j, k) = D(i, j, k) \tag{4}$$

and is iteratively accumulated in the pyramidal search windows, starting from the origin, as illustrated in Fig. 13 and described in the following.

**Accumulation and Reverse Pyramid Influence Zone**  The accumulation proceeds iteratively through intermediate voxels $(i', j', k')$, whose $PD$ is updated from immediately adjacent voxels in the previous $Y$-slice (i.e., $j' - 1$ for the forward direction or $j' + 1$ for the backward direction) according to

$$PD(i', j', k') = D(i', j', k') +$$
$$\max_{\substack{m,n \in \{-1,0,1\} \\ \text{(if contributing)}}} PD(i' + m, j' \mp 1, k' + n), \tag{5}$$

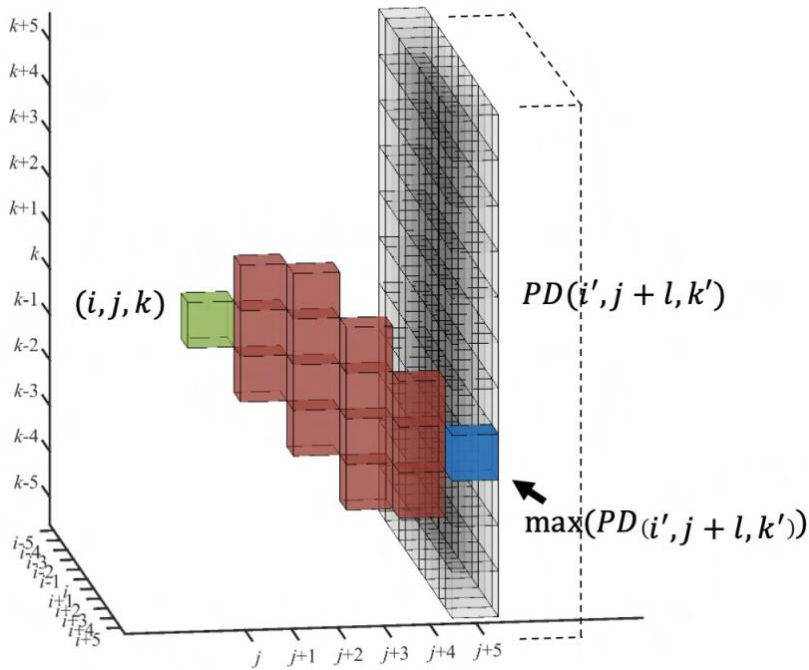

**Fig. 14.** Possible density accumulation paths between two voxels.(Reproduced from [66])

where $\mp$ denotes the minus for Equation (2) and plus for Equation (3) and only neighbors

$m, n \in \{-1, 0, 1\}$ within the above search pyramid are contributing (see Figure 4 of [68] for an illustration). This scheme allows only voxels in a reverse pyramid to influence the blue target $(i^{'}, j + l, k^{'})$. The "if-contributing" condition in Equation (**??**) forces an intersection of the reverse influence pyramid with the above search pyramid, yielding the red accumulation zone in Figure 14.

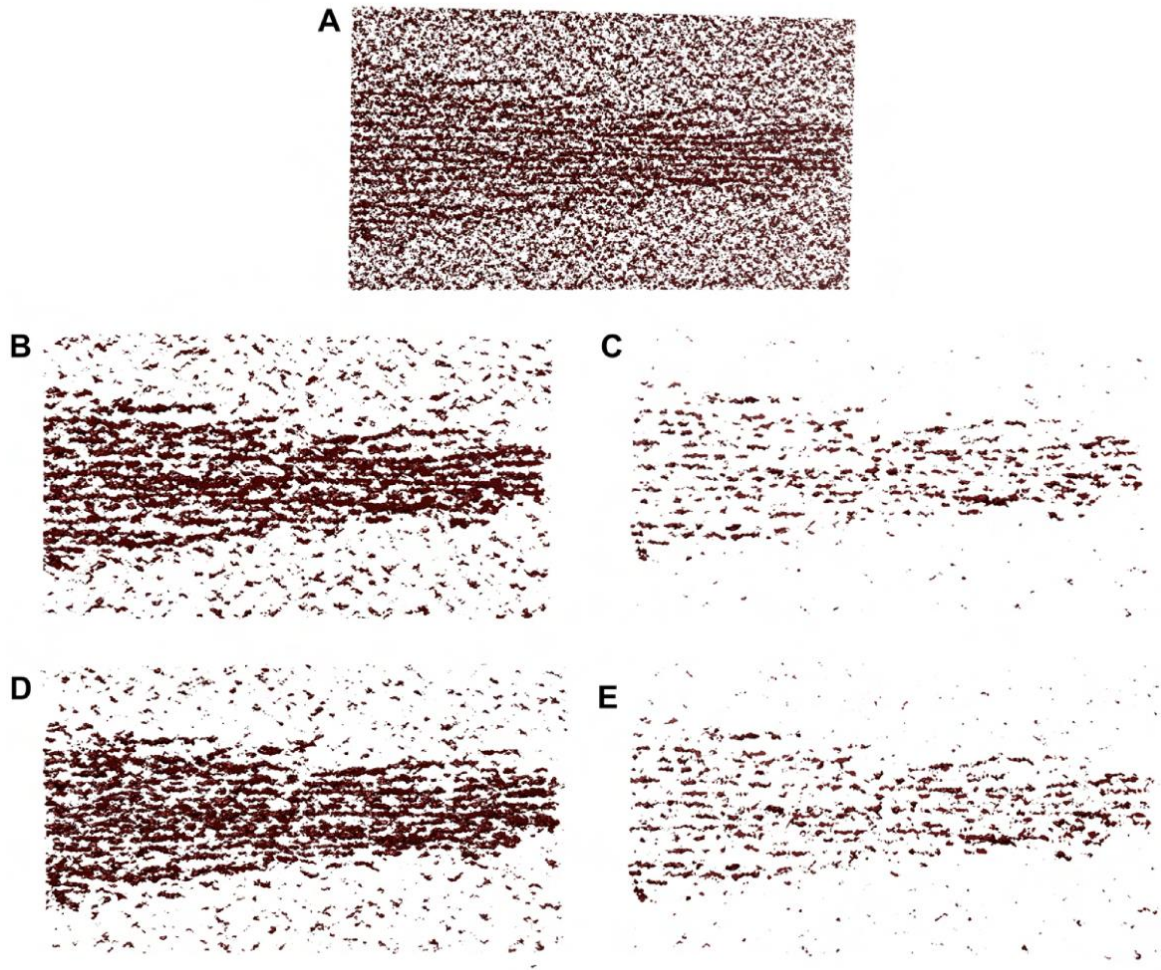

**Fig. 15.** Comparison of the unfiltered and various *CPD*-filtered maps using $l = 5$ voxels. The same 10-voxel-wide slab of is shown, except that it is rendered at the mean + standard deviation isolevel to emphasize the noise. (A) Original unfiltered map with a noise level of 0.8 . (B) The map filtered by multiplication of *FPD* and *BPD* (C) The map filtered by addition/arithmetic mean (D) The map filtered by the geometric mean (E) The map filtered by the minimum. (Reproduced from [66])

**Combining Forward and Backward Path Densities for Filament Pattern Enhancement**  In the subsequent stage, the $FPD(i, j, k; l)$ and $BPD(i, j, k; l)$ values are combined to form a single map, where $(i, j, k)$ acts as the center point and $FPD(i, j, k; l)$ and $BPD(i, j, k; l)$ are sampled from the two opposite directions. It is expected that if the voxel $(i, j, k)$ or its close neighbors are located on a filament segment, it will have high values for both $FPD(i, j, k; l)$ and $BPD(i, j, k; l)$ due to the density accumulation schemes.

The two directional $PD$s can be combined using a blending function. Four types of blending functions are explored.

$$CPD(i, j, k; l) = FPD(i, j, k; l) * BPD(i, j, k; l), \text{ or} \tag{6}$$

$$CPD(i, j, k; l) = FPD(i, j, k; l) + BPD(i, j, k; l), \text{ or} \tag{7}$$

$$CPD(i, j, k; l) = \sqrt{FPD(i, j, k; l) * BPD(i, j, k; l)}, \text{ or} \tag{8}$$

$$CPD(i, j, k; l) = \min(FPD(i, j, k; l), BPD(i, j, k; l)). \tag{9}$$

The $CPD$ values are also normalized to a range from 0 to 1 for an easier way of classifying them in subsequent stages of the workflow (Figure 12). Therefore, no normalization constants appear on the right side of the equations.

The original multiplication (Equation (6)) provides a heuristic score to ensure a logical conjunction (*and* gate); only if both $FPD$ and $BPD$ are large will $CPD$ be large as well. In this type of blending, the product of the two densities in the filtered map $CPD$ no longer corresponds to the density of the biological specimen (e.g., the larger dynamic range might amplify inhomogeneous density variations).

The addition (Equation (7)) is identical to the arithmetic mean (without normalization constants). It appears to be a more natural way to combine accumulated (summed) densities $BPD$ and $FPD$. Moreover, the filtered map $CPD$ has the advantage of being proportional to the physical density of the specimen. Addition is similar to a logical disjunction (*or* gate), so filament voxels (with simultaneously high $FPD$ and $BPD$) are rewarded less than by multiplication.

Because Equation (6) is essentially the square of the geometric mean (ignoring normalization), it is possible to take its square root (Equation (8)). The geometric mean (Equation (8)), much like the arithmetic mean (Equation (7)), is a physical density (not density squared), but like Equation (6), it acts as a logical conjunction because filament voxels with simultaneously high $FPD$ and $BPD$ are rewarded more than surrounding noise (albeit at a compressed dynamic range because of the use of the square root).

Finally, the minimum function (Equation (9)) is tested. Like the geometric mean (Equation (8)),

it is a density and acts as a logical conjunction (because filament voxels with simultaneously high *FPD* and *BPD* are rewarded more). However, the dynamic range of the minimum function is not immediately obvious and requires further testing on actual density maps.

### 4.2.2 Candidate Seed Point Selection

It is often convenient to initiate automatic tracing from a given set of seed points. For example, Sazzed et al. [63] required the user to provide a seed point for each filament in a highly regular (hexagonally packed) actin bundle. Rusu et al. [40] used a genetic algorithm to find seeds for bidirectional tracing of isolated, irregular filaments. In particular, the manual placement of seeds is a tedious process and is only feasible when a small number of filaments are present. For hundreds of actin filaments forming loosely organized bundles with variable spacing among them (Figure 10), a manual seed point selection is not practical (and it is also subjective and not reproducible).

The newly developed CSP generation stage of our workflow (Figure 12) involves a spatial subdecomposition of the map into cubes of a user-defined size. For each cube, the voxel with the highest density value is considered a CSP. Here, cubes of $5 \times 5 \times 5$ voxels are used. (The CSP cube length was identical to the path length, $l$, hence providing a natural length scale for coarse-grained seed placement.)

All the local high-density voxels in the spatial decomposition are initially considered CSPs; however, it is not yet known whether any CFS generated from them constitute true filaments because the final traces are determined later (Figure 12). A direct determination of true seed points and corresponding true filament traces is computationally out of reach because of the low signal-to-noise ratio and missing-wedge artifacts present in tomograms [44]. Therefore, as described in the next section, the algorithm first generates a large number of CFSs from all the CSPs. Later, the CFSs pass through several rounds of screening to determine the true filaments (Figure 12).

### 4.2.3 Tracing of Candidate Filament Segments

From each CSP, a path of length $l$ (the same length as above) in the dominant forward direction ($+Y$ axis in this work) is traced to generate a set of short CFSs. Longer filaments will be fused from the short CFSs, if indicated, at a later stage. A CFS is represented by $((i_s, j_s, k_s) \rightarrow (i_e, j_e, k_e))$, where $(i_s, j_s, k_s)$ is the start voxel (i.e., the CSP), $(i_e, j_e, k_e)$ is the end voxel (i.e., the voxel with the maximum FPD after tracing a path of length $l$ from $(i_s, j_s, k_s)$ along the forward pyramid), and $l = j_e - j_s$. The CFS tracing uses forward processing algorithms described earlier(Equations (2), (4) and (5)).

In the current paper, the relatively narrow width of the accumulation zone, as shown in Fig-ure 14, inspired us to test an alternative straight-line density accumulation that is simpler to implement. In this alternative approach, instead of considering the density of the neighbor voxels (Equations (2) and (5)) for creating the CFS (($i_s$, $j_s$, $k_s$) → ($i_e$, $j_e$, $k_e$)) of length $l$, straight lines are considered, where lines are drawn from each of the seed points ($i_s$, $j_s$, $k_s$) to the target points ($i_e$, $j_s + l$, $k_e$). For a specific seed, the target point ($i_e$, $j_e$, $k_e$) is the point on the base of the search pyramid, $i_s − l ≤ i_e ≤ i_s + l$ and $k_s − l ≤ k_e + l ≤ k_s + l$, that exhibits the maximum *FPD* (Equation (2)) among all the base points. The intermediate voxels of each straight line are determined by interpolation of the $X(Y)$ and $Z(Y)$ indices, here using first-order Lagrange interpolating polynomials in $Y$, rounded to the nearest integer.

For each CFS, its normalized *FPD* (*NPD*) score is calculated, with a range between 0 and 1. The *FPD* values range from 0 to $l + 1$ because the densities $D(i, j, k)$ were previously normalized between 0 and 1. Subsequently, they are divided by $l + 1$ to obtain the *NPD*.

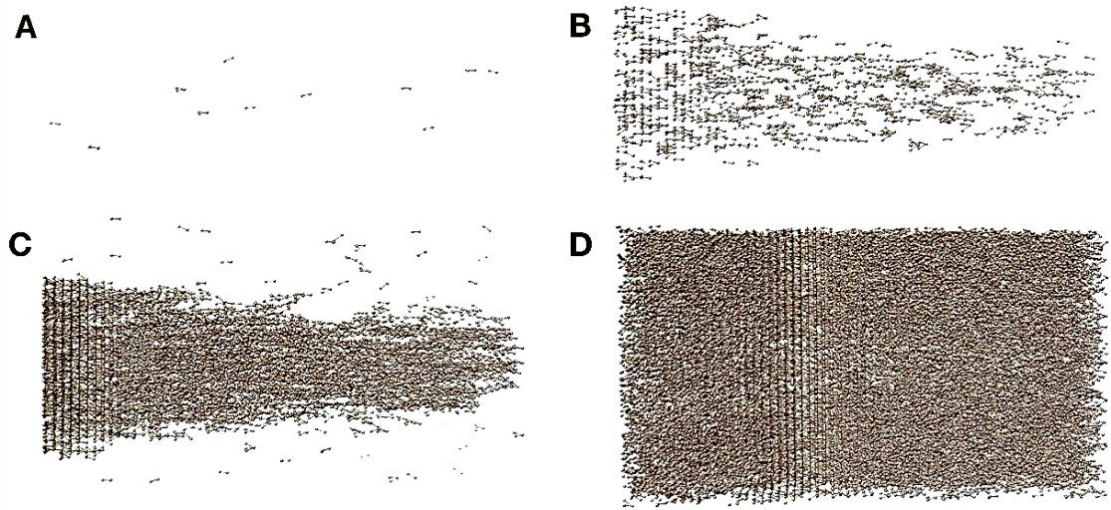

**Fig. 16.** The CFSs represent various bins: Bin 9 (A), Bin 8 (B), Bin 7 (C), and Bin 6 (D). These bins are sorted by the NPD values of the CFSs they contain.(Reproduced from [66])

### 4.2.4 Grouping and Selection of Candidate Filament Segments

Based on their *NPD* scores, the generated CFSs were grouped into 10 bins of a width of 0.1. Therefore, the bin numbers reflect the first floating point digit of the *NPD* values (e.g., bin 10 contains CFSs with *NPD* scores ranging from 0.9 to 1.0). It is observed that segments of true filaments are typically represented by CFSs in high-numbered bins. The gradual shift towards the lower-level bins leads to the appearance of false filament segments. For instance, in the simulated map, true filament segments are primarily represented by CFSs in bins 6–9, whereas false filament segments that are no longer localized in the expected filament region of the map are exhibited by CFSs in bins 5 and lower (see Figure 16)

To identify the bin that starts introducing false CFSs (referred to as threshold bin), the fact that false CFSs in the threshold bin spread to the full volume (Figure 16) are considered because they are mainly picking up noise. By iterating from high to low numbered bins, it is automatically determined at which bin value the CFSs are no longer localized and spread to the full volume. The tomogram is decomposed into $100 \times 100 \times 100$ voxel cubes, which is an intermediate level of detail between the fine CSP grid and global map size. If it is determined that less than 10 CFS midpoints are contained in at least 15% of the cubes, it is concluded that the entire volume is not yet occupied by the CFSs, and the testing proceeds to the next lower bin. In this manner, the approach identifies the uppermost bins that predominantly represent true CFSs. It's important to note that the threshold bin number is not a fixed value; it can vary in different maps depending on the density distribution. For instance, in our simulated tomograms, different indices for threshold bin are observed under varying levels of noise.

Subsequently, in another refinement step, the CFSs of the selected bins are further screened based on backward tracing. Backward tracing helps determine whether a CFS truly represents a filament segment because it is expected that the traces of a filament should be similar in both directions. Specifically, the CFS endpoint from forward tracing is selected and retraced backward using *BFD*. False CFSs are then excluded based on their dissimilar forward and backward trace orientations (angle tolerance: $30°$). Because the algorithm is sufficiently fast to generate CFSs, this retracing step does not introduce any significant computational overhead.

### 4.2.5 Fusion of Filament Segments

The final stage of filament formation employs multiple strategies to fuse surviving CFSs into longer individual filaments:

*Connecting CFS by collinearity:* This step considers the collinearity to connect adjacent filaments that exhibit the same orientation and represent fragments of the same filament. A pair of CFS that are collinear or nearly collinear (0 to $6°$ angle) and are very close (or connected) along the primary axis of the filament (distance between 0 to 10 voxels) are assumed to represent the same filament and are consequently merged to generate a single, longer CFS.

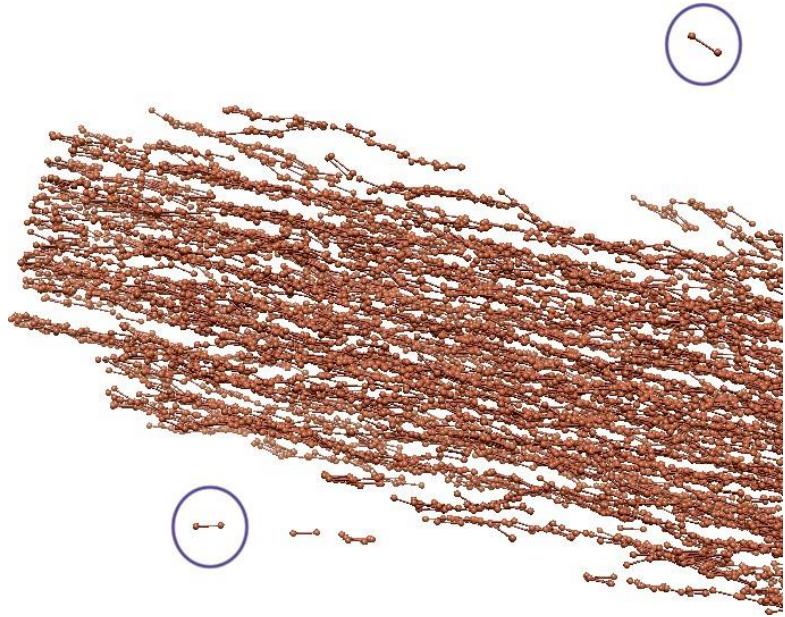

**Fig. 17.** Examples of isolated CFSs (inside blue circle) that can be removed considering the frequency of CFSs in the local neighborhood.(Reproduced from [66])

*Removal of isolated CFS:* An additional screening step automatically excludes spatially distant CFSs that are arbitrarily generated because of the presence of noise in the tomogram. A small number of false CFSs can exhibit moderate to high *NPD* scores. If these CFSs are indeed caused by erratic, local noise and not by nearby filaments (that are populated by other CFSs), these spurious CFSs can be detected and excluded by considering the mutual separation of CFSs in a local region (Figure 7 in [68]). To determine whether a CFS is false (isolated), its center is first computed, and then, the number of other CFSs present within a sphere of a radius of 20 voxels are determined. If the number of neighboring CFSs is less than three, the CFS is considered false and excluded accordingly.

*Extending and fusing the CFS:* Because of the noise present in the cryo-ET map, a filament may not exhibit a homogeneous density distribution along its length. Therefore, it is possible that the fragments of the filament (i.e., CFSs) fall into multiple bins; some may even fall below the threshold density and are excluded in the above screening. To bridge between such weaker segments of the same filament, the surviving CFSs are extended as follows: Each CFS of length *l* is gradually extended in the forward direction by repeatedly adding a new segment of length *l* but only if this new segment has an *NPD* score of at least that of the threshold bin. This process continues until the *NPD* value of the newly generated segment falls below that of the threshold bin or until a maximum length of $5l$ is reached (this empirical limit caps the number of overlapping traces, which will need to be reduced below, but the exact multiplier of $l$ has little effect on the final results).

The current set of CFSs is then fused using a directional traversing algorithm. Specifically, all voxels on the CFSs that survived the previous steps as filament voxels (FVs) are labeled. Starting from one FV as a seed, the framework iteratively traverses in the main filament direction ($+Y$ in this work) by considering that the possible range of movement along the $X$-axis and $Z$-axis is half of the $Y$-axis movement. For example, using relative grid indices for an FV (0,0,0) and connection range of value 2, the algorithm first checks whether voxel (0,1,0) is an FV; if not, it checks whether any of the voxels (0,2,0), (0,2,1), (0,2,−1), (1,2,1), (1,2,−1), (−1,2,−1), (1,2,0), or (−1,2,0) are FVs. If a new FV is found, it is selected as a second voxel of the FS, and the search continues until no FV is found, which marks the end point of that FS. Then, a step that involves traversing again from another FV, which does not belong to any of the existing FSs yet, and follow the same procedure. This process continues until all FVs are assigned to their final FSs.

*Excluding redundant filament segments:* This final refinement step excludes short FSs that overlap with a longer FS along the dominant axis of filaments. By discarding the spurious FS, this step can also help distinguish true filaments from noise artifacts. FSs that have more than 90% voxels in common with a longer FS are automatically discarded (Figure 8 in [68]).

## 4.3 RESULTS AND DISCUSSION

### 4.3.1 Filamentous Pattern Enhancement

Figure 15 illustrates a comparison between the original map and maps after DP-enhancement of the filament pattern. In the original density map (Figure 15A), the distinction between the density levels of filament and noise is not always obvious; thus, computationally separating them may not be possible. In the CPD maps (Figure 15B–E), the filamentous patterns are enhanced and more distinguishable from the noise.

The final FS results of one typical case are compared with the ground truth manual tracing in Figure 18. The persistence length of pure actin filaments is on the order of 10 $\mu$ m, which is more than three orders of magnitude longer than the length $l$=5 voxels, or 4.7 nm. Nevertheless, some of our traces in Figure 18 clearly follow curved paths that are picked up by the tracing algorithm on short scales that justify the use of l=5. This is not surprising because the curved filaments on this scale have been detected by manual tracing (Figure 10), and there may be a biological interpretation—for example, because of the cross-linking of filaments and the decoration of filamentous actin with other proteins.

As of the different normalized CPD distributions, it is hard to predict from Fig. 15 which denoising strategy will enhance the filament density most relative to the noise level. Therefore, a quantitative analysis has been performed of the tracing performance of these four cases in the following section. Quantitative validation is also important for ruling out any errors introduced by the positional and directional granularity of our approach. (Given the short CFS length l=5 voxels and restriction of CFS end points to voxel positions on the 3D grid, our approach is limited to about 90 degrees directional and 1 voxel = 0.947 nm positional granularity).

### 4.3.2 Evaluation of Tracing Results

To assess the performance of the proposed tracing framework, an *F1* score–based statistical measure is employed. The criteria for the *F1* score calculation are similar to those used by [69], except predicted FVs are dilated by one voxel to balance recall and precision values (see below). The ground truth FVs determined by the manual annotation [11]) are compared with those of the automatically traced filaments. True positive (TP), false-positive (FP), and false-negative (FN) voxels are defined as follows:

*True Positive*: A predicted FV is considered a TP prediction if a true FV exists within 3 voxels in any direction.

*False Positive*: A predicted FV is classified as an FP voxel if no true FV exists within 3 voxels in any direction.

*False Negative*: A FV in the ground truth map is considered an FN if no predicted FV is found within 3 voxels in any direction.

Based on the computed *TP*, *FP*, and *FN* values, the recall ($R$), precision ($P$), and $F1$ ($F1$) scores are calculated using the following equations:

$$R = \frac{TP}{TP + FN} \tag{10}$$

$$P = \frac{TP}{TP + FP} \tag{11}$$

$$F1 = \frac{2 \times P \times R}{P + R} \tag{12}$$

Table 2 shows the precision, recall, and *F1* scores of the proposed framework when applied to the simulated tomograms of various levels of noise and using different processing methods. As a performance reference, the exclusive DP-based approach (Table 2, left) provides a very high *F1* score of 0.97 for a simulated map of a noise level of 0.4. The very high recall score (0.99) suggests that the framework identifies almost all the filaments present in the simulated tomogram at this noise level. As the noise level increases, it negatively affects the performance, even though the obtained *F1* scores can still be considered very good. Even at a noise level of 1.0, the DP-based approach shows an *F1* score of 0.86, suggesting that it is capable of detecting filaments in the worst-case map (see the Methods section), albeit with some minor inaccuracies. With the current settings, the precision score is slightly lower than the recall score in the pure DP approach (Table 2, left) because of the remaining FPs, as can be expected given the noisy nature of the tomogram. Nevertheless, the observed F1 scores from 0.86–0.95 (for noise levels 0.6–1.0) are quite high for a density-based structure prediction.

**Table 2.** A performance comparison of the proposed DP-based framework with the line-based approach for tracing actin filaments without density enhancement preprocessing at various levels of noise. UND = undefined because the FP and TP values are both zero.(Reproduced from [66])

| Noise | DP-Tracing w/DP-Enhancement | | | Line-Tracing w/DP-Enhancement | | | DP-Tracing w/o Enhancement | | |
|---|---|---|---|---|---|---|---|---|---|
| | Pre. | Rec. | *F1* | Pre. | Rec. | *F1* | Pre. | Rec. | *F1* |
| 0.4 | 0.945 | 0.994 | 0.969 | 0.591 | 0.988 | 0.740 | 0.963 | 0.909 | 0.935 |
| 0.6 | 0.923 | 0.978 | 0.950 | 0.603 | 0.962 | 0.742 | 0.952 | 0.878 | 0.913 |
| 0.8 | 0.848 | 0.965 | 0.903 | 0.568 | 0.940 | 0.709 | UND | 0 | UND |
| 1.0 | 0.828 | 0.898 | 0.861 | 0.575 | 0.813 | 0.674 | UND | 0 | UND |

Table 2 also provides a comparison of the proposed DP-based framework to two alternative workflows, which are explained in the earlier section, one with an alternative filament tracing based on straight-line density accumulation and one without the filamentous pattern enhancement (denoising).

Substituting the line-based tracing in the workflow (Table 2, center) significantly lowers the precision and *F1* scores compared with the pure DP approach, mainly because of a significant increase in the number of FP filaments. The results suggest that (at least with the current settings) any efficiency gain by the simpler approach comes at too high of a performance cost. Because of the apparent over-interpretation of the noise, the line-based tracing is not implemented in the more expensive preprocessing stage, where noise suppression is crucial.

For a better-quality map (0.4 and 0.6 noise level), the denoising (Table 2, left) provides only a modest benefit (compared with Table 2, right) and may not be necessary. However, at higher noise levels the detection breaks down (Table 2, right). Unless denoising is used, no filaments are detected for noise levels $\geq 0.8$ (at least with the same parameter settings as in the other cases).

Table 3 shows the performance of the proposed DP-based framework with the alternative addition/arithmetic mean (Equation (7)), geometric mean (Equation (8)), and minimum (Equation (9)) blending functions. The results correspond to those of multiplication (Equation (6)) shown in Table 2 in the left column. As can be seen in Table 3 (center, right), the performance of both geometric mean and minimum degrades at higher noise levels. This is because the compressed dynamic range of the density distributions impedes the binning and associated screening of CFSs, such that many FN filaments remain undetected, which then lowers the recall. Only the arithmetic mean blending manages to achieve similar *F1* scores. However, the high value at noise level 1.0 is likely an outlier because the trend at lower noise levels is irregular, as exemplified by low recall values at noise level 0.8 (which exhibits binning problems because of the lower dynamic range). At least with the current binning approach, multiplication remains the most consistently well-performing blending function across all noise levels.

**Table 3.** Performance of DP-based tracing incorporating three different blending techniques: addition/arithmetic mean (Equation (7)), geometric mean (Equation (8)), and minimum (Equation (9)). (Reproduced from [66])

| Noise | Addition | | | Square-Root | | | Minimum | | |
|---|---|---|---|---|---|---|---|---|---|
| | **Pre.** | **Rec.** | *F1* | **Pre.** | **Rec.** | *F1* | **Pre.** | **Rec.** | *F1* |
| 0.4 | 0.945 | 0.994 | 0.969 | 0.950 | 0.982 | 0.966 | 0.946 | 0.99 | 0.968 |
| 0.6 | 0.948 | 0.806 | 0.806 | 0.967 | 0.250 | 0.397 | 0.95 | 0.35 | 0.514 |
| 0.8 | 0.926 | 0.686 | 0.788 | 0.936 | 0.176 | 0.296 | 0.938 | 0.444 | 0.602 |
| 1.0 | 0.939 | 0.879 | 0.907 | 0.874 | 0.138 | 0.239 | 0.904 | 0.139 | 0.241 |

### 4.3.3 Algorithm Run Times

The tracing part of our framework is computationally highly efficient, but the preprocessing step (which is a requirement for noisy maps, as indicated in our *F1* score analysis) takes more time. On an Apple MacBook Pro with a 2.6 GHz Intel Core i7 processor, it takes around 3 min to trace filaments in a simulated tomogram with a size of 283 $\times$ 664 $\times$ 269 voxels. In contrast, the denoising of all voxels by the preprocessing stage is slower, taking about 5 h. The tracing stage of the filaments is much more efficient than the denoising stage because the coarse-grained selection of the CSPs is based on the highest density voxels in the spatial subdivision. For $l = 5$, the CSP generation selects one voxel out of $5 \times 5 \times 5 = 125$ voxels, whereas the preprocessing acts on all the voxels in the map. Moreover, in the preprocessing, each voxel is traced twice: once in the forward direction and once in the backward direction, whereas in the tracing stage, only the forward direction is required.

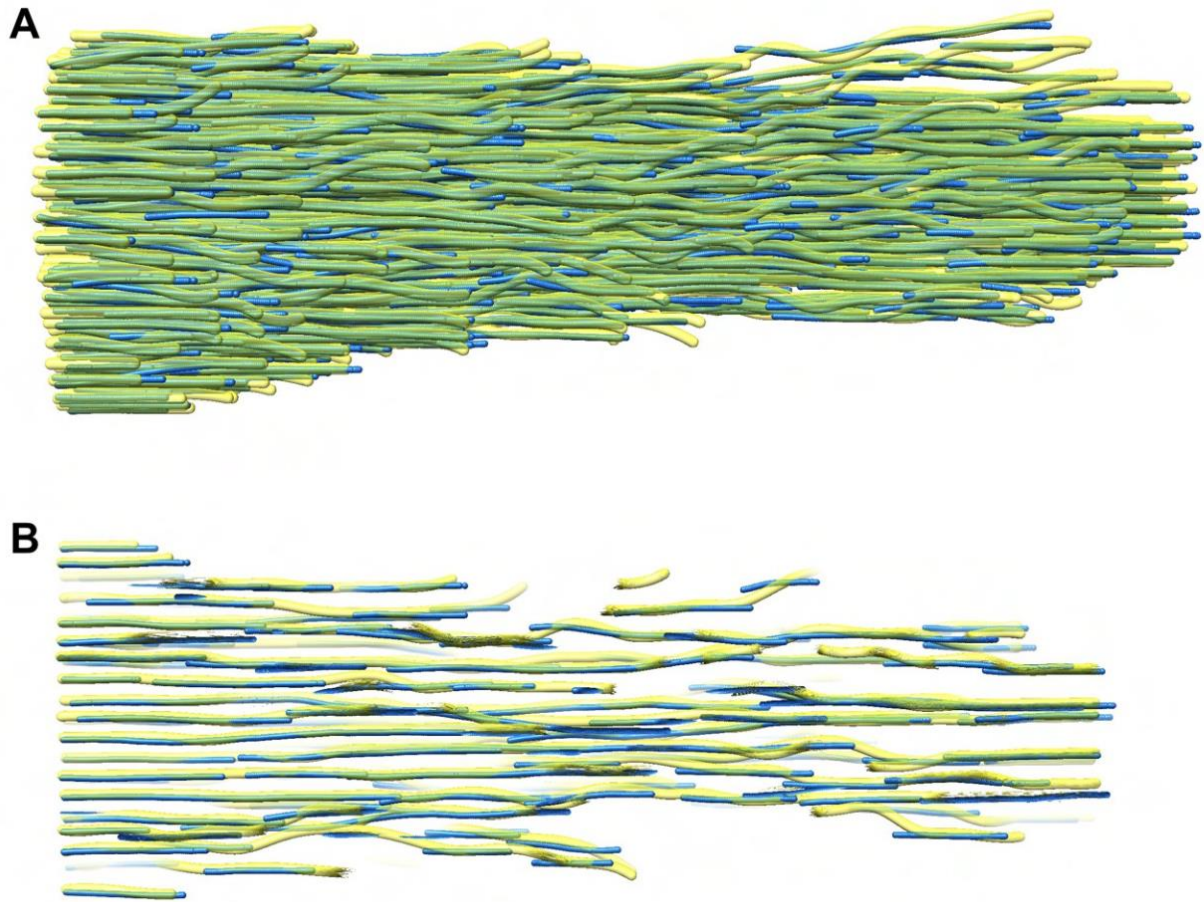

**Fig. 18.** Automatically detected FSs (solid blue; this work) superimposed by the ground truth manually traced filaments (transparent yellow, [11]). (A) The green areas in this full-view rendering indicate good agreement because of the subtractive color mixing. (B) A slab consisting of 10 $Z$-slices, which is taken from the center portion of (A), provides a detailed view of the individual filaments. The FSs (blue) have been obtained from the 0.60 noise level simulated map (Fig. 11B), which is closest to the automatically matched noise, here by using DP-based enhancement with multiplication, DP-based CFS tracing, and $l = 5$.(Reproduced from [66])

## 4.4 SUMMARY

Spaghetti Tracer, a fast and fully automatic framework, has been proposed for tracing semiregular actin filaments. The proposed DP-based approach is a spatial domain technique that is applied directly to the voxels of the 3D tomogram. The algorithm's performance has been quantitatively validated through the use of simulated maps based on a known model.

The neighborhood-based density accumulation scheme (Figure 14) enables robustness in tracing because clean filament densities are not visible that could be picked up by a thin line accumulation). The pyramidal search window ensures that the $PD$s are large when filaments pass through the tip of the pyramid while providing some robustness against noise because individual voxel densities are replaced with $PD$s. The tracing assumes that filaments are oriented in a mean direction and bundled, even though individual filaments may deviate by up to 45° from the main direction because of the search pyramid. The key advantage of our current methodology lies in its capability to accurately analyze dense filament bundles while maintaining the ability to track individual curved filaments.

Among the various blending functions, the multiplication of forward and backward path densities provides for an efficient filter lifting the filament density above the noise level. Our results (Table 2) show that such denoising is crucial for the detection of filaments in lower-quality maps. There are alternative denoising filters already in use in tomography [70, 71, 72], but generally, the earlier filters have made no assumptions about the shape of the biological structures. For example, the earlier work of Starosolski et al. [70] also considered a path density–based filtering, but numerically expensive random walks were required to sample the density map isotropically. In contrast, our bidirectional filter is designed for filaments that are mainly oriented in the mean direction of the bundle, where it is possible to take advantage of this known direction to develop a more efficient approach.

The tracing stage of our framework is very fast, it only takes minutes facilitated by the spatial coarse graining of the CSPs. However, the denoising of all tomogram voxels is a current bottleneck and still takes several hours on a standard computer. Note that in both the forward and backward directions, density is accumulated from the origin (initialized in Equation (4)). Therefore, DP is performed only locally for each voxel, and the full density map needs to be processed exhaustively, which is expensive.

# CHAPTER 5

# UNTANGLING IRREGULAR ACTIN CYTOSKELETON ARCHITECTURES IN TOMOGRAMS OF THE CELL WITH *STRUWWEL TRACER*

## 5.1 INTRODUCTION

Shape, motility, and transport within a eukaryotic cell are based on an extensive actin cytoskeleton [73]. To visualize actin filaments in their native state, researchers commonly employ cryo-electron tomography (cryo-ET), a specialized imaging technique that enables 3D insight into the internal cellular structure. Cryo-ET has recently shown that actin filaments provide a "missing evolutionary link" between archaea and complex eukaryotic life forms, such as animals and plants [74]. The cryo-ET by Pilhofer's lab at ETH Zurich supports an emerging hypothesis that extensive cytoskeletal actin structures arose first in the Asgard archaea, before the appearance of the first eukaryotes on Earth, and could therefore have contributed to the emergence of complex organisms [75]. Moreover, the observed cytoskeletal protrusions in the archaea suggest a detailed mechanism for eukaryogenesis, in which a primordial Asgard archaeon (the closest known relative of eukaryotes) interacts with the predecessor of the bacterial endosymbiont by means of the actin-powered protrusions and eventually endogenizes it [75, 74].

Cryo-ET involves capturing a series of 2D images of a specimen at extremely low temperatures, preserving its frozen state while maintaining its hydrated environment. This enables detailed imaging of actin filaments without disrupting their native structure. These 2D projections are subsequently aligned using various computational techniques, collectively referred to as 3D reconstruction, to generate the 3D tomogram. The 3-5 nm resolution tomograms obtained by cryo-ET typically present considerable noise, induced by the limited electron dose used in the image acquisition, and they exhibit directional artifacts from the absence of certain view directions ("missing wedge" in Fourier space [44]), caused by the limited tilt range of the specimen holder in the microscope.

Determining the organization of actin filaments is highly important as they are responsible for a variety of cellular processes, such as muscle contraction, transport, and cell motility [73], and therefore it can aid in characterizing experimental, dynamic, or pathological changes in the cytoskeleton [69]. Actin filaments are thin and flexible. They assemble to form diverse organizational patterns, from hexagonal close-packed bundles in the shaft of hair cell stereocilia [63], to directionally biased semi-ordered strands in the stereocillia taper region [44, 66] to irregular, randomly oriented actin networks in filopodia [40]. The three types of actin organization can be likened to the appearance of a person's hair in various stages of tidiness (brushed, windblown, unkempt), and each type requires specialized algorithms for their computational interpretation. Our recent algorithm development was mainly focused on ordered or semi-ordered patterns in stereocilia, where filaments form strands with a mean direction that can be exploited in directional denoising or deconvolution [63, 44, 66]. In many cells, however, such as in the ancestral archaea [74], the filaments are randomly oriented and form highly irregular networks, which defy any directionally biased denoising approaches. In this work, a novel approach, *Struwwel Tracer* is presented, to quantitatively organize such important disorganized actin networks. (The tool was named after Heinrich Hoffmann's famously shaggy Struwwelpeter character.)

The presence of noise, artifacts, and other structures, such as actin-binding proteins, can obfuscate the filaments and make the tracing difficult. Distinguishing true filaments from noise or other structures in tomograms requires careful examination and expertise that ultimately make manual tracing of actin filaments a labor-intensive task that requires significant time and effort [11, 40, 43]. In addition, manual tracing is often subjective in nature and relies heavily on the annotator's interpretation. The reliability and reproducibility of the annotation procedure are greatly compromised by factors such as low contrast, various types of artifacts, and inherent uncertainty. This is particularly evident when the procedure is applied directly to whole cell tomograms (i.e., without any subtomogram averaging) with typical resolutions below 2 nm and varying in orientation due to the missing wedge effect. Moreover, when dealing with large data sets containing 2D image stacks, the repetitive nature of the annotation task increases the chance of error, such as an inadvertent skipping between neighboring filaments. To mitigate these potential errors and to make the tracing objective and reproducible, it becomes necessary to develop automated approaches that can be independently validated.

Over the years, a number of automatic methods have been developed to trace actin filaments [40, 43, 44, 1, 76, 77]; however, the existing methods have certain limitations:

1. A significant portion of these approaches can be ruled out for the present work because they are only applicable to actin filaments observed in relatively clean light or confocal microscopy images [47, 78].

2. Among the cryo-ET-related methods, some are only applicable for the tracing of well-ordered actin filaments [63, 44, 66]. Among these, one noteworthy tool is our *Spaghetti Tracer* approach [66]. *Spaghetti Tracer* introduced a paradigm shift in the tracing of semi-regular actin filaments because it is a dynamic programming-based method at the voxel level that does not require an expensive missing wedge correction, template convolution, or de-convolution. Therefore, it yields a substantial improvement in time efficiency over template convolution [40, 43] or deconvolution methods [44], enabling fast and accurate tracing of such filament arrangements. (The accuracy of *Spaghetti Tracer* was validated in a rigorous statistical analysis, achieving F1 scores of 0.86-0.95 on phantom tomograms at experimental conditions.) The success of *Spaghetti Tracer* motivated us to extend its capabilities to randomly oriented actin filaments with *Struwwel Tracer*.

3. There are very few earlier algorithms that are agnostic of the relative orientations and distances of the actin filaments, so that they can trace central lines of irregular filaments individually without leveraging information of adjacent filaments or requiring a mean direction. *Volume Tracer* [40] utilized an expensive genetic-algorithm-based search employing a population of cylindrical templates (combined with a bi-directional tracing) to detect randomly oriented filaments in *Dictyostelium discoideum* filopodia. The genetic algorithm was implemented as part of our group's free, open-source *Situs* and *Sculptor* packages, but it required extensive computational time on the order of several days when applied to a complete tomogram, without guaranteeing convergence (leading to false negatives when a user is impatient). Co-author Rigort also developed a similar template-matching method independently [43]. This approach was implemented in *Amira*, a commercial software requiring a paid license and limiting any algorithmic modifications by third parties or end users.

4. Recently, a number of deep-learning-based approaches were proposed for the segmentation of diverse biological assemblies, including actin [46, 79]. For example, [46] presented a deep-learning-based segmentation approach for a voxel-level classification of shapes of interest in the tomogram and integrated the approach into the *EMAN2* [28] package. However, these segmentation tools are generic in nature and are not specifically designed for filamentous shape structures. Besides, they require users to annotate training data and fine-tune the deep learning model, which could be a laborious process. These segmentation approaches only provide a voxel-level density segmentation, without any tracing of central lines, thus, they cannot be considered complete filament tracing frameworks. Recent studies that have used these segmentation methods subsequently required tools from the commercial *Amira* software for additional processing (such as the above template matching) [1, 76, 77].

In summary, the current art of identifying actin filament networks in cryo-ET has several drawbacks, including a fragmentation of tools and packages, the prerequisite to buy commercial software (*Amira*), complex manual interventions (e.g., manual annotation for segmentation or the training of deep learning networks), and computational expense. Our newly proposed approach can address all these drawbacks. *Struwwel Tracer* is an efficient, accurate, free, open-source tool for tracing randomly oriented filaments in actin networks.

## 5.2 MATERIALS AND METHODS

Filaments in an irregular actin network, such as those found in filopodia, lamellipodia, or stress fibers, are particularly challenging to trace due to their arbitrary orientations and potential overlap or branching with other filaments. Here, simulated phantom tomograms are generated from ground truth traces to provide a controlled basis for our method development and validation. Subsequently, various stages of the proposed tracing framework are described, shown in Figure 19, to address such challenges in the tracing of actin filaments with arbitrary orientations. First, how the framework detects local seed points in the map are defined, from which it generates short candidate filament segments (CFSs). Next, numerical details are provided on how path densities are accumulated by exploring all potential filament paths within $45°$ search pyramids from the $x$, $y$, and $z$ directions. After placing CFSs across the map based on the maxima of path densities, a manual segmentation can be efficiently performed by visualizing an intermediate "pruning map" in a third-party molecular graphics program. Finally, algorithmic details for how the surviving CFSs are iteratively fused into longer, curved filaments based on their relative orientations and gap spacings after extension are provided.

### 5.2.1 Simulated Phantom Tomograms

Our group previously introduced *TomoSim* [67], a tomogram simulation approach that leverages pre-existing filament traces to generate authentic phantom tomograms whose noise color, strength, and missing wedge were matched to experimental tomograms. To establish a reliable benchmark for validating *Struwwel Tracer*, actin traces of a *Dictyostelium discoideum* filopodium is utilized. A spline interpolated from the filament trace in Supplementary Data 5 from Rusu et al. [40] was volumized using a Gaussian filter with a full width at half maximum of 9 nm with a voxel spacing of 1.912 nm.

The simulation of a tomogram aims to replicate the noise and missing wedge artifacts typically encountered in experimental maps. The simulated tomogram does not incorporate non-filamentous biological features, such as membranes. Consequently, the simulated tomogram serves as a well-defined ground truth specifically tailored to assess the accuracy of any filament tracing framework. An experimental tomogram obtained from Supplementary Data 2 from Rusu et al. [40] is used for noise-color modeling and signal-to-noise ratio (SNR) calibration. A wedge is masked in the frequency domain to emulate a tilt range of $\pm 45°$.

The signal strength is matched to the experimental tomogram's SNR using the color-matched noise profile, and used as the basis for testing our tracing against a range of noise levels. Noise levels are quantified as the noise intensity ratio relative to the experimental noise. For example, a noise level of 0.5 would have half the noise intensity of the experimental tomogram and twice the SNR. The experimental map used for the calibration is the raw tomogram prior to any denoising or local normalization (which is typically performed on a tomogram prior to tracing [40]). Therefore, a noise level of 1.0 is a worst-case scenario that is not encountered in tracing practice. To simulate the effect of the missing denoising step, a range of five equally spaced noise levels from 0.35 to 0.95 are generated, as normalized by the worst-case level (noise is not calibrated on the processed tomogram [40] to avoid any subjective bias from that study, and instead used a range of reduced noise levels as calibrated from the raw tomogram). Note that the absolute noise scale in simulated tomograms is not strictly defined anyway because of the hidden misalignment of unknown true filament densities with the prescribed ground truth traces [67]; therefore, the relative noise strengths are more relevant than the absolute scale.

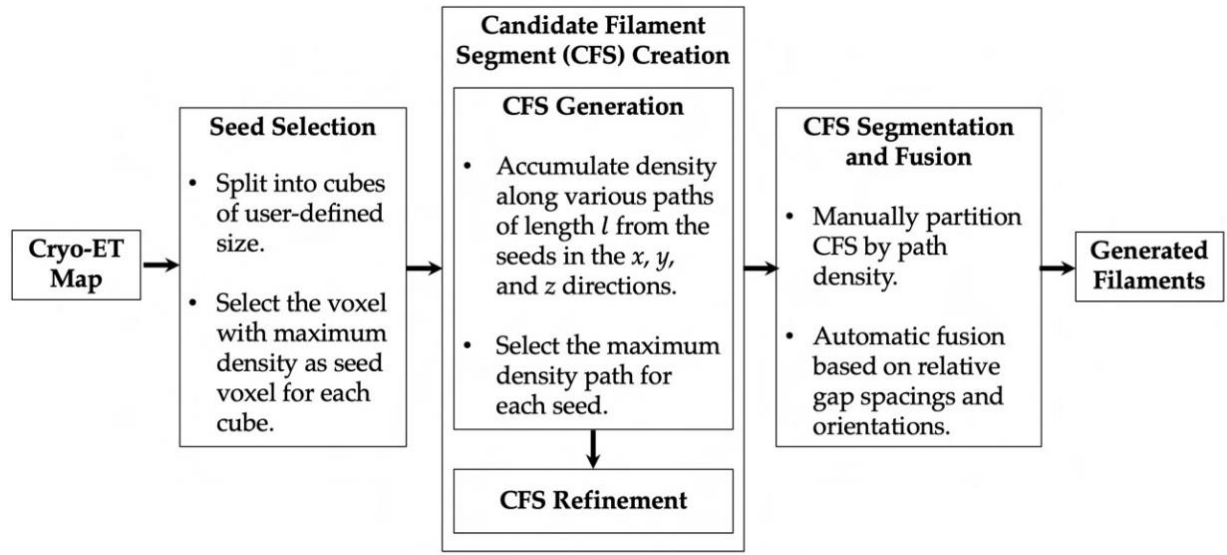

**Fig. 19.** Overall flow diagram of *Struwwel Tracer*.(Reproduced from [80])

## 5.2.2 Automatic Seed Selection

The detection of filaments in cellular tomograms is initiated from suitably chosen high-density seed points. Manual filament tracing relies on user-provided seed points, which were considered as the ground truth in previous work [63]. However, due to the labor-intensive tagging, such manual seed point selection is relatively sparse and introduces uncertainty and a subjective bias. An automated approach is less concerned with the expense of the selection and can consider a denser oversampling of potential seed points, but it might introduce false positives. Therefore, instead of considering seed points as definitive ground truth, they are utilized solely as starting points for the CFSs, which will be screened and refined later.

For an automatic seed point selection on a 3D grid, the high-density voxels within a local neighborhood are identified. To achieve this, the tomogram is sub-partitioned into non-overlapping 3D cubes along each axis. The voxel with the highest density value within each cube is then designated as a seed voxel. In semi-regular filaments [66], the cube side length was chosen to be identical to the CFS length $l$ in the CFS creation stage (below), which provides a natural length scale for the coarse-grained seed placement. However, in irregular actin networks, it is expected the final seeds to be more irregularly distributed in the search cubes, so a denser sampling with grid spacing $l/2$ was implemented in *Struwwel Tracer*. For this research, a CFS length of $l = 10$ voxels is used, resulting in seed search cubes of dimensions $5 \times 5 \times 5$ voxels. (At a voxel spacing of 1.912 nm in the phantom tomograms above, five voxels correspond approximately to the 9 nm actin filament diameter, so the seed point density should afford a complete detection of filaments).

It is assumed that seed points with local high density are necessary but not sufficient prerequisites for true filaments. Identifying true seed voxels (that are part of a true filament) is difficult at this stage. This difficulty arises from the low SNR and the presence of missing wedge artifacts or other structures in tomograms. The determination of whether a density segment that originates in a seed belongs to a filament or not is performed in a later stage of the approach. Consequently, the short filament segments originating from seed voxels are not referred to as predicted filament segments, rather as *candidate* filament segments (CFSs).

## 5.2.3 CFS Creation

For this stage of the approach, (Fig. 19) comprises two sub-steps, CFS generation and CFS refinement.

**CFS Generation**

A CFS originates in the seed voxel $(i, j, k)$ and terminates in the end voxel $(i', j', k')$, which is determined for each seed $(i, j, k)$ by the following equations. The length $l$ of the CFS in this study is user defined and is equivalent to the infinity norm of the resulting CFS vector, $l = max\{i' - i, j' - j, k' - k\}$. As mentioned previously, a value of $l = 10$ voxels is used in this work.

On a cubic grid, let the local density at each voxel $D$ be globally normalized to a range of 0 to 1. Three separate path densities, denoted as $PD_{x,y,z}$, are initialized at the seed voxel $(i, j, k)$ for the three Cartesian $x$, $y$, and $z$ axes:

$$PD_{x,y,z}(i, j, k) = D(i, j, k) \tag{13}$$

These path densities are then accumulated for each Cartesian axis over $l$ voxels in the forward (positive) direction within search pyramids that have the seed $(i, j, k)$ as the vertex and can deviate by up to $45°$ from the corresponding axis. The search pyramids, shown in green ($x$), red ($y$), and blue ($z$) in Figure 20, will be formally defined below. The pyramidal search approach was inspired by the *Spaghetti Tracer* algorithm [66]. However, unlike the earlier method, which was focused on a single dominant direction, no assumption is made about CFS directionality with *Struwwel Tracer*. Therefore, three Cartesian search pyramids will be combined in the following, to cover all possible orientations.

At each intermediate voxel $(i'', j'', k'')$ within a search pyramid, the corresponding path density (initialized in Eqn. 13) is then accumulated from up to nine contributing neighbor voxels from a previous cross-section. To illustrate, when tracing in the $y$ direction (red in Fig. 20), the path density $PD_y(i'', j'', k'')$ is accumulated from the preceding $y$-slice, $j'' - 1$ by mathematical induction. The accumulation scheme, as shown in more detail in Figure 4 of [68], is applied to all intermediate voxels within the search pyramids:

$$PD_y(i'', j'', k'') = D(i'', j'', k'') + \max_{\substack{m,n \in \{-1,0,1\} \\ \text{(if contributing)}}} PD_y(i'' + m, j'' - 1, k'' + n). \tag{14}$$

Note that only neighbors $m, n \in \{-1, 0, 1\}$ within the search pyramid are considered. Therefore, as described in [66], the accumulation zone from the seed $(i, j, k)$ to the final endpoint $(i', j', k')$ is actually an intersection of two overlapping pyramids, i.e., path densities are accumulated only in a relatively localized region. Similarly, the $PD_x(i'', j'', k'')$ and $PD_z(i'', j'', k'')$ are accumulated in their respective directions, with indices permuted accordingly:

$$PD_x(i'', j'', k'') = D(i'', j'', k'') +$$
$$\max_{\substack{m,n \in \{-1,0,1\} \\ \text{(if contributing)}}} PD_x(i''-1, j''+m, k''+n), \tag{15}$$

$$PD_z(i'', j'', k'') = D(i'', j'', k'') +$$
$$\max_{\substack{m,n \in \{-1,0,1\} \\ \text{(if contributing)}}} PD_z(i''-m, j''+n, k''-1). \tag{16}$$

The final forward path densities for green ($x$), red ($y$), and blue ($z$) search pyramids in Figure 20 are then computed on the forward-facing square bases of the pyramids:

$$FPD_x(i, j, k; l) = \max_{\substack{j-l \leq j' \leq j+l \\ k-l \leq k' \leq k+l}} PD_x(i+l, j', k'), \tag{17}$$

$$FPD_y(i, j, k; l) = \max_{\substack{i-l \leq i' \leq i+l \\ k-l \leq k' \leq k+l}} PD_y(i', j+l, k'), \tag{18}$$

$$FPD_z(i, j, k; l) = \max_{\substack{i-l \leq i' \leq i+l \\ j-l \leq j' \leq j+l}} PD_z(i', j', k+l). \tag{19}$$

The $FPD_{\{x,y,z\}}(i, j, k; l)$ values in equations 17, 18, and 19 range from 0 to $l + 1$ (number of voxels in the CFS). For the segmentation in the next stage of the algorithm, the maximum between the three $FPD$ values is divided by $l + 1$ to obtain a single normalized $NPD$ value that ranges from 0 to 1:

$$NPD(i, j, k; l) = \frac{\max\limits_{u \in \{x,y,z\}} FPD_u(i, j, k; l)}{l + 1}. \tag{20}$$

The combined 3D search region (roughly a hemisphere of directions comprising the three search pyramids in Fig. 20) accounts for all possible (unsigned) filament orientations. The final endpoint $(i', j', k')$ of the CFS is then defined as the point that exhibits the maximum $FPD_u$, according to the winning $u$-axis that defines the $NPD(i, j, k; l)$ in Equation 20.

The path-based density accumulation addresses the challenge posed by substantial noise levels in the tomogram, as discussed in prior work [68, 66]. Rather than solely focusing on the density of individual voxels, or on density measurements along predetermined axes, the algorithm aims to detect an elongated high-density path with a specific infinity norm $l$ and an orientation determined by the maximum attainable path density.

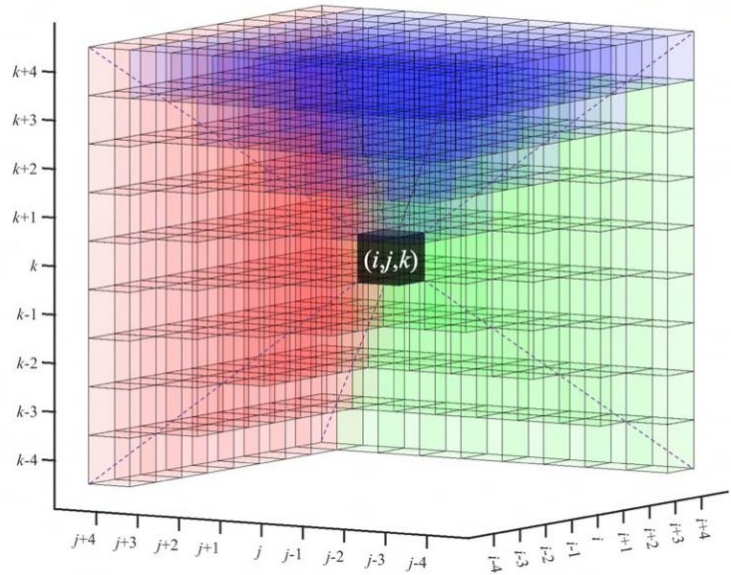

**Fig. 20.** Pyramidal density search regions along different axes based on equations 17 (green), 18 (red), and 19 (blue), starting from the seed voxel (black) at position ($i,j,k$). For a clearer illustration a CFS length (infinity-norm) $l = 4$ voxels is used for the search pyramid dimensions, whereas $l = 10$ was used elsewhere in this paper (see text).(Reproduced from [80])

## CFS Refinement

The forward CFSs are subjected to an additional screening step based on backward tracing. From the end point $(i^{'}, j^{'}, k^{'})$ of each forward CFS, a backward CFS with new end points $(i^{*}, j^{*}, k^{*})$ is constructed by applying the above tracing algorithm in reverse. It is expected that for a high-density CFS that is part of a true filament, the backward tracing will follow a comparable path in the reverse. Conversely, forward and backward CFS characterized by low density or noise, which are unlikely to be part of any filament, would diverge from each other. To determine the similarity of paths between forward and backward tracing, the angle between the points $(i, j, k)$, $(i^{'}, j^{'}, k^{'})$, and $(i^{*}, j^{*}, k^{*})$ is calculated, and a fixed angle threshold of $20^{\circ}$ is employed to screen out inconsistent forward and backward CFSs. An example of the intermediate results after the refinement stage is shown in Figure 21. The surviving CFSs (Fig. 21B) exhibit the preferred orientation of filaments in the actin-rich interior of the cell, whereas the exterior of the cell (lower left) yields unstructured CFS patterns that are not true filaments and will need to be filtered out in the following.

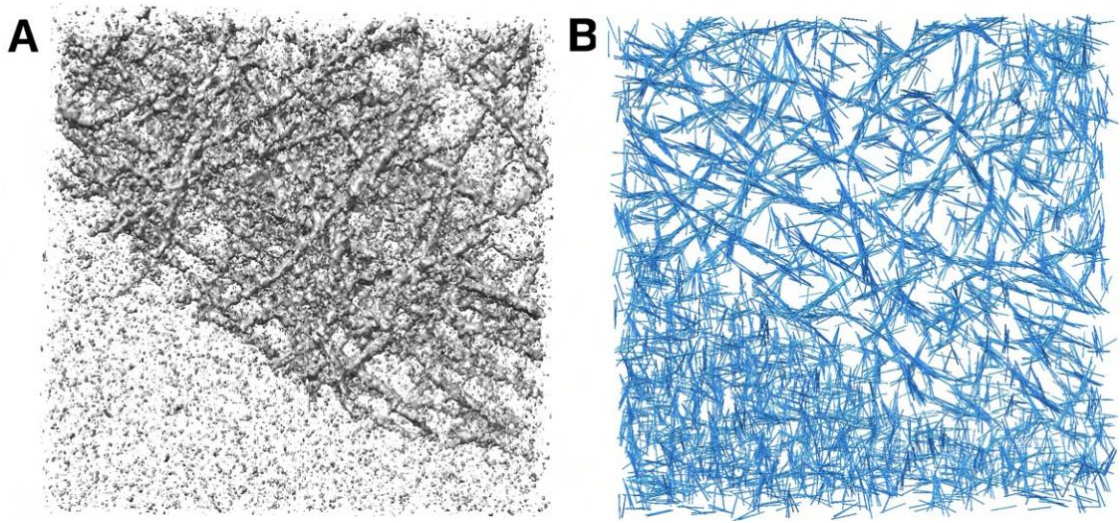

**Fig. 21.** (A) Density map of the simulated *Dictyostelium discoideum* filopodium tomogram at noise level 0.50 (shown at an isocontour level of mean + 1.5 × standard deviation). Note that the cell membrane was not simulated. (B) Filament segments of length $l = 10$ voxels, generated by applying the forward path-based density accumulation and followed by backward tracing refinement. All molecular graphics in this work was created with UCSF Chimera [65].(Reproduced from [80])

### 5.2.4 CFS Segmentation

Figure 21B illustrates the segmentation problem in the irregular actin networks that are the focus of this work. In earlier work on ordered actin filament bundles (with a dominant orientation), it is found that it is possible to screen out spurious CFSs (such as in the lower left corner) using an automated binning method that finds a suitable *NPD* threshold above which CFSs remain contained in the true filament region. CFSs with *NPD* values below this threshold were then eliminated automatically. A similar *NPD* threshold segmentation is proposed. However, irregular actin network CFSs exhibit low *NPD* contrast (there is no separate directional denoising stage as in regular filaments [66]) and cannot be assumed that irregular networks are localized in a specific region. Therefore, it would be desirable to fine-tune the threshold as a continuous parameter instead of prescribing discrete bin values. Therefore, a novel way is introduced to allow users to perform such a continuous thresholding quickly and conveniently with a third-party molecular graphics program such as VMD [81], Sculptor [38], or UCSF Chimera [65]. Our approach outputs a so-called "pruning map" that contains the *NPD* values masked by the CFS locations. Voxels within close proximity (i.e., one voxel) of the CFSs are assigned the *NPD* value (Eqn. 20) of the CFS. When a voxel can draw *NPD* values from multiple CFSs, the highest value is retained.

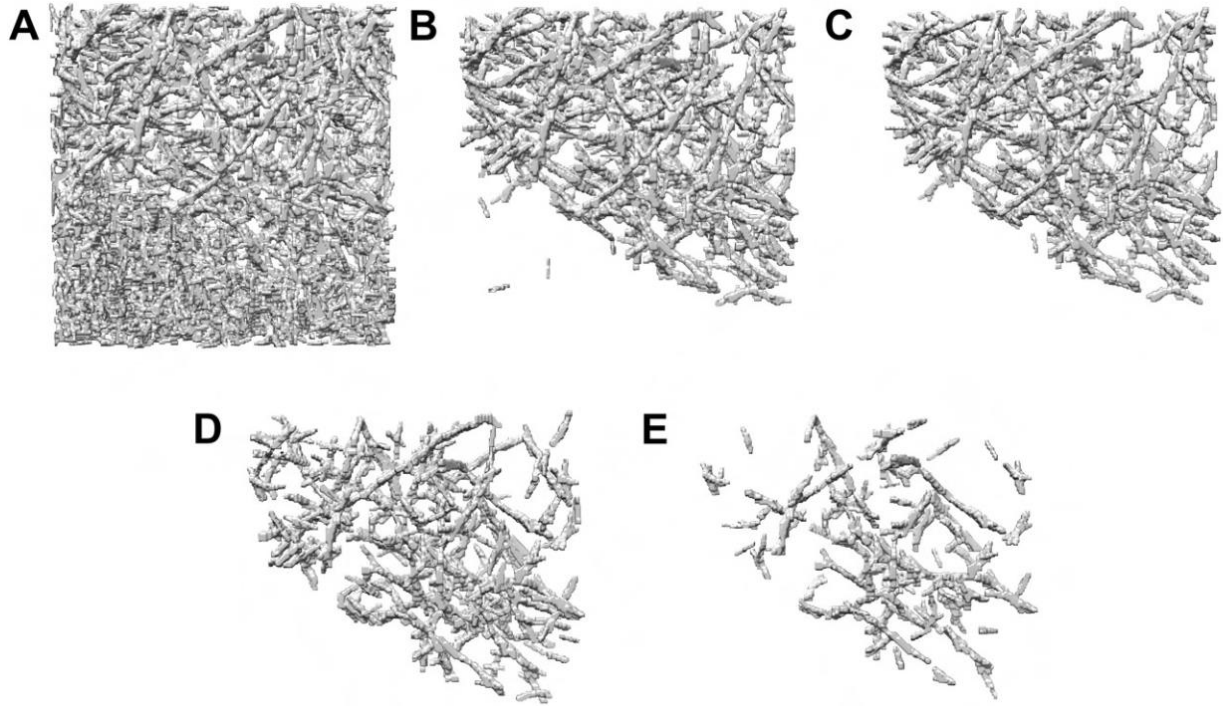

**Fig. 22.** Pruning map of the simulated tomogram in Figure 21A (noise level: 0.50) at different isocontour levels: (A) 0.40, (B) 0.45, (C) 0.50, (D) 0.55, and (E) 0.60.(Reproduced from [80])

In Figure 22, the pruning map returned by *Struwwel Tracer* is shown at different isocontour levels. By adjusting the level in the histogram window, users can visually determine the threshold in the pruning map (i.e., the threshold *NPD*) that represents likely filament segments. An overly low threshold would result in numerous false positives (Fig. 22A), whereas an overly high threshold would lead to a significant number of false negatives (Fig 22E). Visual examination indicates that isocontour values within the range depicted in Figure 22C and D are likely to represent the filaments accurately. Note that the subsequent steps of the algorithm are not very sensitive to the selected threshold value because of the screening and gap filling that still follow. The main purpose of the visual segmentation is to suppress spurious filaments, as is shown in Figure 22A, so a reasonable guess of the threshold parameter suffices in practice.

## 5.2.5 CFS Fusion

In the filament fusion stage, multiple geometric strategies are employed to combine short, linear CFSs (Fig. 21B) into curved, longer filaments. To account for the irregular organization of actin networks, the following fusion steps were specifically designed for *Struwwel Tracer* and are different from the collinearity test developed earlier for oriented filaments [66].

### Fusion Based on Physical Proximity

The initial fusion step considers the relative spacing and orientation of neighboring CFSs. Adjacent filaments that exhibit similar orientations (default angle tolerance of CFS center lines: $30^{\circ}$) will be connected. Whether two adjacent CFSs overlap or touch each other is determined based on their spacing along the center lines of the CFSs (with a default gap tolerance of 10 voxels). CFSs that meet the criteria are combined by connecting the end point of a CFS to the starting point of the successive CFS. Because it is not required that two CFSs exhibit perfectly matched orientations and positions before they are merged, the center line of the fused CFSs is smoothed through a process involving sampling, interpolation, and the removal of redundant points.

### Fusion by Extension

In addition to the proximity fusion of short CFSs, it is tested whether a one-time extension of CFSs by a length $l$ can extrapolate CFSs such that they make contact with another CFS using the same angle and spacing criteria. The extension step aims to fill up any noise-induced gaps present in the filaments. Due to the high noise levels and missing wedge artifacts in the tomograms, it is possible that true filaments exhibit regions of density at the level of the noise, which may be excluded by a high threshold during the visual inspection of the pruning map (Fig. 22). It is therefore examine whether the one-time extension by a length $l$ can help to fuse more CFSs. Similar to the previous step, the center lines is smoothed of fused CFSs by sampling, interpolation, and removal of redundant points.

## 5.3 RESULTS

To evaluate the performance of *Struwwel Tracer*, a comprehensive statistical F1 score analysis is performed on simulated tomograms of a *Dictyostelium discoideum* filopodium with a known ground truth. Different levels of noise were added to simulate realistic imaging conditions. Ad-ditionally, the effectiveness of *Struwwel Tracer* is demonstrated on an experimental tomogram of a mouse fibroblast lamellipodium that was interpreted for the first time by filament tracing. Finally,the implementation details and computation times of our software are reported.

### 5.3.1 Statistical Evaluation of Tracing Accuracy in Simulated Tomograms

Earlier works [68, 66, 69] have established a rigorous statistical evaluation protocol for testing the accuracy of filament tracing approaches. This protocol is based on simulated phantom tomograms generated from known filament traces under realistic conditions that match the noise profile and missing wedge properties of an experimental tomogram (see Materials and Methods). To evaluate *Struwwel Tracer*, the precision, recall, and F1 scores are cpomputed based on the observed agreement.

The ground truth filament voxels of simulated *Dictyostelium discoideum* filopodium maps (see Materials and Methods) are compared with the filament voxels predicted by the tracing. Due to the inherent high noise levels present in cryo-ET images, previous studies [69] have highlighted the limitations of conducting one-to-one voxel-level comparisons between ground truth and predictions. Therefore, here, a more pragmatic approach is used; rather than direct voxel-level comparison, the comparison based on a neighborhood range [69, 66] of three voxels are performed. The true positive (TP), false positive (FP), and false negative (FN) voxels are then computed in the following way:

*True Positive* If a ground truth filament voxel is found within a $3 \times 3 \times 3$ voxel neighborhood of a predicted voxel, it is considered TP.

*False Positive* If no ground truth voxel is found within a $3 \times 3 \times 3$ voxel neighborhood of a predicted voxel, it is considered FP.

*False Negative* If no predicted voxel is found within a $3 \times 3 \times 3$ neighborhood of the ground truth voxel, it is considered FN.

The recall ($R$), precision ($P$), and F1 scores are then computed according to

$$R = \frac{TP}{TP + FN},$$ (21)

$$P = \frac{TP}{TP + FP},$$ (22)

$$\text{F1} = \frac{2 * R * P}{R + P}.$$ (23)

Table 4 presents the precision, recall, and F1 scores obtained by *Struwwel Tracer* in simulated tomograms with varied levels of noise. *Struwwel Tracer* achieves a high F1 score of 0.90 when applied to the lowest noise map tested. The high precision score of 0.97 indicates that the framework recognizes mostly true filaments in the simulated tomogram (FP predictions are negligible). As the noise level increases, the performance degrades a bit due to slightly lower recall values (i.e., a small number of true filaments are missed); however, the obtained F1 scores still remain above 0.8. Given the inherent noise in the tomogram, the recall scores range slightly below the precision scores since missing filaments densities result in FNs.

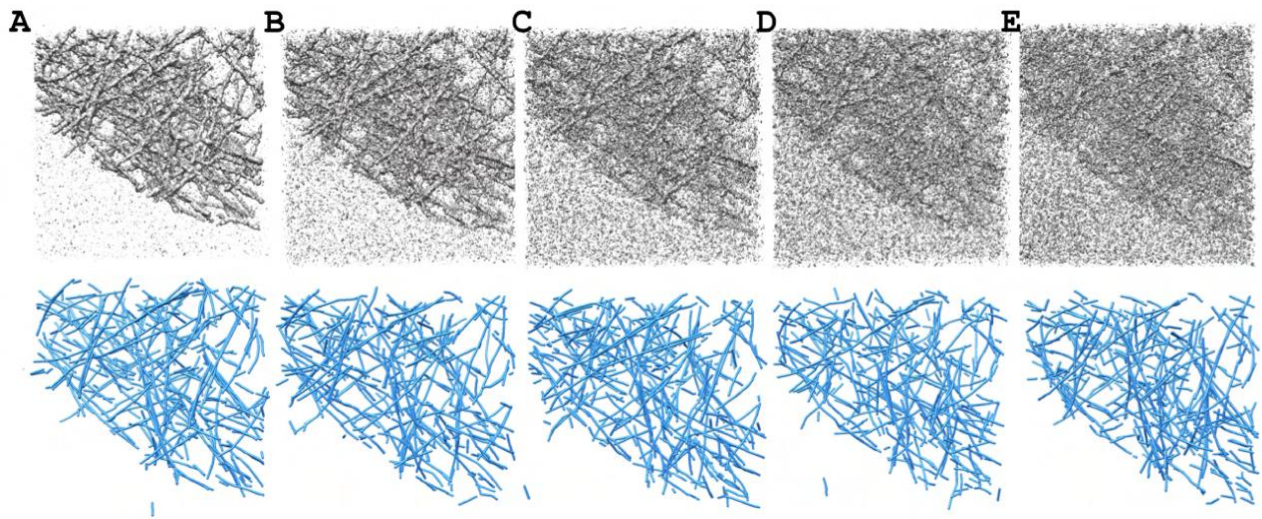

**Fig. 23.** The simulated tomograms (top row) used in this study (isocontour level: mean + 1.5 × standard deviation) and the corresponding *Struwwel Tracer* predictions (bottom row). The simulated tomograms are shown at the following noise levels (see Materials and Methods): (A) 0.35, (B) 0.50, (C) 0.65, (D) 0.80, and (E) 0.95.(Reproduced from [80])

**Table 4.** A performance comparison of the proposed *Struwwel Tracer* approach for simulated tomograms of a *Dictyostelium discoideum* filopodium at various noise levels.(Reproduced from [80])

| Noise level | Precision | Recall | F1 score |
|:-----------:|:---------:|:------:|:--------:|
| 0.35 | 0.97 | 0.85 | 0.90 |
| 0.50 | 0.97 | 0.81 | 0.88 |
| 0.65 | 0.96 | 0.84 | 0.89 |
| 0.80 | 0.95 | 0.79 | 0.87 |
| 0.95 | 0.95 | 0.78 | 0.85 |

Nevertheless, the observed F1 scores ranging from 0.85 to 0.90 (illustrated in Figure 23 for noise levels 0.35 to 0.95) present a remarkable level of accuracy for a density-based structure prediction. The high *Struwwel Tracer* F1 scores are only slightly below those observed earlier with *Spaghetti Tracer* (0.86-0.95) on much more ordered actin filaments [66].

### 5.3.2 Measuring Filament Center Lines in a Previously Untraced Experimental Tomogram

The *Struwwel Tracer* is applied to an experimental tomogram of a mouse fibroblast lamellipodium [82] that was deposited by the authors in the Electron Microscopy Data Bank (EMDB) [83] as EMD-11870. This specific map has not been computationally traced due to their focus on Arp2/3 protein branch junctions in [82], so a measurement of actin center lines could provide complementary information to the paper. (Tracings were shown for similar lamellopodia tomograms in [1], but only EMD-11870 is publicly available).

Since this is a previously untraced map without any ground truth (i.e., no annotated model), in Figure 24 a visual comparison is provided to demonstrate the excellent agreement between the density and the predicted filament center lines. EMD-11870 [82] is a larger tomogram that contains randomly oriented and branched actin filaments. For our demonstration in Figure 24A, a 312 × 320 × 20 sub-region of the tomogram is selected in which filamentous patterns of actin can be clearly appreciated. The *Struwwel Tracer* prediction results are illustrated in Figures 24B and C.

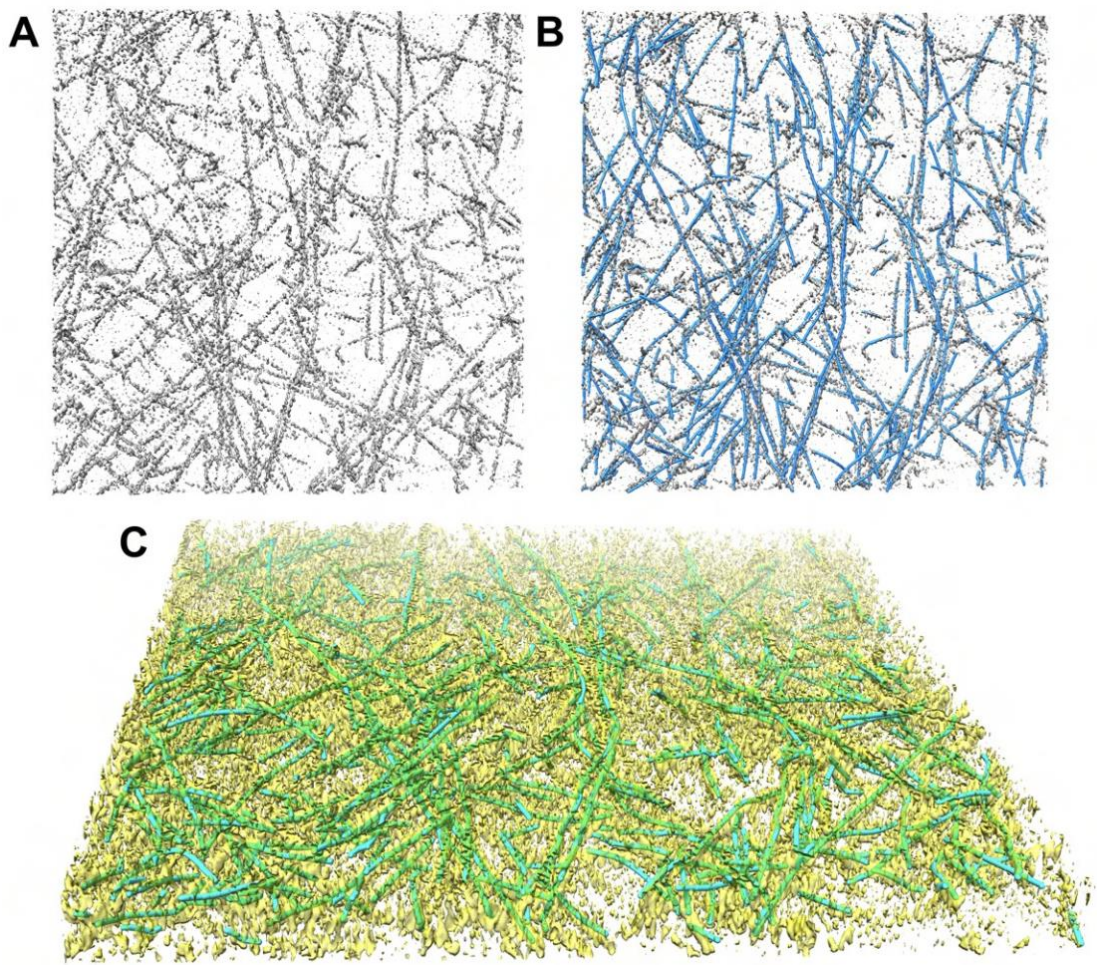

**Fig. 24.** (A) A 312 × 320 × 20 subtomogram cropped from EMD-11870 [82] map indices *i*=100-411, *j*=200-519, and *k*=74-93. The isocontour level in the rendering is 0.4518 (mean + 1.5 × standard deviation). (B) The map density in (A) overlayed by the *Struwwel Tracer* predicted filament center lines (blue). (C) 3D perspective rendering of the map density in (A) (yellow) overlayed by the *Struwwel Tracer* predicted filament center lines (green).(Reproduced from [80])

### 5.3.3 Computation Time and Manual Intervention

In terms of computational efficiency, *Struwwel Tracer* demonstrates outstanding performance, surpassing the earlier tracing method [40] (which took days of computational time) by several orders of magnitude in speed. On an Apple MacBook Pro equipped with a 2.6-GHz Intel Core i7 processor, it is observed that for a simulated tomogram of size 200 × 200 × 71 voxels, *Struwwel*

*Tracer* takes approximately three minutes of run time. Given the ongoing developments in cryo-ET, it is expected that hundreds of tomograms can soon be acquired within a few days [1]. *Struwwel Tracer* would be able to match this processing speed.

In addition, the proposed method does not require extensive manual intervention, which is the case for interactive and deep-learning-based tools for which users either need to tune many parameters or label training data and train the model for prediction. For the *Struwwel Tracer* approach, user intervention is needed only once in the CFS segmentation step to select an approximate value of the threshold in the pruning map. This requires opening the pruning map in visualization software and takes at most a few minutes.

### 5.3.4 Software Implementation and Dissemination

The newly developed *Struwwel Tracer* approach is set to be seamlessly integrated into *Situs*, a widely used, open-source software package for biological image interpretation. *Struwwel Tracer* will be disseminated with *Situs* version 3.2 as a new command line tool named *strwtrc* (following the seven-letter naming convention). To ensure full compatibility with *Situs*, *strwtrc* has been implemented using C/C++, the primary language of *Situs*. A notable advantage of *strwtrc* is that it does not rely on any third-party libraries, making it self-contained and efficient. Moreover, *strwtrc* is fully compatible with both Linux and MacOS operating systems.

In addition, *strwtrc* is designed to be user friendly and accessible, requiring minimal programming language proficiency and no prior coding experience. The number of user-defined parameters is intentionally kept to a minimum (Table 5), making it easier for users to operate the software. The default parameters are designed to work well for most cases; however, it is important to note that they may not be optimal for every possible data set. A summary of all parameters is shown in Table 5. A comprehensive user manual and tutorial will be included with the *Situs* package to guide users through the individual stages of the approach.

**Table 5.** Command line parameters of the *strwtrc* program implemented in *Situs* and corresponding stages of the approach.(Reproduced from [80])

| Parameter Name | *strwtrc* Argument | Description | Default Value | Program Stages |
|---|---|---|---|---|
| **Required Parameter** | | | | |
| Threshold | -thr | Threshold for partitioning the CFS by the normalized path density. | N/A (user-defined based on the pruning map) | CFS segmentation |
| **Optional Parameters** | | | | |
| Length | -len | Length (infinity- norm) of the CFS in voxel units. Internally, this also defines the spacing of the cubic grid for placing seed points (half this value, see text), and the extension length of the CFS (same value) | 10 | Automatic seed selection, CFS generation, and CFS fusion |
| Gap Spacing | -gap | Maximum gap spacing, in voxels, tolerated while fusing adjacent CFSs | 10 | CFS fusion |
| Fusion Angle | -ang | Maximum angle, in degrees, tolerated while fusing adjacent CFSs | 30 | CFS fusion |

## 5.4 DISCUSSION AND CONCLUSION

The release of *Struwwel Tracer* marks the culmination of several years of development effort focused on actin filament tracing. This effort extends and completes the set of efficient solutions for hexagonal, close-packed bundles (*BundleTrac* [63]) and semi-regular bundles with dominant direction (*Spaghetti Tracer* [66]) to irregular networks of randomly oriented actin filaments. *Struwwel*

*Tracer* also completes the paradigm shift that began with *Spaghetti Tracer*, when we first used a dynamic, programming-based method at the voxel level that does not require an expensive missing wedge correction, template convolution, or deconvolution. Therefore, both *Spaghetti Tracer* and *Struwwel Tracer* yield a substantial improvement in time efficiency over earlier template convolution or deconvolution approaches. This development was guided by a rigorous optimization of performance using a statistical F1 score analysis enabled by simulated phantom tomograms.

The proposed framework incorporates a directional path search in all three Cartesian directions ($x$, $y$, and $z$) to ensure a comprehensive coverage of filament orientations. The algorithm identifies the path with the highest density, generating short filament segments that are subsequently combined to form the final filament traces. Evaluation using F1 scores on simulated tomograms proves the algorithm's high efficacy in filament tracing. Visual inspection of the results further confirms its agreement with experimental tomograms. Our approach is robust and fast, works with simple parameter settings, and can deliver a comprehensive performance on simulated and experimental data sets. Moreover, in contrast to deep-learning-based approaches that rely on substantial processing power, such as dedicated GPUs, the implementation in the *Situs* package does not necessitate such resources. It can seamlessly operate on any conventional laptop or desktop computer with a standard CPU.

In addition to the high efficacy in tracing, the proposed framework is also robust as it is, in principle, capable of tracing various types of filaments, whether they are randomly oriented networks or regular (with a dominant direction). However, our earlier tools (*Volume Tracer* [40], *BundleTrac* [63]), *ConDe* [44], and (*Spaghetti Tracer* [66]) have unique features that were optimized for their specific applications (e.g., an additional capability to detect alpha-helices in *Volume Tracer*, the use of hexagonal bundle templates in *BundleTrac*, the ability to prescribe diverse shape templates in *ConDe*, and an integrated directional denoising that takes advantage of an ordered filament arrangement in *Spaghetti Tracer*). A comprehensive F1 score comparison of various tools and conditions could be performed in a future review.

As the study of the actin cytoskeleton is of increasing importance in biology, the proposed tool will become useful to structural biologists in need of free, open-source software solutions. Although this manuscript focuses primarily on the computational aspects of actin filament tracing, the developed framework can be applied by any experimental lab to newly acquired experimental tomograms of the actin cytoskeleton. Functionality for measuring actin filament length distributions and the angles between filaments, and for detecting branch junctions [43, 82], can be included in future work.

# CHAPTER 6

# SEGMENTATION OF SECONDARY STRUCTURES IN EXPERIMENTALLY-DERIVED 3D CRYO-EM IMAGES USING DEEP CONVOLUTIONAL NEURAL NETWORKS

## 6.1 INTRODUCTION

Proteins serve a wide array of functions in living organisms, such as aiding in molecule transport, providing structural support, and boosting the immune system. They are composed of a combination of twenty naturally occurring amino acids. However, it's the unique arrangements of these amino acids that give rise to their distinct three-dimensional (3D) structures, ultimately determining their functions. Thus, understanding a protein's 3D configuration is crucial for comprehending its biological role.

Currently, one of the leading techniques for revealing the atomic arrangement of proteins is cryo-electron microscopy (cryo-EM). This method involves rapidly freezing samples in liquid-nitrogen-cooled liquid ethane and then imaging them with an electron microscope at extremely low temperatures. The increased accessibility of cryo-electron microscopy (Cryo-EM) has greatly advanced the study of various molecules, including the generation of 3D density maps of proteins, also known as 3D density images, with resolutions ranging from 3 Å to 20 Å [84]. Density maps below a 5 Å resolution typically allow for the discernment of a protein chain's backbone, enabling the derivation of near-atomic structures, although achieving high-accuracy structures typically necessitates a density map resolution closer to 3 Å. Conversely, when dealing with density maps above a 5 Å resolution (i.e., medium resolution map), extracting the atomic structure directly becomes challenging due to the limited level of molecular details.

Protein secondary structures, such as $\alpha$-helices and $\beta$-sheets, are the most distinguishable characteristics in a medium-resolution cryo-EM density map, even though amino acids are not discernible at such a resolution. In most medium-resolution maps, an $\alpha$-helix resembles a cylinder, and a $\beta$-sheet appears as a thin layer of density in a medium-resolution map, although the general shape characters may be affected by their sizes and the density from local environment of the molecule.

Over the years, various computational methods have been developed to derive atomic structures from density maps. The resolution of a cryo-EM density map is an effective estimator of the density map's quality. When the resolution reaches about 3.5 Å, density features become distinguishable for tracing the backbone of a protein chain [85]. Phenix [86], a popular software for determining atomic structures, was initially developed for X-ray data and was subsequently extended for cryo-EM density maps of high resolution [86]. For medium resolution density map, a number methods have been developed to detect secondary structure elements such as $\alpha$-helices and $\beta$-sheets from cryo-EM density maps [51, 87, 52, 55, 88, 54, 89]. These approaches rely on various algorithmic and classical image processing techniques. In practice, accurate detection of secondary structure is challenging because it is affected by the poor and non-uniform shape characteristics of a helix or a $\beta$-strand in medium resolution image [90, 55, 58].

Deep learning-based methods demonstrated potential and performance across a spectrum of image segmentation tasks in diverse domains, including biological and medical fields in early 2010s [91, 92, 93, 91]. Building on this inspiration, our lab introduced several approaches between 2016 and 2018 for protein secondary structure detection [58, 56]. For example, Li et al. [58] proposed a deep learning-based model that utilizes a 3D CNN with inception learning and residual connections. The authors trained and tested their model using 25 simulated cryo-EM images. Similarly, Haslam et al. [56] used a small dataset primarily consisting of simulated data for protein secondary structure detection. The authors utilized 31 atomic protein structures from the Protein Data Bank (PDB), each simulated to a 9Å resolution with a 1Å voxel size. Out of these 31 3D images, 25 were allocated for training, while the remaining six were designated for testing purposes. However, they used a U-net based semantic segmentation architecture for detecting protein secondary structure. Inspired by these earlier successes, I became involved in this project around 2018 with the goal of demonstrating the efficacy of a deep learning-based method on experimental data. This involved incorporating advanced architecture and a larger dataset. This chapter depicts my contributions to protein secondary structure detection, spanning from June 2018 to August 2020.

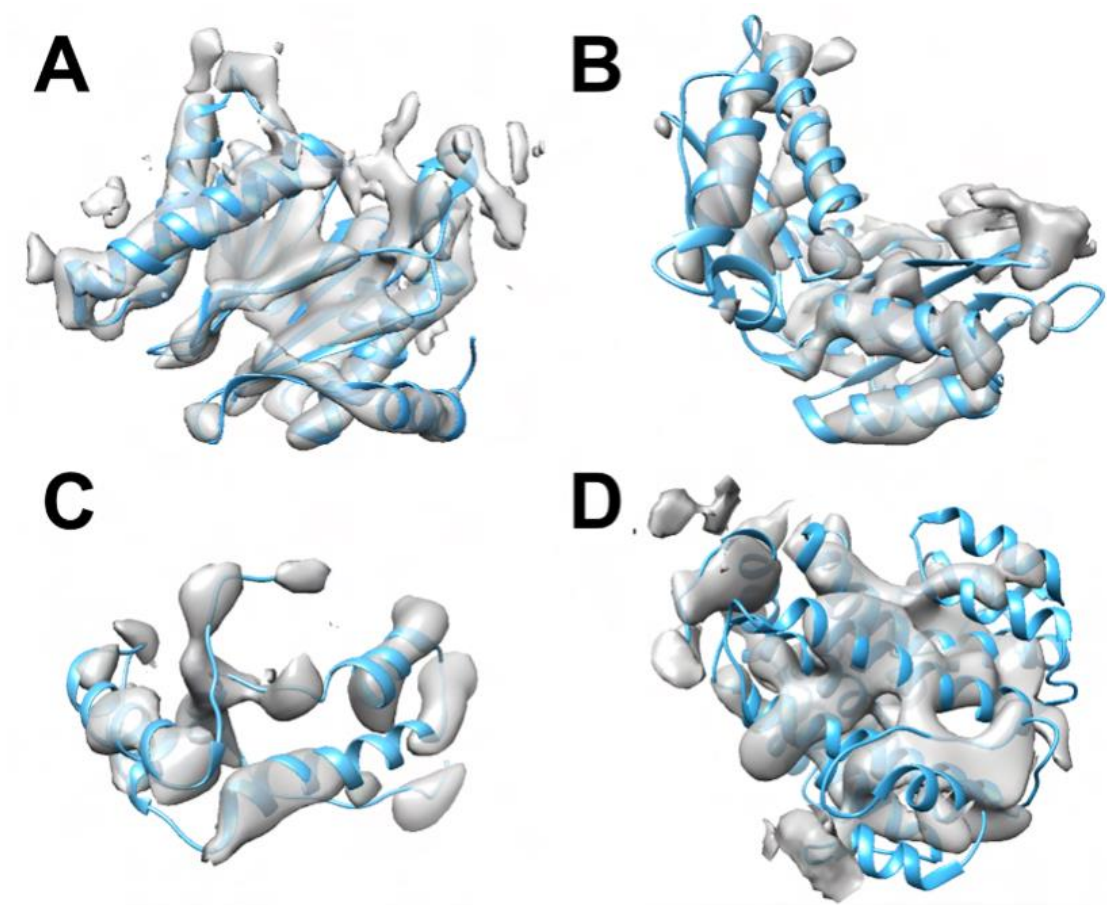

**Fig. 25.** Cryo-EM density maps of variable map-to-model agreements. (A) Chain H of (EMDB ID: 1733, 6.8Å resolution) from Bin 1, (B) Chain C of EMDB ID: 4075, 5.35Å resolution of Bin 2, (C) Chain Q of EMDB ID: 8129, 7.80Å resolution, from Bin 3, (D) Chain W of EMDB ID: 6286, 8.30Å resolution from Bin 4. Atomic models are shown as ribbons.

In this work [1], an effective U-net-based segmentation model has been proposed for the detection of protein secondary structures in medium-resolution cryo-EM 3D images. The proposed method utilizes masked chain images, which are created by first isolating the chains from the protein atomic model. It then utilizes the coordinate information of these chains to extract and mask corresponding regions in the density map. Note that employing chain-based cropping ensures the presence of entire secondary structures in the image, facilitating more effective learning for the model. However, it's essential to acknowledge that this approach comes with high memory requirements, especially when dealing with larger chain sizes. For labeling individual voxels in the density map, the STRIDE secondary structure annotation tool [94] was employed. The network was trained using 1,268 chains isolated from approximately 550 cryo-EM density maps. To address the class-imbalance issue, various types of loss functions, including weighted cross-entropy and dice loss, were considered. Results from a test set comprising 33 cryo-EM density maps show an overall F1 score of 0.76 for helix detection and 0.60 for $\beta$-sheet detection at the residue level.

---

[1]Timothy Baker, a master's student in our lab from 2018 to 2020, made significant contributions to this work, particularly in integrating the dice loss with the existing cross-entropy loss function and creating a distributed version of the code

It's important to note that this project is currently ongoing, and my contributions to it have paved the way for subsequent research and publications within our research group. These contributions involve the development of the publicly available tool DeepSSETracer [95], seamlessly integrated with ChimeraX [96]. Another noteworthy work stemming from this project is the work of Deng et al.[97]. Both studies employed advanced techniques such as curriculum learning, where the model starts with high-quality and easy-to-learn data, gradually incorporating lower-quality data. To prevent forgetting what has been learned in earlier phases during subsequent training phases, the Gradient of Episodic Memory (GEM) technique [98] was employed.

## 6.2 MATERIALS AND METHOD

### 6.2.1 Data Quality Evaluation and Selection

All cryo-EM density maps were downloaded from EM Data Resource (EMDB) [2] with a requirement of resolution between 5-10 Å and a corresponding atomic structure available in Protein Data Bank (PDB). The atomic structures of individual protein chains were used as the envelope to extract the density region of the chains, after which for each chain, a pair representing its atomic structure and corresponding density map were created. Since it is common to see multiple copies of the same chain in a cryo-EM density map, duplicated copies were removed to reduce training bias. The Needle-Wunsch algorithm was used to align a pair of sequences, and near identical chains (i.e., with more than 70% shared sequence identity) were removed. No chain in the testing data shares more than 35% sequence identity with any chain in the training set.

It is known that the quality of density maps varies and is crucial for the agreement between the density and the atomic structures [61]. To exclude low-quality data during training, a screening process was performed, discarding chains with poor matches between their atomic structures and density maps. The quality of each extracted density map (of a chain) was evaluated by its cylindrical fit of helices [61]. More specifically, an F1 score was calculated from the fit for each helix in a chain, and the average F1 score over all helixes was subsequently computed and used to score the quality of the density map. The mean and standard deviation of scores obtained from 4935 chains were calculated, 0.55 and 0.1, respectively, and were used to create four bins of chains. Bin 1 is the group of chains with the highest score, in other words, the best-estimated agreement between atomic structures and their corresponding density maps. Each chain in Bin 1 has a score of at least 0.7 which is 1.5 standard deviation away from the overall mean. It is observed that the larger difference between precision and recall (resulting from the cylindrical fit) often suggests a worse agreement between atomic structure and its density map. Therefore, to create bins 2, 3, and 4, which have changes with progressively deteriorating quality,the difference between precision and recall scores are used in addition to the quality score. For example, Bin 2 contains chains with a score between 0.6 and 0.7, and less than 0.15 precision-recall difference. Bin 3 contains both chains with a score that is also between 0.6 and 0.7, but with more than 0.15 precision-recall difference, and chains with a score between 0.55 and 0.6, and less than 0.15 precision-recall difference. Bin 4 contains all remaining chains in the dataset.

---

[2]https://www.ebi.ac.uk/pdbe/emdb/

In this work, only chains from the first three bins were used for training and evaluation, totaling at 1,350, including 238, 612, and 500 chains from bins 1, 2, and 3, respectively. The total 1,350 chains were partitioned into three disjoint subsets, used for training (N = 1,268), testing (N = 33), and validation (N = 49), respectively.The splitting of training, testing, and validation is performed in such a way that it ensures a similar presence of chains representing different bins in these three sets. Each density map is resampled to 1 per voxel. STRIDE was used to annotate secondary structures, for which H, G, and I were considered as helix residues and B, b, and E were considered as $\beta$-sheet residues. Voxels within 3 from C$\alpha$ atom of the helix and $\beta$-sheet residues were labeled as helix voxel and $\beta$-sheet voxels, respectively. The rest of the voxels were labeled as background.

Please note that the dataset used in this study represents a different set compared to the later work of DeepSSETracer. In DeepSSETracer, a more sophisticated approach was employed to identify a distinct testing set. From the initial testing set, chains with unidentified sequences for amino acids (labeled with "UNK") were discarded. It's important to note that although many chains labeled with "UNK" in the PDB file for some of the amino acids, still, contain atomic coordinate information for the chain's backbone. Subsequently, from the remaining testing set, chains with 35% sequence identity with any chain in the training set were eliminated. To ensure structural dissimilarity of the testing chains concerning training chains having unknown sequences (i.e., UNK), the TM-align score was computed. TM-align scores were used to assess the structural similarity between different chains. For a chain to be included in the final test set, it was required to have a TM-align score lower than 0.5 in at least one of the two TM-align scores, which were normalized with respect to the lengths of the two sequences. The final test set comprises 28 chains, each of which is sufficiently different from any other chain in the test set as well as from every chain in the training set, either through sequence identity or structural similarity screening.

### 6.2.2 Architecture

The architecture of the proposed framework is designed to process the entire chain image simultaneously instead of splitting it into multiple fixed-size patches. A fixed patch size can split secondary structures at any position, which may impact the shape information. Chain images in training can vary The architecture of the proposed framework is designed to process the entire chain image simultaneously instead of splitting it into multiple fixed-size patches. A fixed patch size can split secondary structures at any position, which may impact the shape information. Chain images in training can be of varying sizes ranging from 16 to 100 Å in any of the x, y, and z directions, a size that a medium-sized chain typically has. The network for protein secondary structure detection was adapted from the 3D U-Net proposed in [99]. Given that the training set is relatively small, to prevent overfitting and reduce training time, five layers (rather than the nine in [99]) of convolution blocks with dropout integrated were used. Moreover, in order to capture more contextual information and complex patterns, the number of feature channels is doubled in each convolution.

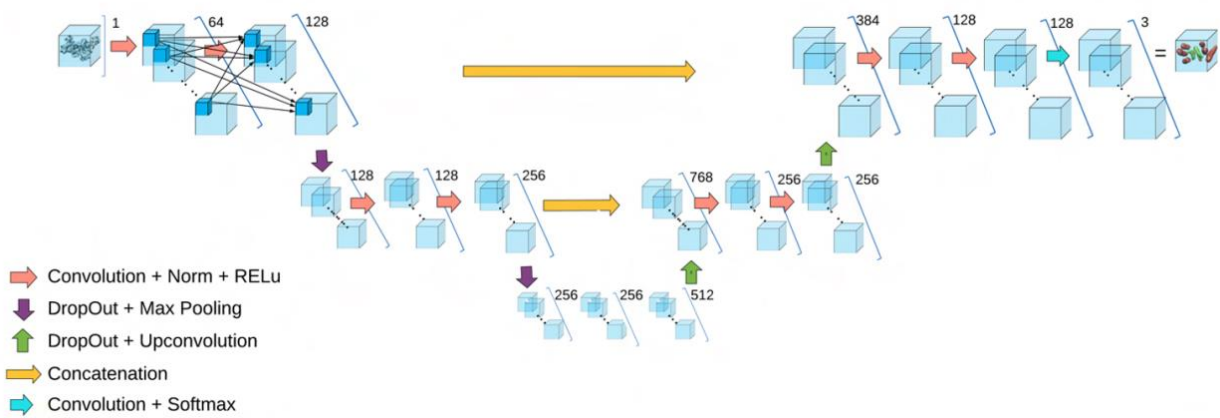

**Fig. 26.** The architecture of the adapted 3D U-Net. Each blue block represents the activation map resulted from the operations as indicated. The number of channels (filters) are labeled by the blocks.

Like all other U-Net-based architectures, the network consists of a down-sampling (also known as contracting or analysis) path that includes the first two layers and an up-sampling (also known as synthesis) path that includes the last two layers (Figure 26). The third layer (right in the middle) is the bottleneck layer, as it has the lowest resolution. Each layer in the network consists of two convolutions using 3x3x3 kernels (or filters), each followed by a batch normalization (BN) and a ReLu nonlinearity. Each layer in the down-sampling path ends with a 2x2x2 max pooling with a stride of two, reducing the resolution by half in all dimensions. In contrast, each layer in the up-sampling path begins with a 2x2x2 transpose convolution with a stride of two to bring up the resolution by a factor of two. Dropout was added to each layer right before the max-pooling in the down-sampling path or transpose convolution in the up-sampling path. In the last layer, a 1x1x1 convolution is used to decrease the number of output channels to three, corresponding to the three classes, i.e., helix, $\beta$-sheet, and background, respectively. The network contains 6,142,723 total trainable parameters. The receptive field for each voxel in the prediction is 35x35x35.

Cross entropy is known to have problem with class imbalance, i.e., the imbalance in the distribution of the class voxels belong to. More specially, models trained with cross entropy tend to perform poorly on voxels of the class with fewer voxels, even though the overall accuracy is relatively good. Compared to cross-entropy, the dice coefficient is less sensitive to class imbalance (especially the imbalance between foreground and background voxels), but less smooth, thus difficult to optimize.

**Table 6.** The distribution of class labels. The background voxels make up most voxels in the density maps

| Dataset | background | helix | sheet |
|---|---|---|---|
| Overall | 0.9924 | 0.0044 | 0.0032 |
| training | 0.9924 | 0.0042 | 0.0033 |
| validation | 0.9922 | 0.0050 | 0.0027 |
| testing | 0.9925 | 0.0039 | 0.0035 |

There is a significant class imbalance in the 3D Cryo-EM image data used in this work, especially between foreground (helices and $\beta$-sheets) and background voxels (Table 6). To handle such class imbalance, the proposed method combines cross-entropy and dice coefficient. Moreover, as a first component of the objective function, instead of the standard cross-entropy, its weighted version is used.

To prevent overfitting, in addition to the dropout, kernel regularization is incorporated. The Proposed Framework was implemented using Tensorflow API. Adam optimizer [100] was used to optimize the weights in the network. For the hyper-parameters, an extensive search is performed with the help of validation data. The hyper-parameters used to generate the final model are $\lambda_v$=0.3, $\lambda_u$=0.7, $\lambda_d$=0.3, $\lambda_r$=0.00001, dropout rate: 0.5, learning rate: 0.001, batch size: 1 and number of epochs: 40.

### 6.3 RESULTS AND DISCUSSION

#### 6.3.1 Performance Evaluation

To assess the performance of The proposed framework, voxel-level F1 score, residue-level F1 score, and Q3 accuracy are computed for secondary structure detection. The voxel-level F1 score was determined by comparing The proposed framework's individual voxel predictions with the actual labels. To calculate this score, initially precision and recall scores are computed for both helix and sheet structures, utilizing true positives, false positives, and false negatives counts. These precision and recall scores were subsequently used to calculate the F1 score.

The residue-level F1 score was calculated considering the predictions and true labels of the residues (i.e., amino acid). For every $C\alpha$ atom in the atomic model, its class prediction is determined by all the voxels that reside within a $3AA$ radius through a maximum voting procedure. The class predictions of $C\alpha$ atoms were compared with true labels of secondary structures annotated using STRIDE to calculate precision and recall, and finally the F1 score. In addition, the Q3 score was calculated at the residue-level. The three classes included in the Q3 score calculation include helix, $\beta$-sheet, and other residues. An other residue is a non-helix and non-$\beta$-sheet residue, and it is determined if the number of background voxels dominates within a 3Å radius of the of $C\alpha$ atom.

**Table 7.** F1 scores of the proposed model on 33 testing chains from various resolutions and bins

| EMDB_PDB_Chain (resolution ) | Bin | C$\alpha$(H/S/T) | F1 voxel | | F1 residue | |
| --- | --- | --- | --- | --- | --- | --- |
| | | | Helix | Sheet | Helix | Sheet |
| 1657_4v5h_AE(5.80) | 1 | 42/38/150 | 0.60 | 0.50 | 0.69 | 0.58 |
| 1798_4v5m_AE(7.80) | 3 | 42/58/150 | 0.65 | 0.51 | 0.81 | 0.69 |
| 2994_5a21_G(7.20) | 3 | 37/21/133 | 0.58 | 0.53 | 0.75 | 0.53 |

**Table 7 Continued.**

| | | | | | | |
|---|---|---|---|---|---|---|
| 3206_5$fl2$_K(6.20) | 3 | 12/47/106 | 0.40 | 0.45 | 0.39 | 0.64 |
| 3491_5$mdx$_H(5.30) | 2 | 33/0/42 | 0.71 | NA | 0.83 | NA |
| 3594_5$n61$_E(5.80) | 3 | 86/52/212 | 0.61 | 0.54 | 0.71 | 0.68 |
| 3850_5$oqm$_4(5.80) | 2 | 128/49/297 | 0.61 | 0.63 | 0.71 | 0.72 |
| 3850_5$oqm$_g(5.80) | 1 | 81/0/85 | 0.75 | NA | 0.96 | NA |
| 3948_6$esg$_B(5.40) | 2 | 51/0/78 | 0.74 | NA | 0.90 | NA |
| 4041_5$ldx$_H(5.60) | 1 | 199/0/296 | 0.73 | NA | 0.89 | NA |
| 4041_5$ldx$_I(5.60) | 2 | 48/19/176 | 0.52 | 0.40 | 0.64 | 0.57 |
| 4078_5$lms$_D(5.10) | 2 | 86/18/208 | 0.63 | 0.37 | 0.72 | 0.40 |
| 4107_5$luf$_M(9.10) | 3 | 322/0/439 | 0.53 | NA | 0.70 | NA |
| 4141_5$m1s$_B(6.70) | 2 | 81/158/366 | 0.52 | 0.50 | 0.56 | 0.64 |
| 4182_6$f42$_G(5.50) | 3 | 16/66/180 | 0.37 | 0.40 | 0.43 | 0.61 |
| 5036_4$v69$_AD(6.70) | 3 | 78/19/205 | 0.58 | 0.34 | 0.68 | 0.38 |
| 5942_3$j6x$_25(6.10) | 3 | 25/5/70 | 0.48 | 0.00 | 0.58 | 0.00 |
| 5943_3$j6y$_80(6.10) | 3 | 12/8/52 | 0.35 | 0.10 | 0.53 | 0.22 |
| 6149_3$j8g$_W(5.00) | 3 | 19/48/94 | 0.53 | 0.67 | 0.72 | 0.89 |
| 6446_3$jbi$_V(8.50) | 2 | 116/0/131 | 0.74 | NA | 0.95 | NA |
| 6456_3$jbn$_AL(4.70) | 3 | 89/8/211 | 0.69 | 0.38 | 0.82 | 0.30 |
| 6810_5$y5x$_H(5.00) | 2 | 38/10/100 | 0.53 | 0.56 | 0.78 | 0.58 |
| 7454_6$d84$_S(6.72) | 3 | 65/34/142 | 0.51 | 0.45 | 0.65 | 0.55 |
| 8016_5$gar$_O(6.40) | 2 | 65/0/80 | 0.64 | NA | 0.81 | NA |
| 8128_5$j7y$_K(6.70) | 1 | 71/0/93 | 0.78 | NA | 0.88 | NA |
| 8129_5$j8k$_AA(7.80) | 2 | 187/60/446 | 0.59 | 0.41 | 0.71 | 0.50 |
| 8129_5$j8k$_D(7.80) | 2 | 170/41/384 | 0.60 | 0.29 | 0.70 | 0.40 |
| 8130_5$j4z$_B(5.80) | 1 | 63/9/154 | 0.67 | 0.44 | 0.75 | 0.42 |
| 8135_5$iya$_E(5.40) | 2 | 88/44/210 | 0.65 | 0.60 | 0.78 | 0.70 |
| 8357_5$t4o$_L(6.90) | 1 | 106/0/160 | 0.69 | NA | 0.83 | NA |
| 8518_5$u8s$_A(6.10) | 1 | 114/13/208 | 0.78 | 0.55 | 0.94 | 0.46 |
| 8693_5$viy$_A(6.20) | 1 | 68/0/133 | 0.67 | NA | 0.75 | NA |
| 9534_5$gpn$_A(5.40) | 2 | 59/0/88 | 0.71 | NA | 0.89 | NA |
| **Weighted Average** | | 2697/825/5879 | 0.63 | 0.48 | 0.76 | 0.60 |

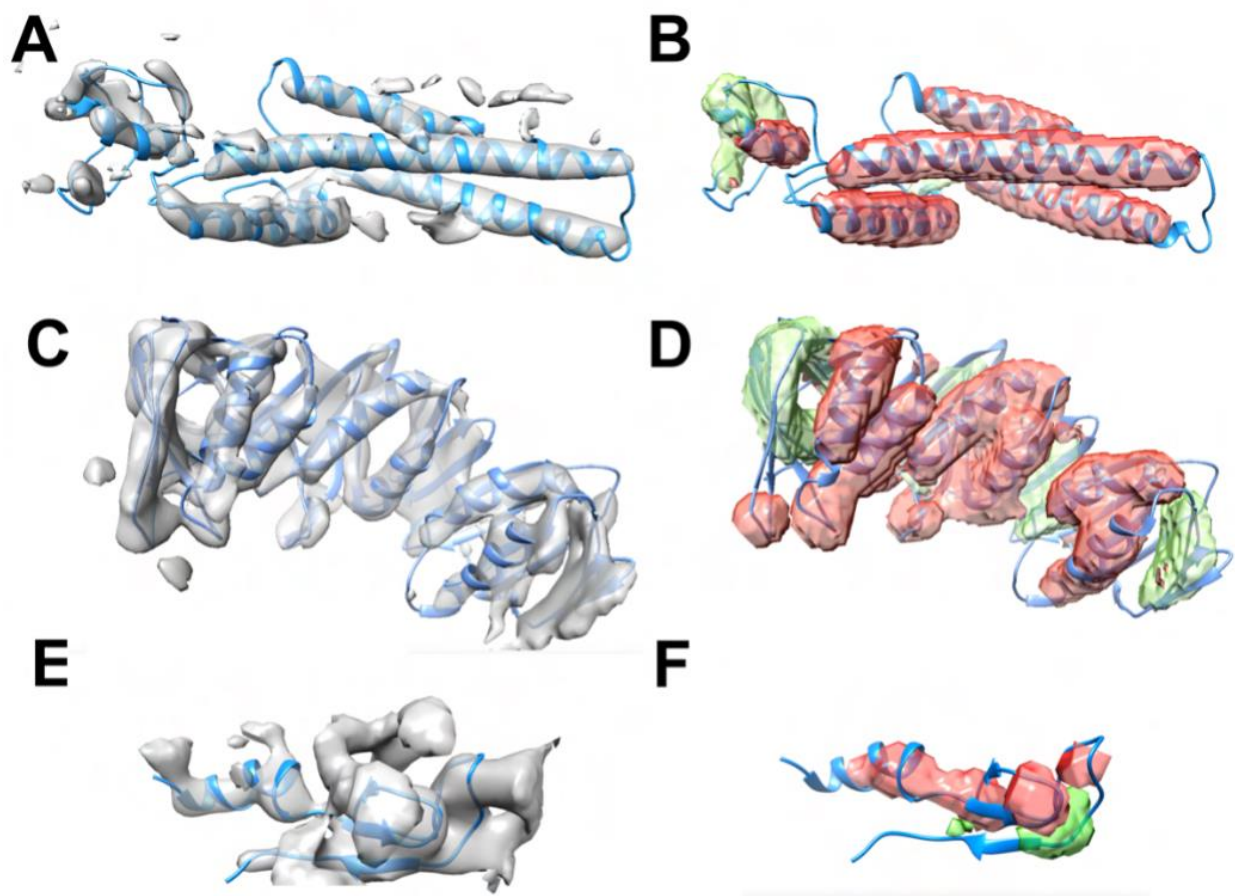

**Fig. 27.** Predictions by the proposed framework for 3 density maps with variable level of accuracy: Top panel (chain: 4089_5ln3_D, highly accurate), middle panel (chain: 6452_AH, moderate accuracy)and bottom (chain: 5943_80 , low accuracy). Panel A, C, and E show the STRIDE annotated model along with density maps while Panel B, D, and F show predictions from the proposed framework.

### 6.3.2 Quality

The performance of the proposed framework was evaluated using 33 test cases, of which 8, 13, and 12 are from Bin 1, Bin 2, and Bin 3 respectively (Table 8). Three examples of secondary structures segmented from cryo-EM density maps are shown with varying levels of detection accuracy (Figure 27). In the first example (top row of Figure 27), the component of cryo-EM density map (EMDB-8518) that corresponds to chain A of 5u8s (PDB ID) was partitioned in Bin 1, the bin with the best cylindrical fit between the density and atomic structure for helices. The chain's atomic structure 114 C$\alpha$ atoms in helices, and 13 C$\alpha$ atoms in a $\beta$-sheet (Table 7). the proposed framework was able to detect all the helices and partial a $\beta$-sheet region with a voxel-level F1 score of 0.78 and 0.55 for helix and $\beta$-sheet detection respectively (Table 7). The second example (second row of Figure 27) involves a much larger chain (4141_5m1s_B) that contains 366 amino acids, out of which 158 are in $\beta$-sheets (Table 7). This component of the cryo-EM map was partitioned in Bin2 with the second level of cylindrical fit, and the proposed framework shows a voxel-level F1 score of 0.52 and 0.50 for helix and $\beta$-sheet detection respectively. A voxel-level F1 score suggests the ability to distinguish voxels in a 3-dimensional image, although the ultimate goal is to recognize the atomic structure of the image. A relaxed measurement of the F1 score at the residue level also serves the purpose, since it is likely that not every helix voxel is needed to be recognized to distinguish a helix backbone. Therefore, a residue-level F1 score is generally higher than a voxel-level F1 score. In this case, the proposed framework shows a residue-level F1 score of 0.56 and 0.64, for which the helix detection is slightly lower than that of the first example, and the $\beta$-sheet detection is slightly higher. Several factors might have contributed to a different performance in the helix detection. The two cases were estimated in two different bins which suggests a difference in cylindrical fit. The first example has longer helices and they appear to show a more uniform cylindrical nature of the helix (Figure 27). The better recognition of $\beta$-sheet in the second example might be from larger $\beta$-sheets and the challenge of accurately detecting a small $\beta$-sheet, i.e. of 13 amino acids. Note that the partition of bins is only based on helix properties. Although the cylindrical fit indirectly suggests the fit at $\beta$-sheet regions for some cases, it does not necessarily indicate the complexity for $\beta$-sheets for all cases. Ideally, estimation is needed for $\beta$-sheet regions.

**Table 8.** The average helix F1 score of each bin calculated using the 33 testing chains

| Chain Quality (based on Helix) | #Num of chains | Average F1 score (voxel) | Average F1 score (residue) |
|---|---|---|---|
| Bin 1 | 8 | 0.71 | 0.84 |
| Bin 2 | 13 | 0.63 | 0.77 |
| Bin 3 | 12 | 0.52 | 0.65 |

The third example is the worst-performing example of the 33 test cases. The component cryo-EM map (EMD-5943_8 *j6y*_80) was one of the 12 test cases in Bin 3. In this case, chain 80 of 3j6y (PDB ID) contains 12 helix residues and 8 $\beta$-sheet residues with a total length of 52 amino acids. Visual inspection shows that the density region of the helix shows less cylindrical character, unlike the two other examples. The $\beta$-sheet region also loses the thin surface nature existing for many $\beta$-sheets, even though the entire cryo-EM map (EMD-5943) has a decent resolution of 6.1Å, which roughly suggests the overall quality of the image rather than for local regions. For this case, a residue-level F1 score of 0.53 and 0.22 were shown for helix and $\beta$-sheet detection (Table 7).

While it's commonly assumed that the accuracy of secondary structure detection is influenced by the quality of the density map, there has been limited quantitative research on this aspect. Among the 33 test cases, the relationship between helix F1 scores and the Bin IDs are examined that were derived by cylindrical fit scores. Even though the proposed framework and cylindrical quantification are completely different approaches, the pattern of average F1 scores is clearly linked to the estimated bins. For example, the average F1 score of Bin 2 was calculated using the F1 scores of the 13 cases in Bin 2. The average voxel-level F1 score decreases from 0.71 to 0.63, and to 0.52 from Bin 1, to Bin 3 due to decreased cylindrical fit estimation. This suggests that the detection at the voxel level could have about 0.1 difference, on average, in F1 scores depending on what bins of testing data are used. Training data does not include bin information, and in fact 500 of the total 1350 data points belong to Bin 3. Interestingly, even though Bin 3 dominates over Bin 1 and Bin 2 in the dataset, the performance of helix detection is still the worst in Bin 3. This suggests that either the density pattern of helices in Bin 3 is less consistent. Cylindrical fit shows a lower level of agreement between an atomic model and density at the helix region for Bin 3. It could be either the lower quality of the density region or the interpretation of the atomic model to cause the disagreement in cylindrical fit.

To evaluate the detection accuracy concerning the Q3 score, residue-level class label accuracy of the helix, $\beta$-sheet, and 'other' class are computed. Note that the 'other' class label of a C$\alpha$ atom indicates the dominance of background voxels in its 3 neighborhood. The overall accuracy of each class is determined by a weighted average. For instance, to calculate overall helix accuracy, each chain is weighted by the number of helix C$\alpha$ atoms it contains. Thus, a chain with a higher number of helix C$\alpha$ atoms will have more influence on the overall helix detection accuracy calculation.

### 6.3.3 Comparison of Accuracy Between the Proposed Framework and SSETracer

SSETracer is one method to detect both helices and $\beta$-sheets from medium-resolution cryo-EM density maps [55]. It is publicly available as a plugin to Chimera [65], a popular visualization tool for cryo-EM density data and atomic structures. The recognition of secondary structures in SSETracer is based on an evaluation of the skeleton, local structural tensor, local thickness, and density values. The skeleton is derived from the density map using Gorgon [101]. Although SSE-Tracer can provide good accuracy, the challenge is that results are often sensitive to user-given parameters, as for many image-processing-based methods. One of the important parameters is the density threshold. To examine the effect of density thresholds, 20 thresholds are sampled equally spaced between the mean and the 5-standard deviation of the density map, so that two thresholds are about 25% of a standard deviation. An automation script was created to collect the skeleton of each sample threshold from Gorgon using Autohotkey (https://www.autohotkey.com), since Gorgon is a user-interactive program. The automation script executes SSETracer using the same sample threshold as used for deriving the skeleton. The residue-level F1 scores were calculated from the detected helix and $\beta$-sheet voxels for each sample threshold.

The examination of F1 scores using 20 density thresholds shows that the density threshold is an important parameter affecting detection accuracy. The threshold needed to detect helices is generally higher than that for the detection of $\beta$-sheets. This is because helices are generally denser regions than $\beta$-sheets. Although SSETracer has the potential to detect secondary structures quite accurately, it is not trivial in practice for a non-experienced user due to the choice of a density threshold. For each test case, the best of the 20 helix F1 scores was examined (column 5 of Table 9), as well as the average of three non-zero F1 scores (BestNbr F1 score, column 7 of Table 9) that were obtained from three consecutive thresholds near the one producing the highest F1 score. For the 12 tested cases, the average best F1 score reduced from 0.81 to 0.73 for helix detection and from 0.58 to 0.47 for $\beta$-sheet detection, if additional two consecutive density thresholds near the best threshold were considered. This experiment shows that SSETracer is sensitive to the change of about 25% of a standard deviation in density threshold for some cases, even near the best threshold. It is also generally not advisable to expect a universal best threshold for all test cases since a specific threshold needs to be considered for each test case. This is shown in our experiment of using the same threshold for all test cases, such as 3.25 and 2.5 standard deviations (column 9, 10 and column 11, 12 in Table 9) of each density map, the average F1 score for the 12 cases is greatly reduced.

**Table 9.** Comparison of residue-level F1 scores of the proposed framework and SSETracer. The number of C$\alpha$ atoms are shown for H: helix, S: $\beta$-sheet, T: total chain. SSETracer (Best): the highest F1 score of 20 F1 scores obtained using 20 samples of density threshold. SSETracer (BestNbr): the average of three F1 scores obtained using three density thresholds in the neighborhood of the one that produced the best F1 score; F1_SSETracer (Std-3.25), F1_SSETracer (Std-2.5): the F1 scores obtained using density threshold of 3.25 standard deviation and 2.5 standard deviation of the density map respectively.

| EMDB_PDB_Chain (resolution) | C$\alpha$ (H/S/T) | Proposed Model | | SSETracer | | | | | | | |
|---|---|---|---|---|---|---|---|---|---|---|---|
| | | | | Best | | BestNbr | | Std-3.25 | | Std-2.5 | |
| | | H | S | H | S | H | S | H | S | H | S |
| 5036_4v69_AD (6.70) | 78/19/205 | 0.68 | 0.38 | 0.71 | 0.34 | 0.69 | 0.21 | 0.63 | 0 | 0.72 | 0.09 |
| 5942_3j6x_25 (6.10) | 25/5/70 | 0.58 | 0.00 | 0.8 | 0.29 | 0.67 | 0.19 | 0.8 | 0 | 0.61 | 0 |
| 5943_3j6y_80 (6.10) | 12/8/52 | 0.53 | 0.22 | 0.93 | 0.61 | 0.67 | 0.30 | 0.52 | 0 | 0.58 | 0.62 |
| 6149_3j8g_W (5.0) | 19/48/94 | 0.72 | 0.89 | 0.70 | 0.75 | 0.58 | 0.66 | 0.71 | 0 | 0.33 | 0.48 |
| 6446_3jbi_V (8.50) | 116/0/131 | 0.95 | NA | 0.93 | NA | 0.91 | NA | 0.92 | NA | 0.82 | NA |
| 6810_5y5x_H (5.00) | 38/10/100 | 0.78 | 0.58 | 0.76 | 0.63 | 0.73 | 0.26 | 0.76 | 0 | 0.46 | 0.64 |
| 7454_6d84_S (6.72) | 65/34/142 | 0.65 | 0.55 | 0.76 | 0.69 | 0.55 | 0.54 | 0.48 | 0.45 | 0.25 | 0.46 |
| 8016_5gar_O (6.40) | 65/0/80 | 0.81 | NA | 0.94 | NA | 0.82 | NA | 0.83 | NA | 0.7 | NA |
| 8128_5j7y_K (6.70) | 71/0/93 | 0.88 | NA | 0.95 | NA | 0.94 | NA | 0.93 | NA | 0.94 | NA |
| 8130_5j4z_B (5.80) | 63/9/154 | 0.75 | 0.42 | 0.79 | 0.63 | 0.61 | 0.40 | 0.74 | 0 | 0.79 | 0.64 |
| 8135_5iya_E (5.40) | 88/44/210 | 0.78 | 0.70 | 0.68 | 0.45 | 0.67 | 0.43 | 0.66 | 0.43 | 0.33 | 0.45 |
| 9534_5gpn_Ae (5.40) | 59/0/88 | 0.89 | NA | 0.77 | NA | 0.63 | NA | 0.78 | NA | 0.42 | NA |
| **Weighted Average** | | 0.79 | 0.62 | 0.81 | 0.58 | 0.73 | 0.47 | 0.75 | 0.19 | 0.61 | 0.43 |

Results from the 12 test cases show that the efficacy of the proposed methodology is close to the best performance of SSETracer for helix detection, with an average F1 score of 0.79 versus 0.81 and there is no need for a user to sample many thresholds. Without knowing the ground truth in practice, the best performance is not known when SSETracer is applied. The greatest advantage of proposed method is that it does not require a user to choose parameters, and it offers robust accuracy. It is also observed that the proposed framework shows a much better F1 score than SSETracer for $\beta$-sheet detection. The average F1 score for $\beta$-sheet is 0.62 for the proposed framework and 0.58 for SSETracer. $\beta$-sheets are generally harder to detect accurately than helices due to their lower density and more dynamic shapes. the proposed framework appears to have more robust accuracy for $\beta$-sheet detection.

### 6.3.4 Secondary Structure Detection Using Emap2sec

The design of the proposed framework and Emap2sec share similarities and differences. Both methods utilize CNN architectures to train on cryo-EM density maps of medium resolution. The Emap2sec architecture was designed to train on entire cryo-EM maps that may contain tens of chains [59]. As a way to handle large maps in training, patches of 11Å cube size are sent to the network as inputs. Only voxels with density values higher than a threshold are considered as center voxels in the network. Both practices reduce the amount of training when large density maps are used. the proposed framework is designed for individual chain components that are often much smaller than the entire cryo-EM map. Instead of using patches, the input to the network allows density maps of different sizes from 16Å to 100Å in any of the x, y, and z directions. Using the entire density map of a chain eliminates artifacts of chopping secondary structures. Although the training limits density maps to within 100 Å in any dimension, the limit does not apply to the dimension in testing. The size of density maps in testing can be much bigger once a model is developed in training. Emap2sec uses two phases in the architecture, while the proposed framework uses one phase that is adapted from U-net. An effective loss function was developed to reduce the class imbalance problem that is related to using a larger dimension size in input.

Regardless of differences in the design, both methods are capable of detecting the overall regions of most helices and $\beta$-sheets. For the Emap2Sec, publicly available code and data from Code Ocean (https://codeocean.com/capsule/4873360/tree) were used. A standard procedure was used according to the tutorial, with stride size=2, vm=5, and a density contour suggested by the density map. the proposed framework uses 1Å per voxel to represent the segmented volume. It is observed that the output voxel points from Emap2sec are sparse, and our attempt to use finer sampling points failed. In this experiment, it is observed that the prediction of helixes (red surface, purple dots) mostly overlap for longer helices, but have some differences in some shorter ones (circles in Figure 28). Using detected helix voxels and $\beta$-sheet voxels, residue-level accuracy was calculated for helix C$\alpha$ atoms and $\beta$-sheet C$\alpha$ atoms. As an example, a helix C$\alpha$ atom is determined using maximum voting of voxel labels within 3Å-radius from the C$\alpha$. For the case *EMD_6149_3 j8gw* (Figure 28 B), the residue-level helix accuracy is 0.47 and 0.21 for the proposed framework and Emap2sec respectively. The difference in accuracy may reflect the missed helix in the detection and sparseness of voxels produced using Emap2sec. However, the current implementation of both methods limits a more rigorous comparison.

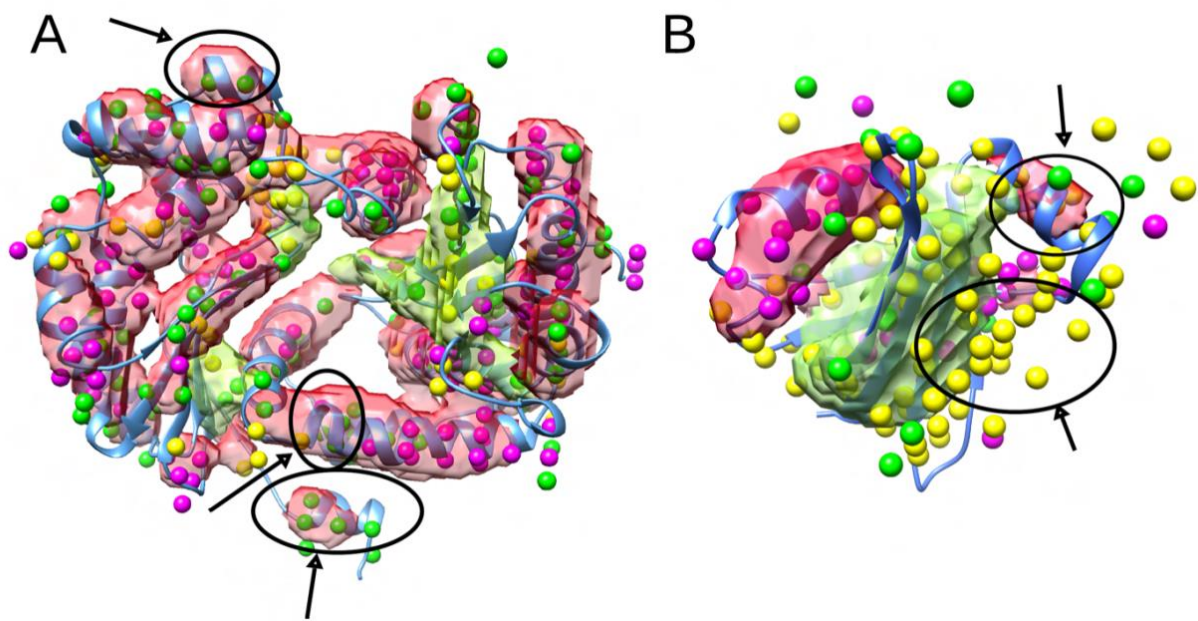

**Fig. 28.** Two examples of segmented secondary structure voxels using the proposed framework and Emap2sec. The component cryo-EM density map (EMD_8129_5j8k_AA in (A) and EMD_6149_3j8g_W in (B)) was used for the two methods to detect secondary structures. The segmented helix volume (transparent red surface) and $\beta$-sheet volume (transparent green surface) were detected using the proposed framework. The detected helix voxels (purple dots), $\beta$-sheet voxels (yellow dots), and other voxels (green dots) were obtained using Emap2sec. Some differences are shown in circles and arrows.

.

## 6.4 CONCLUSION

The relative position of secondary structures provides important constraints for deriving the atomic structure of the protein. In this study, a 3D convolutional neural network (CNN) based architecture has been presented for the segmentation of secondary structure elements from medium-resolution cryo-EM images. Instead of using consecutive patches of a fixed size, the proposed network utilizes density images cropped and masked at the chain level. An effective loss function has been designed to address the class imbalance issue in this problem. The network was trained and tested using over 1300 chains extracted from experimentally derived cryo-EM density maps.

Results from 33 test cases show overall residue-level F1 scores of 0.76 and 0.60 for the detection of helices and $\beta$-sheets, respectively. A comparison between the proposed framework and SSETracer shows that the proposed framework offers robust detection performance which is comparable with the best performance of SSETracer without the need to choose parameters. Although density features of secondary structures, such as $\alpha$-helices and $\beta$-sheets, are distinguishable in cryo-EM density maps at medium resolution, precise detection of secondary structures from such maps is challenging. One of the challenges arises from the disparity between the shape features of density maps and the fitting of atomic structures within them. In order to facilitate training and testing, the measure of the cylindrical fit of helices is used to group components of density maps into different bins. It is found that detection accuracy, reflected by the F1 score, may differ by about 0.1 on average from the highest-quality bin to the next. This suggests that rigorous evaluation of secondary structure detection needs to measure quantitatively the quality of the density map included in the testing data.

## CHAPTER 7

## CYLINDRICAL QUALITY ASSESSMENT OF ALPHA-HELIX IN MEDIUM-RESOLUTION CRYO-EM DENSITY MAP

### 7.1 INTRODUCTION

The number of atomic structures that are derived from cryo-electron microscopy (cryo-EM) density maps has increased rapidly over recent years. As of December 1, 2019, there were 824 structures modeled from cryo-EM density maps at reported resolution values in the medium-resolution range (5–10 Å) compared to 3,144 structures derived from resolutions better than 5 Å [102, 103]. The quality of a map produced by an experimental cryo-EM laboratory improves over time as more data is collected. Consequently, medium-resolution maps are routinely created in the initial stages of a cryo-EM imaging project before the specimen preparation protocols are tuned to perfection. Lower-resolution regions can also be present in overall high-resolution maps due to conformational flexibility or libration of the specimen. For these reasons, medium-resolution maps are often the first and only observations available of a new system. Early insights into an unfamiliar system hold considerable biological significance, compelling researchers to analyze them at the atomic level. Consequently, investigators often endeavor to construct models despite the inherent risks and inaccuracies involved. Our current study aims to evaluate the accuracy of model fitting, particularly in light of the growing availability of deposited map and model pairs within the challenging medium-resolution range.

Various structural modeling methods have been developed to utilize known atomic structures as initial templates. Some structures were derived from medium-resolution density maps through rigid-body fitting [104, 65] and real space refinement. Other structures were derived using comparative modeling methods and flexible fitting. It remains quite difficult and risky to model atomic structures de novo for most proteins, owing to the resolution limitations. However, in many cases, secondary structure elements, such as $\alpha$-helices and $\beta$-sheets, can be assigned with confidence into medium-resolution maps. Our understanding of the quality of density patterns and corresponding atomic models in medium-resolution maps is limited. Without knowledge of reliable density patterns, it is risky to overfit atomic models on erroneous density or to underutilize reliable density patterns in model building.

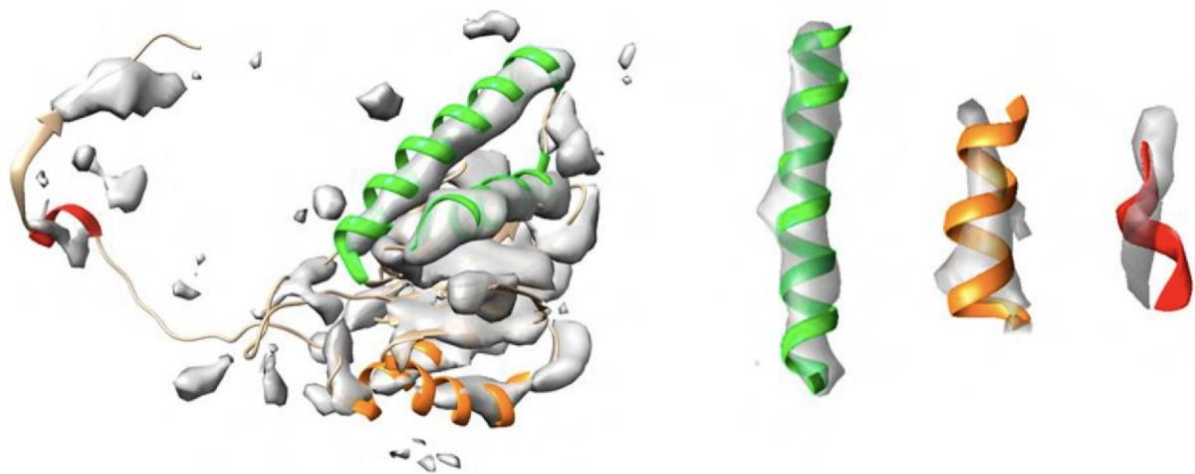

**Fig. 29.** Illustration of different levels of map/model similarity exhibited by helices in the same map. The surface representation of the density map (EMDB ID 4089, gray, corresponding to Chain 2) is superimposed on Protein Data Bank (PDB) ID 5ln3 Chain 2 (ribbon). Helices with different levels of similarity are indicated by green, orange, and red, respectively, from strong similarity to poor similarity (with separate views, on the right, for three examples).(Reproduced from [61])

Among secondary structural features, $\alpha$-helices often appear as cylindrically shaped density regions, and $\beta$-sheets appear as thin layers of density in medium-resolution cryo-EM maps. Although the location of $\alpha$-helices is commonly detectable, density strength varies considerably at helix regions. The density of some helices appears to fill almost all of the helix backbone region expected of the atomic structure (green helices in Figure 29) when visualized at a threshold tuned to the surface of secondary structural features. Other helix backbone regions appear to be partially filled (orange helices in Figure 29); but for some outliers, the orientation of the helix appears to be different from the direction of the corresponding density region (red helix in Figure 29). Variations in density strength can be explained by the limitations of systematic image processing, such as image alignment and 3D reconstruction artifacts, or structural flexibility of molecules. To represent a spatially heterogeneous level of detail that is often observed in cryo-EM maps, the concept of local resolution has recently been introduced. While local resolution provides important information about the quality of specific regions of a map, it is not clear how the concept can be best utilized in model building. Due to the effects of noise, density sharpening, correction of microscope parameters, and 3D reconstruction, individual voxels may not be reliable; moreover, the voxel spacing is typically much finer than indicated by the medium-spatial resolution.

Therefore, this work focuses on assessing the local quality of a map at the level of secondary structure elements. The level of detail of a similarity measure based on secondary structure elements, such as helices, is, by design, intermediate between individual voxels and the global map. The features are large enough to be robust, but the similarity measure is still local compared to the global cross-correlation coefficient [105].

The reproducibility and discriminative ability of a local similarity score are important during the model-building process, because a global score, such as the cross-correlation (CC), varies with the size of the map volume under consideration. Consequently, an earlier application of CC scores in the Phenix program [106] required a local region mask to evaluate individual residues. An alternative tool, EM-Ringer [107], evaluated side chains only and produced a single score to represent the quality of the fit, but the approach did not generalize to the secondary structure level. The closest related approach, in terms of applicability to a secondary structure, was the Z-score. A Z-score was calculated from 27 CC scores using $\pm$ zero or one voxel shifts in the x, y, and z directions, respectively. The approach did not take specific shape information into account, and by design, it depended on the granularity of the map. In contrast, the new F1 score proposed below uses a simple geometric metric and is invariant when zooming in to the level of residues.

Despite the challenges associated with interpreting medium-resolution density maps, over 800 atomic models have already been derived from such data (based on the statistics from 2020), mostly through fitting an existing X-ray structure followed by refinement. Since each atomic model often involves multiple chains, and each chain, in turn, often has multiple helices and $\beta$-sheets, the large number of map/model pairs provides a solid data set for quantitative analysis. Since helices are the more distinctive secondary structural feature and are visible across the entire medium-resolution range, this study proposes a quantitative score to evaluate cylindrical similarity for helix regions in a medium-resolution density map. Owing to the diverse 3D reconstruction and modeling approaches used in cryo-EM and a resulting multitude of potential algorithmic and experimental limitations, the similarity score was designed to be agnostic of the origin of a specific mismatch.

## 7.2 DATASET

On September, 2018, 654 medium-resolution cryo-EM density maps are downloaded with corresponding atomic models from the EMDB and PDB, respectively. For a protein with multiple copies of the same chain sequence, only one was included to eliminate redundancies. The final data set consisted of 3,247 protein chains. The cryo-EM density region corresponding to each chain was extracted from the entire density map using UCSF Chimera with a radius of 5 Å around each atom. Because the method was designed to measure helices, chains without helices were excluded. Chains that lay completely outside any molecular density were also excluded.

## 7.3 METHODOLOGY

### 7.3.1 Cylindrical Similarity Score

The density distribution of a helix closely resembles a cylinder at medium resolution, with highest densities found near the central axis of the helix. The similarity of an atomic model was quantified using two suitably chosen template cylinders, derived from the central line of a helix using C$\alpha$ atoms (Figure 30 C). The central line was produced using the AxisComparison tool [64] from an atomic model in PDB format. In the AxisComparison method, every four consecutive C$\alpha$ atoms of a helix are averaged to generate initial central points, which are interpolated to produce a smooth line (Figure 30 C). The radius was selected as 2.5 Å and 4 Å for the inner and outer cylinders, respectively, to approximate the radius of the helix backbone and a typical radial size of an $\alpha$-helix. The 3 Å and 5 Å values for inner and outer cylinder radii, respectively, were also explored, but it is found that the use of 4 Å radius for the outer cylinder reduced the density overlap from surrounding non-helix density. At each density threshold, the number of helix density voxels within the inner cylinder, VxInner, measured the (ideally large) volume of the intersection between the helix density and the model (Figure 30 C). The number of helix density voxels between the inner and outer cylinder, VxOut, measures the volume of identifiable helix voxels outside of the helix backbone model (denoted ExDen in Figure 30 C). The number of inner cylinder voxels that have lower densities than the threshold, measures the volume where the atomic helix backbone model was not supported by sufficient map density (denoted ExMod in Figure 30 C).

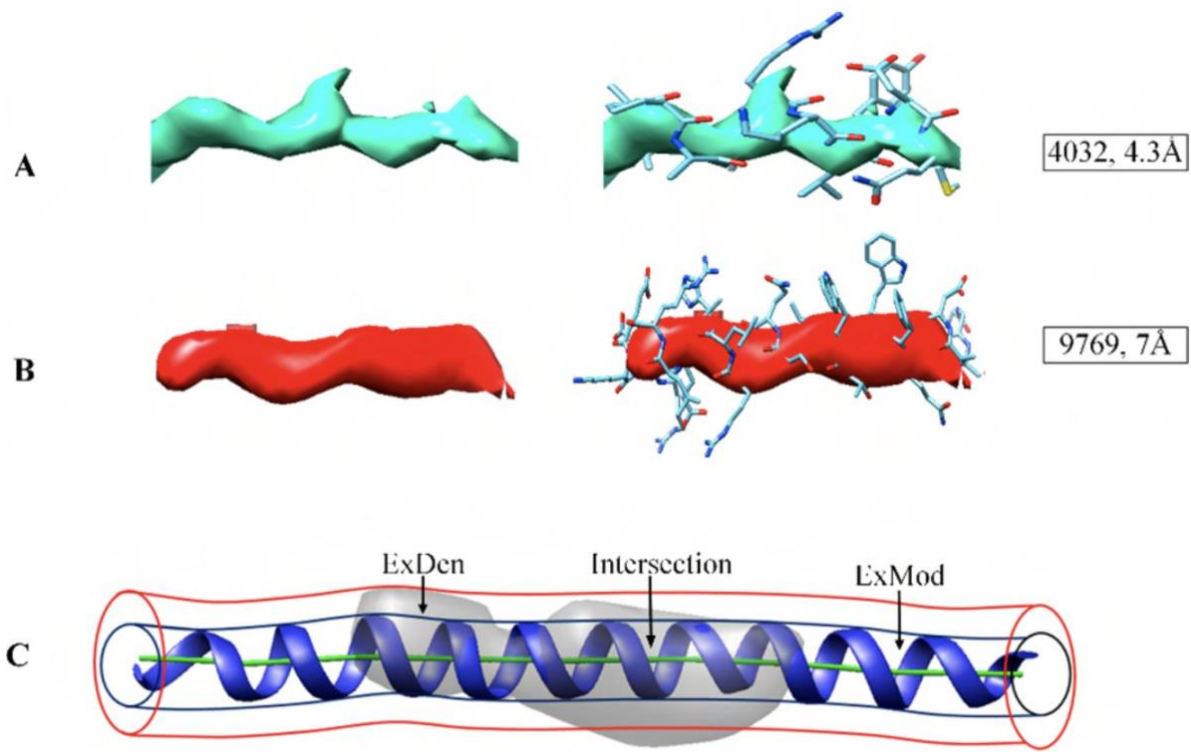

**Fig. 30.** Examples of helix densities at different resolutions and their cylindrical similarity: (A) density for a helix in EMDB ID 4032 with 4.3 Å resolution superimposed on the atomic model; (B) density for a helix in EMDB ID 9769 with 7 Å resolution superimposed on the atomic model; (C) two template cylinders of 2.5 Å and 4 Å radii, respectively, were used to measure the cylindrical similarity. Part (C) shows the intersection of map and model and two mismatched regions, ExDen and ExMod (where density is not accounted for by the model or vice versa).(Reproduced from [61])

$$F_1 = \frac{2 \times Pden \times Rmod}{Pden + Rmod} \tag{24}$$

$$Pden = \frac{Intersection}{Intersection + ExDen} = \frac{VxInner}{VxInner + VxOut} \tag{25}$$

$$Rmod = \frac{Intersection}{Intersection + ExMod} = \frac{VxInner}{VxInner + \overline{VxInner}} \tag{26}$$

Density thresholds were sampled between the mean and the maximum of the density map, and the threshold that maximized the F1 score (eq 24) was selected automatically.

The F1 score is a commonly used metric for assessing the performance of a machine learning model. Our adaptation of the F1 score (eq 24) was adapted from the standard interpretation in statistics where the F1 score is the harmonic mean of precision and recall. Although it is typically used to compare ground truth and prediction, the F1 score in our implementation measured similarity between two sets: the density region (above a threshold) in the vicinity of a helix and the region of a helix backbone model. Pden (eq 25) represents the accuracy of the helix density, i.e., the percentage of agreed map/model volume among the total volume of map relevant to the helix. Rmod (eq 26) represents the accuracy of the model, i.e., the percentage of agreed map/model volume among the total volume relevant to the helix backbone model. F1 scores lie between zero and one, where one indicates perfect similarity between the two sets.

When the resolution is high, such as at about 4 Å, the coil of the helix backbone starts to become visible in cryo-EM density maps ( Figure 30 A): the helix backbone exhibits a higher density than the side chains. The higher density of the helix backbone thus requires a density threshold for visualization that obscures the expected side chain regions (Figure 30 A). At medium resolution, however, instead of the coil of the backbone, only a cylindrically shaped density is observed (Figure 30B): the highest density voxels are often located near the central axis of the cylinder. Two cylinders are used to measure the cylindrical character of the density along a helix that is represented by the central axis of a helix atomic model. The inner radius of 2.5 Å was designed to capture the backbone density of a helix: a helix with good map/model similarity is expected to have a density threshold at which the density is primarily located within the inner cylinder of 2.5 Å radius.

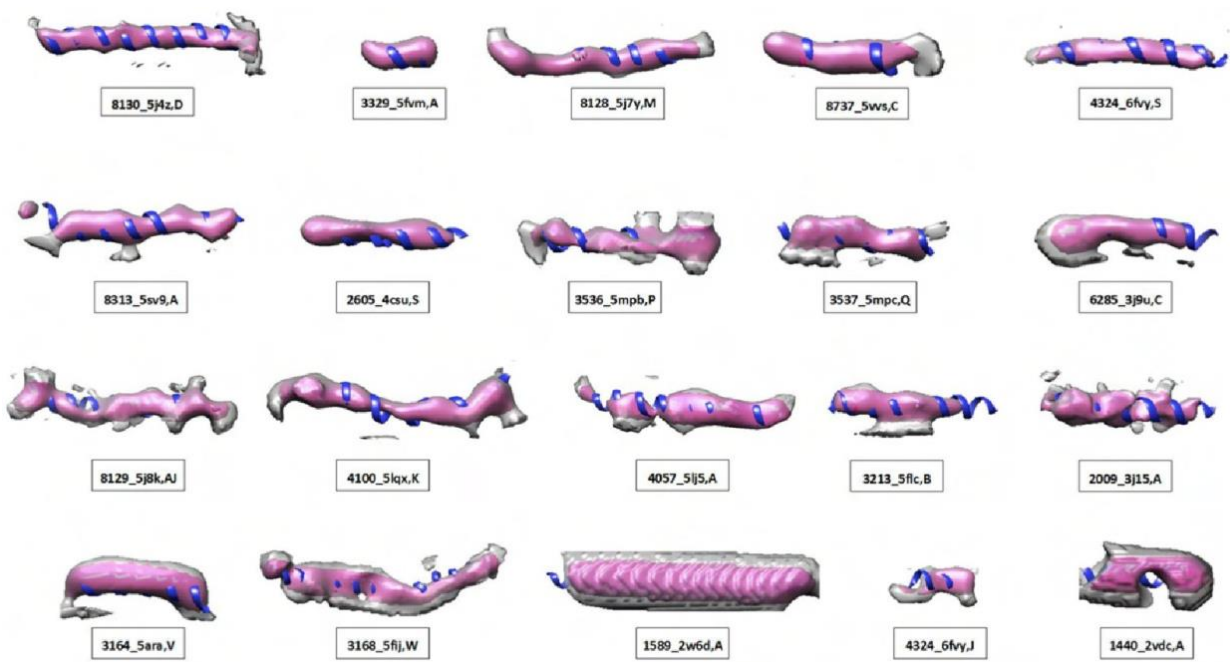

**Fig. 31.** Twenty examples of helical map/model pairs with F1 similarity scores (see Table 10). The atomic structure of a helix (ribbon) is superimposed on the density extracted using 4 Å radius (pink) and 7 Å radius (gray, to visualize the immediate neighborhood), respectively, around the central axis of the helix. Each helix density is displayed using the threshold that optimizes the F1 score. Panels are sorted by the F1 score from top left to lower right (see details in Table 10).(Reproduced from [61])

The cylindrical similarity of helices was evaluated for 30,994 helices from 3,247 protein chains corresponding to maps with resolutions between 5 Å and 10 Å. Each helix region in Figure 31 was shown using the automatically selected density threshold that maximizes the F1 score. The observed F1 scores varied considerably, and those helices with higher F1 scores appear more cylindrical through the entire length of the model. The F1 score also appears to stratify the data as intended. As an example, the helix located on chain D of PDB ID 5j4z between amino acid segment 131 to 161 appeared to have a good cylindrical shape (top left in Fig 31). The maximum F1 score of the helix was 0.863 when the density threshold was at 0.159, at which the inner cylinder was maximally filled, while the density volume between inner and outer cylinder was minimized.

**Table 10.** Cylindrical similarity evaluation scores (i.e., F1 score) of 20 helices.(Reproduced from [61]

| ID | Ch | segment | F1 | P | R | T | R |
|---|---|---|---|---|---|---|---|
| 8130 (5j4z) | D | 131-161 | 0.863 | 0.863 | 0.863 | 0.159 | 5.8 |
| 3329 5fvm | A | 884-891 | 0.833 | 0.851 | 0.817 | 0.129 | 6.7 |
| 8128 5j7y | M | 63-85 | 0.794 | 0.75 | 0.845 | 0.137 | 6.7 |
| 8737 5vvs | C | 26-39 | 0.751 | 0.741 | 0.76 | 0.039 | 6.4 |
| 4324 6fvy | S | 306-325 | 0.741 | 0.817 | 0.678 | 0.03 | 6.1 |
| 8313 5sv9 | A | 161-188 | 0.738 | 0.769 | 0.709 | 165.048 | 5.9 |
| 2605 4csu | S | 43-60 | 0.717 | 0.696 | 0.739 | 0.105 | 5.5 |
| 3536 5mpb | P | 334-356 | 0.698 | 0.613 | 0.809 | 0.039 | 7.8 |
| 3537 5mpc | Q | 253-273 | 0.688 | 0.681 | 0.695 | 0.046 | 7.7 |
| 6285 3j9u | C | 597-616 | 0.675 | 0.625 | 0.734 | 0.036 | 7.6 |
| 8129 5j8k | AJ | 16-47 | 0.668 | 0.574 | 0.797 | 0.096 | 7.8 |
| 4100 5lqx | K | 46-75 | 0.662 | 0.592 | 0.75 | 0.147 | 7.9 |
| 4057 5lj5 | A | 841-870 | 0.646 | 0.591 | 0.712 | 0.035 | 10 |
| 3213 5flc | B | 1736-1762 | 0.632 | 0.658 | 0.608 | 0.085 | 5.9 |
| 2009 3j15 | A | 176-198 | 0.621 | 0.604 | 0.64 | 0.132 | 6.6 |
| 3164 5ara | V | 29-47 | 0.618 | 0.515 | 0.774 | 0.192 | 7.4 |
| 3168 5fij | W | 138-182 | 0.597 | 0.465 | 0.831 | 0.159 | 7.4 |
| 1589 2w6d | A | 5-44 | 0.537 | 0.392 | 0.854 | 0.243 | 9 0 |
| 4324 6fvy | J | 131-141 | 0.452 | 0.365 | 0.592 | 0 | 6.1 |
| 1440 2vdc | A | 619-633 | 0.41 | 0.326 | 0.553 | 35.314 | 9.5 |

[a] Columns from left to right are (A) EMDB ID, PDB ID; (B) chain ID, amino acid segment of the helix, the best F1 score; (C) Pden, the accuracy of helix density; (D) Rmod, the accuracy of helix backbone model; (E) the density threshold that maximizes the F1 score; and (F) resolution of the density map

When a helix was not clearly distinguishable from the surrounding density in some spots, such as in EMDB ID 3536 chain P segment 334–356 (middle of the second row in Figure 31), Pden was lower since the indistinguishable density was reflected in VxOut. In this case, the accuracy of the helix density was 0.613, and the F1 score was 0.698 (Table 10 ). When a misalignment occurred between helix density and its atomic model, such as in the helix in EMDB ID 3164 chain V from amino acid 29 to 47 (first example in the bottom row in Figure 31), the shifted density contributed to reduced VxInner and increased VxOut. The resulting F1 score was 0.618 in this case (Table 10).

The histogram of helix F1 scores shows that the most frequent F1 scores were at about 0.55 with about 1,600 helices (Figure 32 A). Populations with F1 scores less than 0.55 sharply reduced as the score decreased. Three examples with F1 scores lower than 0.55 show a poor match with 0.452, 0.537, and 0.410 F1 scores (Table 10), respectively (right three examples in the last row in Figure 10). The histogram of helix F1 scores suggests that below an F1 score of 0.55, there is a poor map/model similarity of helices.

An averaged $F_1$ score was calculated at a chain level to represent the overall density fitting at multiple helices in a chain. In some chains, it is possible to have helices with different levels of cylindrical similarity. As an example, in Chain 2 of PDB ID 5ln3, the density regions at five helices have F1 scores of 0.761, 0.769 (green in Figure 29, 0.631, 0.700 (orange in Figure 29), and 0.567 (red in Figure 29), respectively. The F1 score of a chain was calculated as a weighted average of F1 scores of all helices, in which the weights were derived by the lengths of helices. Similar asymmetric distribution of F1 scores between the left and the right side of the most popular F1 scores was also observed at the chain level. The most popular score for a chain was between 0.54 and 0.56 (Figure 32 B). The results suggest that chains with F1 scores lower than 0.54 exhibited poor similarity. The scores at the helix level varied greatly, with minimum and maximum F1 scores of 0.171 and 0.848, respectively.

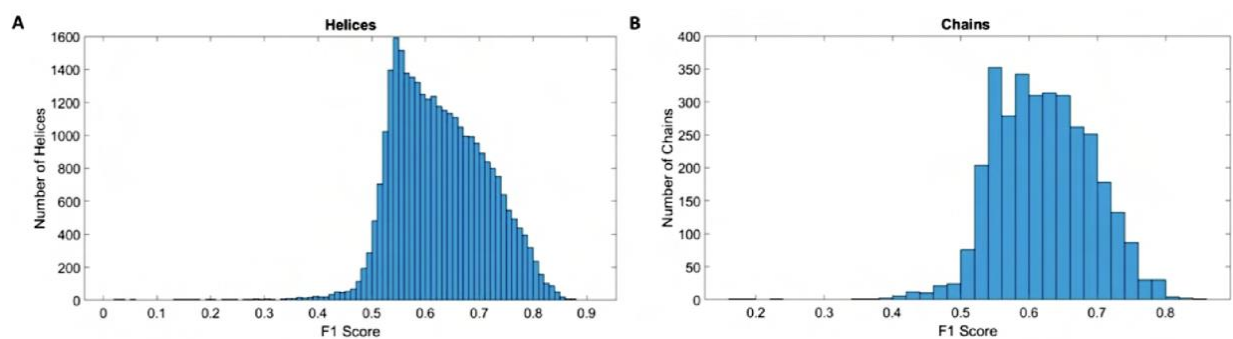

**Fig. 32.** Histograms of F1 scores in medium-resolution cryo-EM maps: (A) the histograms of F1 scores for 30,994 helices and (B) the histograms of F1 scores for 3,247 protein chains.(Reproduced from [61])

The F1 scores are agnostic of the origin of dissimilarities; both errors in maps and in models can lower the score. To compare with another established local measure, local resolution maps using the MonoRes tool [108] are created for the 20 helices that were analyzed using cylindrical similarity (shown in Figure 31). MonoRes focuses exclusively on the map quality, and not on the atomic model, so a similar results is not expected.

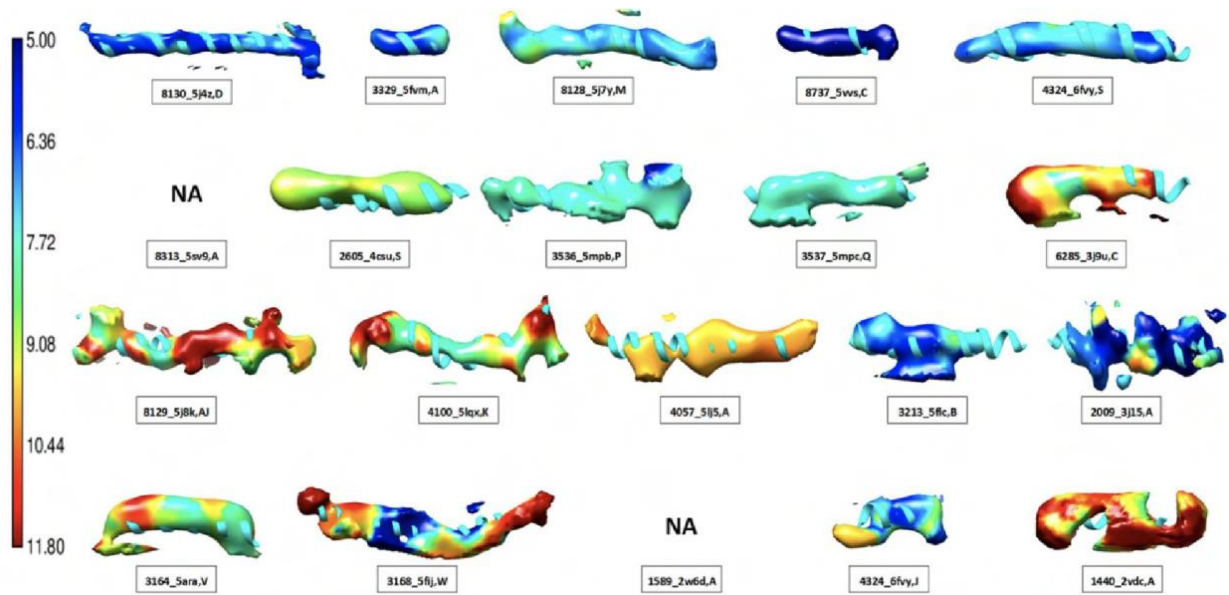

**Fig. 33.** The local resolution of 18 helix regions. Density regions near helices were extracted using a cylinder of 7 Å in radius from the central axis of the corresponding helix model (ribbon). Local resolutions produced using MonoRes [108] were used to color the density according to the resolution bar (left). The EMDB ID, PDB ID, and chain ID are provided for each helix. Refer to Table 10 for their amino acid segment and the F1 score in each case. The threshold that maximizes the F1 score was used in displaying density. Panels are sorted by the F1 score such that the upper-left helix has the highest F1 score and the lower-right has the lowest. (Reproduced from [61])

To visualize local resolutions in the vicinity of a helix, density regions were extracted using 7 Å radial distance from the central axis determined by AxisComparison. MonoRes produced local resolutions for 18 of the 20 EMDB map entries. All 18 cases were displayed at the density thresholds that maximize the F1 score in each case. Although local resolution and cylindrical similarity measure different properties at different levels of detail, the two scores share certain similarities. It is observed that the best F1 scores were often associated with more uniform local resolutions at the helix region. The five highest F1 scores of the 18 cases (top row in Figure 33) had F1 scores between 0.741 and 0.863. The density at the five helix regions had similar local resolutions of about 5 Å to 7.5 Å (darker blue to light blue in Figure 33). This suggests that helices with the best F1 scores tended to have similar local resolutions near the high end of the medium-resolution range. It is also observed that high F1 scores were not always associated with high local resolutions. For example, EMDB ID 2605 chain S (second example of the second row in Figure 33) had a good cylindrically shaped density, with an F1 score of 0.717, whereas voxels on the surface showed about 9 Å resolution (Figure 33). This suggests that cylindrical helix density can be detected accurately in good quality cryo-EM maps, even near the low end of the medium-resolution range. The two cases with the lowest F1 scores (right two on the bottom row of Figure 33) had missing density regions and hence were weak in cylindrical character. For the helix in EMDB ID 1440, which had an F1 score of 0.410, many voxels on the surface exhibited low-resolution values of about 10 Å (Figure 33). In the case of the helix in EMDB ID 4324, the F1 score was 0.452, and voxels with resolutions of about 6 and 10 Å were observed.

The F1 score measures the cylindrical similarity between the density and the model at a helix region. This is reflected in the results, showing that the highest scoring densities, such as the top 5 (row 1 in Figure 33), were cylindrical in shape and associated with higher local resolution that is also spatially homogeneous. Those with lower F1 scores deviated from a cylinder shape and were mostly associated with lower or spatially fluctuating local resolution. F1 scores collectively compare the density with the model of a helix. Local resolution varies from voxel to voxel, and for poor similarity map/model pairs, it is observed that there could be as much as a 5 Å difference in local resolution within the same helix region.

## 7.4 DISCUSSION AND CONCLUSION

Quantifying the local fit at the secondary structure level is important for validating medium (5–10 Å) resolution density maps and their associated atomic models. Earlier local measures focused mainly on the amino acid residue level, which is applicable for higher resolutions where side chain densities are distinguishable. The CC score between the experimental map and a simulated map (based on the model) or the map value at specific side chain atoms was used in earlier work. It is more challenging to measure the local fit in medium-resolution maps since density features at the secondary structure level are less reliable than individual atoms. The proposed F1 score method is a shape-based geometric measure that does not use CC, and hence it avoids the chal-lenge of producing a realistic expected density map for secondary structures. It also ignores minor differences in side chain appearance and focuses on the shape of a helix backbone that resembles a cylinder. This idea is based on our experience that side chain features are not distinguishable at medium resolution.

Cryo-EM density maps at medium resolution provide significant challenges for interpretation. With a growing number of atomic models derived from such density maps, using both template structures and fitting, it is important to understand how well models fit in reliable density regions of such maps. Since helices often show as cylindrical density in medium resolutions, a measure of cylindrical similarity is proposed for each helix using the natural shape of a helix backbone. The proposed method was applied to 30,994 helices in medium-resolution density maps and shown to be effective in distinguishing different levels of cylindrical similarity. The best F1 scores tended to be associated with spatially uniform local resolution of about 5 Å to 7.5 Å. Due to its discriminating power, the F1 score is a potential optimization criterion for the local fitting and refinement of an atomic model at a helix for medium-resolution maps.

In this study, the F1 similarity score was designed to be agnostic of the origin of a specific structural mismatch (regions ExMod or ExDen in Figure 30 C). This simple design of our score is both an advantage and a limitation. It is an advantage because the measure can be used to quickly extract good matching map/model pairs from the EMDB database without being concerned about lower scoring cases. It is a limitation because, at present, regions ExMod or ExDen are not distinguished in Figure 30C for lower levels of similarity (although our Pden and Rmod accuracies in eqs 25 and 26 could provide such a differentiation in future work).

Without knowing the ground truth, it is often difficult to speculate whether the density or model is more correct in the lower-scoring cases. For example, algorithmic or conceptual limitations, such as imprecise local refinement or flexible fitting and the use of a global fitting criterion, can place helix fragments into densities that do not support the placement (ExMod). However, heterogeneity of the biomolecular structure, flexibility, or libration of the specimen could conspire to reduce the alignment of corresponding image intensities such that an aggressively modeled helix might be more accurate than the weak cryo-EM density in that region. Since models are created with very diverse algorithmic approaches, it is unlikely that a single criterion will do justice to all lower-scoring map/model pairs. However, a low F1 score could be a starting point for further study of pathological cases detected by it.

# CHAPTER 8

# SUMMARY OF CONTRIBUTIONS AND FUTURE WORK

## 8.1 SUMMARY OF CONTRIBUTIONS

### 8.1.1 Actin Filament Tracing

Understanding the structure and functionality of the actin filament is highly important as it forms the dynamic cytoskeleton, which gives structural support to cells and links the interior of the cell with its surroundings. However, due to low SNR and missing wedge artifacts of cryo-ET image and dynamic shape and directional characteristics of actin filaments, there remain challenges to accurately determining the structure of actin filaments. The low-resolution cryo-ET image that suffers from a low SNR ratio and missing wedge artifacts limits the efficacy of traditional image processing techniques. Besides, actin filament exhibits multifarious types of shapes and structures, which mandates multiple strategies to successfully locate them.

The existing works related to actin filament tracing have several limitations, including high computational demands, a requirement of heavy manual intervention, and incomplete framework (e.g., only pixel level segmentation); my dissertation aims to address such issues. The primary contributions of this dissertation are the introduction specially designed computational tools for the actin filament tracing of different natures-

- Task 1: Actin bundle detection in shaft region of Stereocilia

- Task 2: Actin filament tracing in tapper region of Stereocilia

- Task 3: Tracing randomly oriented filaments in actin network

Beyond the development of innovative methodologies, the practical utility of these methodologies are ensured by translating them into user-friendly open-source software packages. These softwares provide biophysicists and biologists with powerful and accessible tools for conducting quantitative analyses on various types of actin filaments within 3D cryo-electron tomography images. This dual approach of methodological innovation and software implementation amplifies the impact of our research, making it directly applicable to real-world scientific investigations. Our implementation of filament tracking frameworks within this open-source software offers several key advantages. Firstly, it simplifies the often complex and specialized process of analyzing actin filaments in cryo-electron tomography images, making it accessible to a broader range of researchers and facilitating collaboration across disciplines. Secondly, it allows for the efficient and accurate tracking of filaments in three-dimensional space, providing quantitative insights that were previously challenging to obtain. This enables biophysicists and biologists to not only study the morphology and organization of actin filaments but also delve deeper into their functional roles and interactions within cellular contexts. Overall, our software represents a valuable contribution to the scientific community, empowering researchers to explore the intricate world of actin filaments in three-dimensional cryo-electron tomography images with precision and ease.

### 8.1.2 Segmentation and Quality Assessment of Protein Secondary Structure

In addition to the task of actin filament tracing, the dissertation also delves into research pertaining to protein secondary structures. Specifically, it focuses on identifying two main protein secondary structures: the $\alpha$-helix and the $\beta$-sheet, within medium-resolution cryo-electron microscopy (cryo-EM) images and assessing the quality of $\alpha$-helix.

To solve a significant problem in structural biology, the protein secondary structure detection, a 3D convolutional neural network (CNN) based approach was designed. The accurate segmentation of protein secondary structure elements within Cryo-EM images is pivotal for understanding the atomic structure of proteins. By leveraging the masked chain density in the training data, the proposed framework captures a richer context of protein secondary structure and obtains an impressive F1 scores of 0.76 for helix detection and 0.60 for $\beta$-sheet detection in the testign data demonstrate the accuracy and reliability of the proposed framework. The proposed framework yields similar performance to SSETracer, an algorithmic rule-based tool, which requires user intervention for threshold selection and third-party software for skeletinization. The proposed method is a valuable addition that enhances the capabilities of cryo-EM image analysis and, can, contributes to the broader goal of deciphering the atomic structures of proteins. This is particularly significant in structural biology and drug discovery, where an accurate understanding of protein structures is essential for advancing scientific knowledge and developing therapeutic interventions.

For density map and atomic model validation in the medium-resolution image, traditional methods relying on Cross-Correlation (CC)-based scores often prove less effective. These methods were originally designed for higher resolutions with distinguishable side chain densities, but they face challenges in the medium-resolution range where secondary structure features become less reliable. In addressing this challenge, a novel approach known as the F1 score method is introduced. Unlike CC-based techniques, the F1 score method utilizes shape-based geometric measures, rendering it a valuable tool for the quantification of cylindrical similarity in helical structures, which are commonly observed in medium-resolution density maps. This method excels, particularly in regions where local resolution is spatially uniform, typically ranging from 5 Å to 7.5 Å, and has the potential to serve as a key criterion for optimizing atomic model fitting within helical structures in medium-resolution maps.

## 8.2 FUTURE WORK

To enhance the performance of the Spaghetti Tracer, the central emphasis will be on improving its computational efficiency. The main bottleneck in Spaghetti Tracer is the time-consuming filamentous patterns enhancement step, which requires precise sampling of the tomogram and the computation of forward and backward path densities for all the voxels leading to a slowdown in the overall process.

To improve the efficiency of this step, two approaches can be considered:

- Parallelization: The computational time needed for Spaghetti Tracer can be significantly decreased by incorporating parallelization. This is possible because the computation of a voxel's path density depends solely on the original density within a cuboidal region of dimension, $(2 \times l, 2 \times l, 2 \times l)$, where $l$ is the length of the path. The local path density accumulation scheme for individual voxels enables the ignoring of densities outside their cuboidal regions and allows parallel processing.

- Super voxels: An alternative approach could be exploring the concept of super voxel (i.e., a cube consisting of a collection of voxels) instead of individual voxels. Super-voxels encapsulate multiple voxels to create larger cubes, such as two or three voxels in each direction. Then, for each superpixel, path density can be computed and assigned to all the voxels it contains. Although the path density value received for a voxel will be a bit different compared to its path density values originating from itself, this modification has minimal impact on the final results while potentially speeding up the process. Future research will delve deeper into these techniques to improve the efficiency of the Spaghetti Trace

Within the context of Struwwel Tracer, there remains a need for a comprehensive exploration of the concept of averaging. Particularly, the amalgamation of forward and backward traces presents a significant challenge when dealing with randomly oriented filaments. Subsequent endeavors will be directed toward addressing this issue, with the aim of enhancing the depiction of filamentous patterns in the image.

Regarding the protein secondary structure detection, there is still rooms for improving its efficacy. One strategy could be expanding the size of the dataset used for training and testing. Cur-rently, our work relies on a moderate dataset, but incorporating a more extensive dataset can bolster the robustness and generalizability of the proposed framework. Another promising avenue is integrating secondary structure sequence information into the dataset. By harnessing known secondary structure sequence data, the detection process can benefit from prior knowledge, potentially resulting in more precise secondary structure detection.

For a comprehensive quality assessment of protein secondary structures, expanding the F1-based formula to encompass the assessment of $\beta$-sheet quality could be an important step. $\beta$-sheets constitute a substantial portion of protein secondary structures alongside $\alpha$-helices, and assessing their quality is pivotal for understanding overall protein structure. While the current F1-based formula addresses $\alpha$-helices, extending it to measure the F1-score of $\beta$-sheets would provide a more comprehensive evaluation of protein secondary structure quality.

# REFERENCES

[1] Dimchev, G., Amiri, B., Fäßler, F., Falcke, M. & Schur, F. K. Computational toolbox for ultrastructural quantitative analysis of filament networks in cryo-ET data. *Journal of Structural Biology* **213**, 107808 (2021).

[2] Hudspeth, A. How hearing happens. *Neuron* **19**, 947–950 (1997).

[3] LeMasurier, M. & Gillespie, P. G. Hair-cell mechanotransduction and cochlear amplification. *Neuron* **48**, 403–415 (2005).

[4] Kwan, T., White, P. M. & Segil, N. Development and regeneration of the inner ear: Cell cycle control and differentiation of sensory progenitors. *Annals of the New York Academy of Sciences* **1170**, 28–33 (2009).

[5] Petit, C. & Richardson, G. P. Linking genes underlying deafness to hair-bundle development and function. *Nature neuroscience* **12**, 703–710 (2009).

[6] Barr-Gillespie, P.-G. Assembly of hair bundles, an amazing problem for cell biology. *Molecular biology of the cell* **26**, 2727–2732 (2015).

[7] Pacentine, I., Chatterjee, P. & Barr-Gillespie, P. G. Stereocilia rootlets: Actin-based structures that are essential for structural stability of the hair bundle. *International journal of molecular sciences* **21**, 324 (2020).

[8] Bornschlögl, T. *et al.* Filopodial retraction force is generated by cortical actin dynamics and controlled by reversible tethering at the tip. *Proceedings of the National Academy of Sciences* **110**, 18928–18933 (2013).

[9] Mattila, P. K. & Lappalainen, P. Filopodia: molecular architecture and cellular functions. *Nature reviews Molecular cell biology* **9**, 446–454 (2008).

[10] Lučić, V., Förster, F. & Baumeister, W. Structural studies by electron tomography: from cells to molecules. *Annu. Rev. Biochem.* **74**, 833–865 (2005).

[11] Song, J. *et al.* A cryo-tomography-based volumetric model of the actin core of mouse vestibular hair cell stereocilia lacking plastin 1. *Journal of structural biology* **210**, 107461 (2020).

[12] Pegoraro, A. F., Janmey, P. & Weitz, D. A. Mechanical properties of the cytoskeleton and cells. *Cold Spring Harbor perspectives in biology* **9**, a022038 (2017).

[13] Tang, D. D. & Gerlach, B. D. The roles and regulation of the actin cytoskeleton, intermediate filaments and microtubules in smooth muscle cell migration. *Respiratory research* **18**, 1–12 (2017).

[14] Sun, B. *et al.* Actin polymerization state regulates osteogenic differentiation in human adipose-derived stem cells. *Cellular & Molecular Biology Letters* **26**, 1–17 (2021).

[15] Khan, A. U., Qu, R., Fan, T., Ouyang, J. & Dai, J. A glance on the role of actin in osteogenic and adipogenic differentiation of mesenchymal stem cells. *Stem Cell Research & Therapy* **11**, 1–14 (2020).

[16] Desouza, M., Gunning, P. W. & Stehn, J. R. The actin cytoskeleton as a sensor and mediator of apoptosis. *Bioarchitecture* **2**, 75–87 (2012).

[17] Fuchs, E. & Cleveland, D. W. A structural scaffolding of intermediate filaments in health and disease. *Science* **279**, 514–519 (1998).

[18] Fuchs, E. & Weber, K. Intermediate filaments: structure, dynamics, function and disease. *Annual review of biochemistry* **63**, 345–382 (1994).

[19] Szeverenyi, I. *et al.* The human intermediate filament database: comprehensive information on a gene family involved in many human diseases. *Human mutation* **29**, 351–360 (2008).

[20] Soltys, B. J. & Gupta, R. S. Interrelationships of endoplasmic reticulum, mitochondria, intermediate filaments, and microtubules—a quadruple fluorescence labeling study. *Biochemistry and Cell Biology* **70**, 1174–1186 (1992).

[21] Egelman, E., Francis, N. & DeRosier, D. F-actin is a helix with a random variable twist. *Nature* **298**, 131–135 (1982).

[22] Gan, L. & Jensen, G. J. Electron tomography of cells. *Quarterly reviews of biophysics* **45**, 27–56 (2012).

[23] Beck, M. & Baumeister, W. Cryo-electron tomography: can it reveal the molecular sociology of cells in atomic detail? *Trends in cell biology* **26**, 825–837 (2016).

[24] Al-Amoudi, A. *et al.* Cryo-electron microscopy of vitreous sections. *The EMBO journal* **23**, 3583–3588 (2004).

[25] Villa, E., Schaffer, M., Plitzko, J. M. & Baumeister, W. Opening windows into the cell: focused-ion-beam milling for cryo-electron tomography. *Current opinion in structural biology* **23**, 771–777 (2013).

[26] Lučić, V., Rigort, A. & Baumeister, W. Cryo-electron tomography: the challenge of doing structural biology in situ. *Journal of Cell Biology* **202**, 407–419 (2013).

[27] Castaño-Díez, D., Kudryashev, M., Arheit, M. & Stahlberg, H. Dynamo: a flexible, user-friendly development tool for subtomogram averaging of cryo-EM data in high-performance computing environments. *Journal of structural biology* **178**, 139–151 (2012).

[28] Tang, G. *et al.* EMAN2: an extensible image processing suite for electron microscopy. *Journal of structural biology* **157**, 38–46 (2007).

[29] Himes, B. A. & Zhang, P. emClarity: software for high-resolution cryo-electron tomography and subtomogram averaging. *Nature methods* **15**, 955–961 (2018).

[30] Noble, A. J. & Stagg, S. M. Automated batch fiducial-less tilt-series alignment in Appion using Protomo. *Journal of structural biology* **192**, 270–278 (2015).

[31] Hrabe, T. *et al.* PyTom: a python-based toolbox for localization of macromolecules in cryo-electron tomograms and subtomogram analysis. *Journal of structural biology* **178**, 177–188 (2012).

[32] Scheres, S. H. RELION: implementation of a bayesian approach to cryo-EM structure determination. *Journal of structural biology* **180**, 519–530 (2012).

[33] Nguyen, U. T., Bhuiyan, A., Park, L. A. & Ramamohanarao, K. An effective retinal blood vessel segmentation method using multi-scale line detection. *Pattern recognition* **46**, 703–715 (2013).

[34] Herberich, G. *et al.* Fluorescence microscopic imaging and image analysis of the cytoskeleton. In *2010 Conference Record of the Forty Fourth Asilomar Conference on Signals, Systems and Computers*, 1359–1363 (IEEE, 2010).

[35] Özdemir, B. & Reski, R. Automated and semi-automated enhancement, segmentation and tracing of cytoskeletal networks in microscopic images: A review. *Computational and Structural Biotechnology Journal* **19**, 2106–2120 (2021).

[36] Kervrann, C., Blestel, S. & Chrétien, D. Conditional random fields for tubulin-microtubule segmentation in cryo-electron tomography. In *2014 IEEE International Conference on Image Processing (ICIP)*, 2080–2084 (IEEE, 2014).

[37] Wriggers, W., Milligan, R. A. & McCammon, J. A. Situs: a package for docking crystal structures into low-resolution maps from electron microscopy. *Journal of structural biology* **125**, 185–195 (1999).

[38] Birmanns, S., Rusu, M. & Wriggers, W. Using Sculptor and Situs for simultaneous assembly of atomic components into low-resolution shapes. *Journal of structural biology* **173**, 428–435 (2011).

[39] Stalling, D., Westerhoff, M., Hege, H.-C. *et al.* Amira: A highly interactive system for visual data analysis. *The visualization handbook* **38**, 749–67 (2005).

[40] Rusu, M., Starosolski, Z., Wahle, M., Rigort, A. & Wriggers, W. Automated tracing of filaments in 3D electron tomography reconstructions using Sculptor and Situs. *Journal of structural biology* **178**, 121–128 (2012).

[41] Weber, B. *et al.* Automated tracing of microtubules in electron tomograms of plastic embedded samples of caenorhabditis elegans embryos. *Journal of structural biology* **178**, 129–138 (2012).

[42] Redemann, S. *et al.* C. elegans chromosomes connect to centrosomes by anchoring into the spindle network. *Nature communications* **8**, 1–13 (2017).

[43] Rigort, A. *et al.* Automated segmentation of electron tomograms for a quantitative description of actin filament networks. *Journal of structural biology* **177**, 135–144 (2012).

[44] Kovacs, J. *et al.* Correction of missing-wedge artifacts in filamentous tomograms by template-based constrained deconvolution. *Journal of chemical information and modeling* **60**, 2626–2633 (2020).

[45] Loss, L. A. *et al.* Automatic segmentation and quantification of filamentous structures in electron tomography. In *Proceedings of the ACM Conference on Bioinformatics, Computational Biology and Biomedicine*, 170–177 (2012).

[46] Chen, M. *et al.* Convolutional neural networks for automated annotation of cellular cryo-electron tomograms. *Nature methods* **14**, 983–985 (2017).

[47] Alioscha-Perez, M. *et al.* A robust actin filaments image analysis framework. *PLoS computational biology* **12**, e1005063 (2016).

[48] Herberich, G., Windoffer, R., Leube, R. & Aach, T. 3D segmentation of keratin intermediate filaments in confocal laser scanning microscopy. In *2011 Annual International Conference of the IEEE Engineering in Medicine and Biology Society*, 7751–7754 (IEEE, 2011).

[49] Basu, S., Condron, B., Aksel, A. & Acton, S. T. Segmentation and tracing of single neurons from 3D confocal microscope images. *IEEE journal of biomedical and health informatics* **17**, 319–335 (2013).

[50] Xu, T., Vavylonis, D. & Huang, X. 3D actin network centerline extraction with multiple active contours. *Medical image analysis* **18**, 272–284 (2014).

[51] Jiang, W., Baker, M. L., Ludtke, S. J. & Chiu, W. Bridging the information gap: computational tools for intermediate resolution structure interpretation. *Journal of molecular biology* **308**, 1033–1044 (2001).

[52] Dal Palu, A., He, J., Pontelli, E. & Lu, Y. Identification of alpha-helices from low resolution protein density maps. In *Computational Systems Bioinformatics*, 89–98 (World Scientific, 2006).

[53] Abeysinghe, S. S., Ju, T., Chiu, W. & Baker, M. Shape modeling and matching in identifying protein structure from low-resolution images. In *Proceedings of the 2007 ACM symposium on Solid and physical modeling*, 223–232 (2007).

[54] Baker, M. L., Ju, T. & Chiu, W. Identification of secondary structure elements in intermediate-resolution density maps. *Structure* **15**, 7–19 (2007).

[55] Si, D. & He, J. Beta-sheet detection and representation from medium resolution cryo-EM density maps. In *Proceedings of the International Conference on Bioinformatics, Computational Biology and Biomedical Informatics*, 764–770 (2013).

[56] Haslam, D., Zeng, T., Li, R. & He, J. Exploratory studies detecting secondary structures in medium resolution 3d cryo-EM images using deep convolutional neural networks. In *Proceedings of the 2018 ACM International Conference on Bioinformatics, Computational Biology, and Health Informatics*, 628–632 (2018).

[57] Ma, L., Reisert, M. & Burkhardt, H. RENNSH: A novel¥alpha-helix identification approach for intermediate resolution electron density maps. *IEEE/ACM Transactions on Computational Biology and Bioinformatics* **9**, 228–239 (2011).

[58] Li, R., Si, D., Zeng, T., Ji, S. & He, J. Deep convolutional neural networks for detecting secondary structures in protein density maps from cryo-electron microscopy. In *2016 IEEE International Conference on Bioinformatics and Biomedicine (BIBM)*, 41–46 (IEEE, 2016).

[59] Maddhuri Venkata Subramaniya, S. R., Terashi, G. & Kihara, D. Protein secondary structure detection in intermediate-resolution cryo-EM maps using deep learning. *Nature methods* **16**, 911–917 (2019).

[60] Wriggers, W. & He, J. Numerical geometry of map and model assessment. *Journal of structural biology* **192**, 255–261 (2015).

[61] Sazzed, S., Scheible, P., Alshammari, M., Wriggers, W. & He, J. Cylindrical similarity measurement for helices in medium-resolution cryo-electron microscopy density maps. *Journal of chemical information and modeling* **60**, 2644–2650 (2020).

[62] Kremer, J. R., Mastronarde, D. N. & McIntosh, J. R. Computer visualization of three-dimensional image data using IMOD. *Journal of structural biology* **116**, 71–76 (1996).

[63] Sazzed, S. *et al.* Tracing actin filament bundles in three-dimensional electron tomography density maps of hair cell stereocilia. *Molecules* **23**, 882 (2018).

[64] Zeil, S., Kovacs, J., Wriggers, W. & He, J. Comparing an atomic model or structure to a corresponding cryo-electron microscopy image at the central axis of a helix. *Journal of Computational Biology* **24**, 52–67 (2017).

[65] Pettersen, E. F. *et al.* UCSF Chimera—a visualization system for exploratory research and analysis. *Journal of computational chemistry* **25**, 1605–1612 (2004).

[66] Sazzed, S., Scheible, P., He, J. & Wriggers, W. Spaghetti tracer: A framework for Tracing semiregular filamentous densities in 3D tomograms. *Biomolecules* **12**, 1022 (2022).

[67] Scheible, P., Sazzed, S., He, J. & Wriggers, W. Tomosim: Simulation of filamentous cryo-electron tomograms. In *2021 IEEE International Conference on Bioinformatics and Biomedicine (BIBM)*, 2560–2565 (IEEE, 2021).

[68] Sazzed, S., Scheible, P., He, J. & Wriggers, W. Tracing filaments in simulated 3D cryo-electron tomography maps using a fast dynamic programming algorithm. In *2021 IEEE International Conference on Bioinformatics and Biomedicine (BIBM)*, 2553–2559 (IEEE, 2021).

[69] Rogge, H., Artelt, N., Endlich, N. & Endlich, K. Automated segmentation and quantification of actin stress fibres undergoing experimentally induced changes. *Journal of microscopy* **268**, 129–140 (2017).

[70] Starosolski, Z., Szczepanski, M., Wahle, M., Rusu, M. & Wriggers, W. Developing a de-noising filter for electron microscopy and tomography data in the cloud. *Biophysical reviews* **4**, 223–229 (2012).

[71] Narasimha, R. *et al.* Evaluation of denoising algorithms for biological electron tomography. *Journal of structural biology* **164**, 7–17 (2008).

[72] Moreno, J., Martínez-Sánchez, A., Martínez, J. A., Garzón, E. M. & Fernández, J.-J. To-moEED: fast edge-enhancing denoising of tomographic volumes. *Bioinformatics* **34**, 3776–3778 (2018).

[73] Schoenenberger, C.-A., Mannherz, H. G. & Jockusch, B. M. Actin: from structural plasticity to functional diversity. *European Journal of Cell Biology* **90**, 797–804 (2011).

[74] Imachi, H. *et al.* Actin cytoskeleton and complex cell architecture in an Asgard archaeon. *Nature* **577**, 519–525 (2020).

[75] Rodrigues-Oliveira, T. *et al.* Isolation of an archaeon at the prokaryote–eukaryote interface. *Nature* **613**, 332–339 (2023).

[76] Martins, B. *et al.* Unveiling the polarity of actin filaments by cryo-electron tomography. *Structure* **29**, 488–498 (2021).

[77] Schneider, J. & Jasnin, M. Capturing actin assemblies in cells using in situ cryo-electron tomography. *European Journal of Cell Biology* **101**, 151224 (2022).

[78] Smith, M. B. *et al.* Segmentation and tracking of cytoskeletal filaments using open active contours. *Cytoskeleton* **67**, 693–705 (2010).

[79] Image & Data Analysis Facility, D., Core Reseach Facilities. Yapic. https://yapic.github.io/yapic/ (2022). URL http://https://yapic.github.io/yapic/. Accessed on Jul 5, 2023.

[80] Sazzed, S., Scheible, P., He, J. & Wriggers, W. Untangling irregular actin cytoskeleton architectures in tomograms of the cell with struwwel tracer. *International Journal of Molecular Sciences* **24** (2023).

[81] Humphrey, W., Dalke, A. & Schulten, K. VMD: visual molecular dynamics. *Journal of Molecular Graphics* **14**, 33–38 (1996).

[82] Fäßler, F., Dimchev, G., Hodirnau, V.-V., Wan, W. & Schur, F. K. Cryo-electron tomography structure of Arp2/3 complex in cells reveals new insights into the branch junction. *Nature Communications* **11**, 6437 (2020).

[83] Lawson, C. L. *et al.* EMDatabank unified data resource for 3DEM. *Nucleic Acids Research* **44**, D396–D403 (2016).

[84] Kappel, K. *et al.* Accelerated cryo-EM-guided determination of three-dimensional RNA-only structures. *Nature methods* **17**, 699–707 (2020).

[85] Guo, J. *et al.* Structures of the calcium-activated, non-selective cation channel trpm4. *Nature* **552**, 205–209 (2017).

[86] Adams, P. D. *et al.* PHENIX: a comprehensive python-based system for macromolecular structure solution. *Acta Crystallographica Section D: Biological Crystallography* **66**, 213–221 (2010).

[87] Yu, Z. & Bajaj, C. Computational approaches for automatic structural analysis of large biomolecular complexes. *IEEE/ACM Transactions on Computational Biology and Bioinformatics* **5**, 568–582 (2008).

[88] Rusu, M. & Wriggers, W. Evolutionary bidirectional expansion for the tracing of alpha helices in cryo-electron microscopy reconstructions. *Journal of structural biology* **177**, 410–419 (2012).

[89] Kong, Y. & Ma, J. A structural-informatics approach for mining $\beta$-sheets: locating sheets in intermediate-resolution density maps. *Journal of molecular biology* **332**, 399–413 (2003).

[90] Haslam, D., Zubair, M., Ranjan, D., Biswas, A. & He, J. Challenges in matching secondary structures in cryo-EM: An exploration. In *2016 IEEE International Conference on Bioinformatics and Biomedicine (BIBM)*, 1714–1719 (IEEE, 2016).

[91] Ronneberger, O., Fischer, P. & Brox, T. U-net: Convolutional networks for biomedical image segmentation. In *Medical Image Computing and Computer-Assisted Intervention–MICCAI 2015: 18th International Conference, Munich, Germany, October 5-9, 2015, Proceedings, Part III 18*, 234–241 (Springer, 2015).

[92] Li, R. *et al.* Deep learning based imaging data completion for improved brain disease diagnosis. In *Medical Image Computing and Computer-Assisted Intervention–MICCAI 2014: 17th International Conference, Boston, MA, USA, September 14-18, 2014, Proceedings, Part III 17*, 305–312 (Springer, 2014).

[93] Li, Q. *et al.* A cross-modality learning approach for vessel segmentation in retinal images. *IEEE transactions on medical imaging* **35**, 109–118 (2015).

[94] Heinig, M. & Frishman, D. STRIDE: a web server for secondary structure assignment from known atomic coordinates of proteins. *Nucleic acids research* **32**, W500–W502 (2004).

[95] Mu, Y., Sazzed, S., Alshammari, M., Sun, J. & He, J. A tool for segmentation of secondary structures in 3D cryo-EM density map components using deep convolutional neural networks. *Frontiers in Bioinformatics* **1**, 51 (2021).

[96] Pettersen, E. F. *et al.* UCSF ChimeraX: Structure visualization for researchers, educators, and developers. *Protein Science* **30**, 70–82 (2021).

[97] Deng, Y., Mu, Y., Sazzed, S., Sun, J. & He, J. Using curriculum learning in pattern recognition of 3-dimensional cryo-electron microscopy density maps. In *Proceedings of the 11th ACM International Conference on Bioinformatics, Computational Biology and Health Informatics*, 1–7 (2020).

[98] Lopez-Paz, D. & Ranzato, M. Gradient episodic memory for continual learning. *Advances in neural information processing systems* **30** (2017).

[99] Çiçek, Ö., Abdulkadir, A., Lienkamp, S. S., Brox, T. & Ronneberger, O. 3D U-Net: learning dense volumetric segmentation from sparse annotation. In *Medical Image Computing and Computer-Assisted Intervention–MICCAI 2016: 19th International Conference, Athens, Greece, October 17-21, 2016, Proceedings, Part II 19*, 424–432 (Springer, 2016).

[100] Kingma, D. P. & Ba, J. Adam: A method for stochastic optimization. *arXiv preprint arXiv:1412.6980* (2014).

[101] Baker, M. L. *et al.* Modeling protein structure at near atomic resolutions with Gorgon. *Journal of structural biology* **174**, 360–373 (2011).

[102] Berman, H. M. *et al.* The protein data bank. *Nucleic acids research* **28**, 235–242 (2000).

[103] Lawson, C. L. *et al.* EMDatabank. org: unified data resource for CryoEM. *Nucleic acids research* **39**, D456–D464 (2010).

[104] Emsley, P., Lohkamp, B., Scott, W. G. & Cowtan, K. Features and development of Coot. *Acta Crystallographica Section D: Biological Crystallography* **66**, 486–501 (2010).

[105] Wriggers, W. Conventions and workflows for using Situs. *Acta Crystallographica Section D: Biological Crystallography* **68**, 344–351 (2012).

[106] Afonine, P. V. *et al.* New tools for the analysis and validation of cryo-EM maps and atomic models. *Acta Crystallographica Section D: Structural Biology* **74**, 814–840 (2018).

[107] Barad, B. A. *et al.* EMRinger: side chain–directed model and map validation for 3D cryo-electron microscopy. *Nature methods* **12**, 943–946 (2015).

[108] Vilas, J. L. *et al.* MonoRes: automatic and accurate estimation of local resolution for electron microscopy maps. *Structure* **26**, 337–344 (2018).

# VITA

Salim Sazzed

Department of Computer Science

Old Dominion University

Norfolk, VA 23529

## EDUCATION

2016-2023 (Expected), Doctor of Philosophy, Old Dominion University.

2009-2011, Masters of Science, University of Dhaka.

2004-2009, Bachelor of Science, University of Dhaka.

## PROFESSIONAL EXPERIENCE

2023-Now, Visiting Assistant Professor, University of Memphis.

## PUBLICATIONS

1. Sazzed, S., Scheible, P., He, J., & Wriggers, W., Untangling Irregular Actin Cytoskeleton Architectures in Tomograms of the Cell With Struwwel Tracer , International Journal of Molecular Sciences (IJMS), 2023.

2. Sazzed, S., Scheible, P., He, J., & Wriggers, W.; Spaghetti Tracer: A Framework for Tracing Semiregular Filamentous Densities in 3D Tomograms, Biomolecules, 2022.

3. Sazzed, S.; Scheible, P.; Alshammari, M.; Wriggers, W.; and He, J, Cylindrical Similarity Measurement for Helices in Medium-Resolution Cryo-Electron Microscopy Density Maps, Journal of Chemical Information and Modeling (JCIM), 2020.

4. Sazzed, S., Song, J., Kovacs, J.A., Wriggers, W., Auer, M., and He, J., Tracing actin filament bundles in three-dimensional electron tomography density maps of hair cell stereocilia, Molecules, 2018.

5. Mu, Y.; Sazzed, S.; Alshammari, M.; Sun, J., & He, J., A Tool for Segmentation of Secondary Structures in 3D Cryo-EM Density Map Components Using Deep Convolutional Neural Networks, Frontiers in Bioinformatics, 2021.

6. Song, J.; Patterson, R.; Metlagel, Z.; Krey, J.F.; Hao, S.; Wang, L.; Ng, B.; Sazzed, S.; Kovacs, J.; Wriggers, W., He, J., Barr-Gillespie P. G., and Auer, M. , A cryo-tomography-based volumetric model of the actin core of mouse vestibular hair cell stereocilia lacking plastin 1, Journal of Structural Biology (JSB), 2020.